\documentclass{aa} 

\usepackage{graphicx}
\usepackage{xcolor}

\usepackage{txfonts}
\usepackage{lscape}
\usepackage{natbib,twoopt}
\usepackage[breaklinks=true]{hyperref} 

\bibpunct{(}{)}{;}{a}{}{,}             

\newcommand{\feii}{\mbox{Fe\,{\sc ii}}}
\newcommand{\znii}{\mbox{Zn\,{\sc ii}}}
\newcommand{\crii}{\mbox{Cr\,{\sc ii}}}
\newcommand{\tiii}{\mbox{Ti\,{\sc ii}}}
\newcommand{\siii}{\mbox{Si\,{\sc ii}}}
\newcommand{\suii}{\mbox{S\,{\sc ii}}}

\newcommand{\hi}{\mbox{H\,{\sc i}}}
\newcommand{\hii}{\mbox{H\,{\sc ii}}}
\newcommand{\hh}{\mbox{H}$_2$}
\newcommand{\niii}{\mbox{Ni\,{\sc ii}}}

\newcommand{\cii}{\mbox{C\,{\sc ii}}}

\newcommand{\nii}{\mbox{N\,{\sc ii}}}
\newcommand{\oi}{\mbox{O\,{\sc i}}}

\newcommand{\pii}{\mbox{P\,{\sc ii}}}
\newcommand{\mgii}{\mbox{Mg\,{\sc ii}}}
\newcommand{\mnii}{\mbox{Mn\,{\sc ii}}}
\newcommand{\cuii}{\mbox{Cu\,{\sc ii}}}

\begin{document}

   \title{$\alpha$-element enhancements in the Magellanic Interstellar Medium:\\evidence for recent star formation}

\titlerunning{Chemistry of the Magellanic ISM}

    \author{Annalisa De Cia
          \inst{1,2}
           \and	 
             Julia Roman-Duval\inst{3} 
             \and	 
             Christina Konstantopoulou\inst{2}
             \and	
          Pasquier Noterdaeme\inst{4,5}  
                     \and 
             Tanita Ramburuth-Hurt\inst{2}  
            \and	
          Anna Velichko\inst{2,6}
              \and	
           Andrew J. Fox \inst{7,8}  
           \and	 
             C\'edric Ledoux\inst{9}  
               \and	 
             Patrick Petitjean \inst{5}            
           \and	          
            Iris Jermann\inst{2,10,11}
             \and		              
             Jens-Kristian Krogager\inst{2,12,13}             
          }

   \institute{European Southern Observatory, Karl-Schwarzschild Str. 2, 85748 Garching bei M\"unchen, Germany\\ 
              \email{adecia@eso.org}
              \and
             Department of Astronomy, University of Geneva, Chemin Pegasi 51, 1290 Versoix, Switzerland\\
                           \and
             Space Telescope Science Institute, 3700 San Martin Drive, Baltimore, MD21218, USA\\
                           \and       
                      Franco-Chilean Laboratory for Astronomy, IRL\,3386, CNRS \& Dep. de Astonom\'ia, Universidad de Chile, Santiago, Chile\\        
              \and
              Institut d’Astrophysique de Paris, UMR\,7095, CNRS \& Sorbonne Universit\'e, 98bis bd Arago, 75014 Paris, France\\           
               \and	
             Institute of Astronomy, Kharkiv National University, Sumska 35, Kharkiv, 61022, Ukraine\\ 	    
                      \and
             AURA for ESA, 3700 San Martin Drive, Space Telescope Science Institute, Baltimore, MD 21218, USA\\
              \and
              Department of Physics \& Astronomy, Johns Hopkins University, 3400 N. Charles Street, Baltimore, MD 21218, USA\\
                 \and
             European Southern Observatory, Alonso de C\'ordova 3107, Casilla 19001, Vitacura, Santiago, Chile\\ 
              \and
              Cosmic Dawn Center (DAWN)\\
              \and
              DTU-Space, National Space Institute, Technical University of Denmark, Elektrovej 327, 2800 Kgs. Lyngby, Denmark\\
              \and
             Observatoire de Lyon, 9 avenue Charles Andr\'{e}, 69561 St. Genis Laval, France\\ 
             \and
             Univ Lyon, Univ Lyon1, Ens de Lyon, CNRS, Centre de Recherche Astrophysique de Lyon (CRAL) UMR5574, F-69230 Saint-Genis-Laval, France\\
             }

   \date{ Received Month Day, yyyy; accepted Month Day, yyyy}

  \abstract
   {
Important questions on the chemical composition of the neutral interstellar medium (ISM) in the Large Magellanic Cloud (LMC) and Small Magellanic Cloud (SMC) are still open. It is usually assumed that their metallicity is uniform and equal to that measured in hot stars and \hii{} regions, but direct measurements of the neutral ISM metallicity have not been performed until now. Deriving the metallicity from the observed metal abundances is not straightforward because they also depend on the depletion of metals into dust as well as nucleosynthesis effects such as $\alpha$-element enhancement.}
   {We aim at measuring the metallicity of the neutral ISM in the LMC and SMC, dust depletion, and any nucleosynthesis effects.}
   {We collect literature column densities of \tiii{}, \niii{}, \crii{}, \feii{}, \mnii{}, \siii{}, \cuii{}, \mgii{},  \suii{}, \pii{}, \znii{}, and \oi{} in the neutral ISM towards 32 and $22$ hot stars in the LMC and SMC. We determine dust depletion from the relative abundances of different metals, because they deplete with different strengths. This includes a “golden sample” of sightlines where both Ti and other $\alpha$-elements are available. We fit linear relations to the observed abundance patterns, so that the slopes determine the strengths of dust depletion and the normalizations determine the metallicities. We investigate $\alpha$-element enhancements in the gas from the deviations from the linear fits and compare them with stars.}
   {In our golden sample we find $\alpha$-element enhancement in the neutral ISM in most systems, on average 0.26~dex (0.35~dex) for the LMC (SMC), and Mn under-abundance in the SMC (on average $-0.35$~dex). Measurements of \mnii{} are not available for the LMC. These are higher than for stars at similar metallicities. We find total neutral ISM metallicities that are mostly consistent with hot stars metallicity, on average [$M$/H]$_{\rm tot}=-0.33$ ($-0.83$), with standard deviations of 0.30 (0.30), in the LMC (respectively the SMC). In six systems, however, we find significantly lower metallicities, two out of 32 in the LMC (with $\sim16$\% solar) and four out of 22 in the SMC (3 and 10\% solar), two of which are in the outskirts of the SMC near the Magellanic Bridge, a region known for having a lower metallicity.}
  {The observed $\alpha$-element enhancements and Mn under-abundance are likely due to bursts of star formation, more recently than $\sim1$ Gyr ago, that enriched the ISM from core-collapse supernovae. With the exception of lines of sight towards the Magellanic Bridge, the neutral gas in the LMC and SMC appears fairly well mixed in terms of metallicity.}

   \keywords{ISM: abundances --
                Galaxies: Magellanic Clouds --
                ISM: dust \vspace{-2.0cm}
               }

   \maketitle
%

\section{Introduction}

   \begin{figure*}
   \centering
   \includegraphics[width=\textwidth]{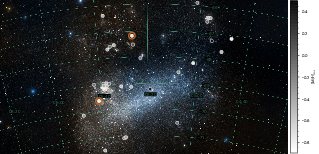}
      \caption{The location of the lines of sight in the LMC. The total metallicities in the neutral ISM along these lines of sight are highlighted with the gray scale. The systems with metallicities that are significantly lower than the nominal value (deviating by $3\sigma$), are highlighted in orange. The metallicities of other lines of sight are consistent with the nominal value from hot stars, within the uncertainties. The background image is from DSS2.}
         \label{fig LMC map}
   \end{figure*}

  \begin{figure*}
   \centering
   \includegraphics[width=\textwidth]{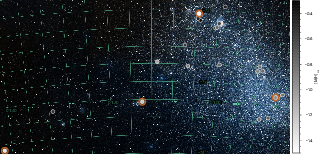}
      \caption{Same as Fig. \ref{fig LMC map}, but for the SMC.}
         \label{fig SMC map}
   \end{figure*}

The chemical evolution of galaxies is intimately related to the evolution of their stars and gas \citep[e.g.][]{Tinsley80}. The chemical properties of the interstellar medium (ISM) of galaxies can be studied in great detail from the rest-frame UV absorption lines of a plethora of metals observable in the spectra of bright background sources, such as stars in the Milky Way, Large Magellanic Cloud (LMC) and Small Magellanic Cloud (SMC).
Different processes shape the observed metal abundances of the neutral ISM: the overall metallicity, the presence of dust, the over-abundance or under-abundance of specific elements due to stellar nucleosynthesis, for example $\alpha$-element enhancement. Large fractions of metals are missing from the observable gas-phase because they are incorporated in dust grains, a phenomenon called dust depletion \citep{Field74,Savage96,Phillips82,Jenkins86,Jenkins09,DeCia16,Roman-Duval21}.

While the observed abundances of all metals naturally depends on the intrinsic overall metallicity, dust depletion affects some elements (called refractory, i.e. with a high condensation temperature) more heavily than others (volatile, with a low condensation temperature). On the other hand, the (subtle) effects of stellar nucleosynthesis can be observed for elements that have specific nucleosynthetic history, such as those produced by core-collapse or Type Ia supernovae (SNe). In addition, ISM ionization can also play a role, although it is less of a concern for gas that has a high enough column density to shield against further ionization and keep it dominated by neutral \hi{}, i.e. the neutral ISM \citep{Viegas95}. The dominant ionization species in this phase are typically \cii{}, \nii{}, \oi{}, \mgii{}, \siii{}, \feii{}, \suii{}, etc. Disentangling these effects from the observed abundance patterns is not straightforward. It is possible to determine the amount of dust depletion from the relative abundances of metals, based on the fact that different metals deplete into dust with different strengths, and without assuming a reference metallicity of the ISM \citep{DeCia16,DeCia21,Konstantopoulou22}. In this way it is possible to determine the total (dust + gas) metallicity, after the correction for dust depletion \citep[e.g][]{DeCia18b,DeCia21}.

The study of the chemical properties of the LMC and SMC are of special interest because it is possible to study in great detail both their resolved stellar populations and their gas, in a lower-metallicity environment than for the Milky Way. In addition, these systems are the largest\footnote{The dynamical masses of the LMC and SMC are $M_{\rm dyn} = 1.9\times 10^{11}$ and $6.5 \times 10^9$ $M_\odot$, respectively \citep{Shipp21,Bekki09}} gas-rich dwarf galaxies close to the Milky Way. The Magellanic Stream of gas highlights the strong interaction between the MCs and the Milky Way \citep[e.g.,][]{Fox10,Donghia16,Lucchini20}. The MCs may have had multiple close encounters in the past four billion years \citep{Bekki04}. Strong gas outflows are observed around the LMC \citep{Barger16}. The metal abundances in the ISM of the MCs have been studied in great detail from UV-spectroscopy, for example by \citet{Tchernyshyov15}, \citet{Jenkins17},  \citet{Roman-Duval21}, including the diversity of relative abundances of individual components along one line of sight \citep{Welty97,Welty99}. These authors estimate the depletions based on the deviations of the observed gas-phase abundances from a fixed reference metallicity (or abundances), which is typically taken from young stars and is assumed to represent the total dust+gas metallicity. This methodology assumes that there are no variations in the gas metallicity and that the abundance ratios are intrinsically solar. With this assumption, any potential variations from the reference abundances is by construction interpreted as due to dust depletion, without disentangling between potential effects of variations in ISM metallicity or nucleosynthesis contribution from specific stellar population (e.g. $\alpha$-element enhancement). This is further discussed in Sect. \ref{sec results} and \ref{sec caveats}. \citet{Konstantopoulou22} study the relative abundances in the neutral ISM in the MCs, together with other environments, and measure the amount of depletion based on the relative abundances of metals and the fact that different metals have different tendencies to deplete into dust. These authors study sequences of dust depletion on the global sample of relative abundances in the MCs, and thus study the overall properties of the sample, without analysing the abundance patterns of individual systems in the MCs.

The $\alpha$ elements are elements such as O, Mg, Si, S, Ca, and Ti, which are mostly produced by $\alpha$-capture in core-collapse SNe \citep{Nomoto06}. On the other hand, the Fe-group elements, including Mn, are mostly produced by Type Ia SNe \citep{Nomoto97}. Enhancement of $\alpha$ elements and Mn under-abundance are often observed in low-metallicity stars in the Milky Way and nearby dwarf galaxies \citep[e.g.][]{McWilliam97,Tolstoy09,deBoer14,Hill19}, and likely the result of star formation, because core-collapse SNe explode before most Type Ia SNe \citep[e.g.][]{Tinsley79}. On the other hand, $\alpha$-element enhancement in the neutral ISM (in the gas, rather than in the stars) are seldom observed in nearby galaxies \citep{Cullen21} and in some gas-rich galaxies \citep[Damped Lyman $\alpha$ Absorbers, DLAs;][]{Wolfe05} with low metal enrichment \citep[][]{Dessauges-Zavadsky02,Dessauges-Zavadsky06,Cooke11,Becker12,Ledoux02,DeCia16}, in the regime where the otherwise strong effects of dust depletion are almost negligible. \citet{Konstantopoulou22} note a potential $\alpha$-element enhancement and Mn under-abundance in the neutral ISM in the Magellanic Clouds (MCs), from small deviations of these systems from the overall depletion properties observed for several other different environments. \citet{Jenkins17} find an underabundance of Mn in the SMC, which they interpret as a stronger tendency of Mn to deplete into dust in the SMC. \citet{Tchernyshyov15}, \citet{Jenkins17},  and \citet{Roman-Duval21} do not find $\alpha$-element enhancements in the neutral ISM in the MCs. However, the formalism used by these authors interprets any deviations of the observed abundances as due to dust depletion by construction.

In this paper we adopt the so called "relative" method to study the ISM, which determines the depletion of the elements based on the relative abundances of metals \citep{DeCia16,DeCia21,Konstantopoulou22}. A different way to estimate the depletions is with the $F*$ method \citep{Jenkins09,Jenkins17}. The development of the $F*$ method enabled the key discovery that the neutral ISM in the Milky Way has a continuum range of dust depletion and allowed substantially more robust estimates of the strength of the dust depletion than what was previously possible. The $F*$ method studies the relations between the observed abundances, [$X$/H], assumes a fixed metallicity of the gas, and no deviations due to nucleosynthesis effects (e.g. no $\alpha$-element enhancements and Mn underabundance), and uses [$X$/H] to trace dust depletion. If there are variations in ISM metallicity, the $F*$ can observe them in the form of a discrepancy between the observed and expected $N({\rm H})$. If there are deviations in the abundances [$X$/H] due to nucleosynthesis, the $F*$ method interprets these variations as due to depletion by construction. Thus, the current formalism of the $F*$ method is not designed to disentangle between the effects of dust depletion, metallicity, and nucleosynthesis effects and cannot purely determine dust depletion in systems with nucleosynthesis effects such as $\alpha$-element enhancements and Mn underabundance.

In this paper we study the chemical properties of the neutral ISM of the LMC and SMC, with particular emphasis on the total (gas + dust) metallicity, dust depletion, $\alpha$-element enhancement and Mn under-abundance in the neutral ISM. In Sect. \ref{sec sample} and \ref{sec method} we describe our sample and methodology, respectively. We present and discuss our results in Sect. \ref{sec results}, include additional caveats in Sect. \ref{sec caveats}, and conclude in Sect. \ref{sec conclusions}.

   \begin{figure*}
   \centering
   \includegraphics[width=\textwidth]{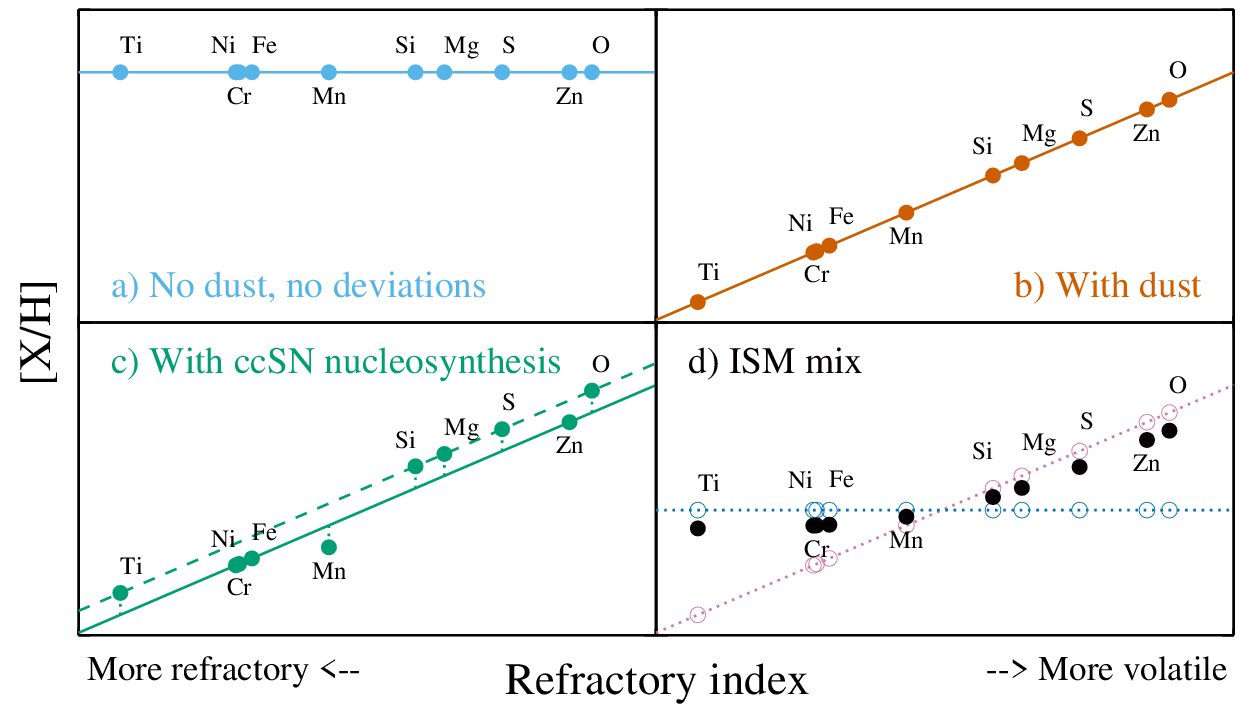}
      \caption{Idealized illustration of different processes that can shape the ISM abundance patterns. The observed abundances are on the $y$-axis, and the $x$-axis represent the tendency of each element to deplete into dust grains. a) If no dust and other processes that may cause additional deviations are at play, the observed abundances reflect the actual metallicity of the gas. b) If there is dust, the refractory elements show lower abundances, because they are depleted. The more dust, the steeper the slope of the linear relation. The $y$-intercept (at no depletion) of the solid line is the total (gas + dust) metallicity. c) Deviations from the linear relation can be observed in the gas for specific elements, in this example $\alpha$-element enhancement and Mn under-abundance due to nucleosynthesis of recent core-collapse SNe. The dashed line shows a fit to the $\alpha$-element data only. In this illustration Zn (and other elements, e.g. P) is excluded from determination of the linear fit and only shown here with the Fe-group elements. Any eventual deviations can be observed in the data. d) The presence along the line of sight of a mix of two gas components (empty circles) with different metallicities and amount of dust can produce bending of the overall observed abundances (filled black circles) measured over the whole line profile. The metallicity derived from the whole line profile could be somewhere in-between the metallicities of the individual components. If the abundances of enough metals with different refractory and nucleosynthetic properties are observed, the effects of the different processes (a-d) can be separated.}
         \label{fig abu}
   \end{figure*}

   \begin{figure*}
   \centering
   \includegraphics[width=0.9\textwidth]{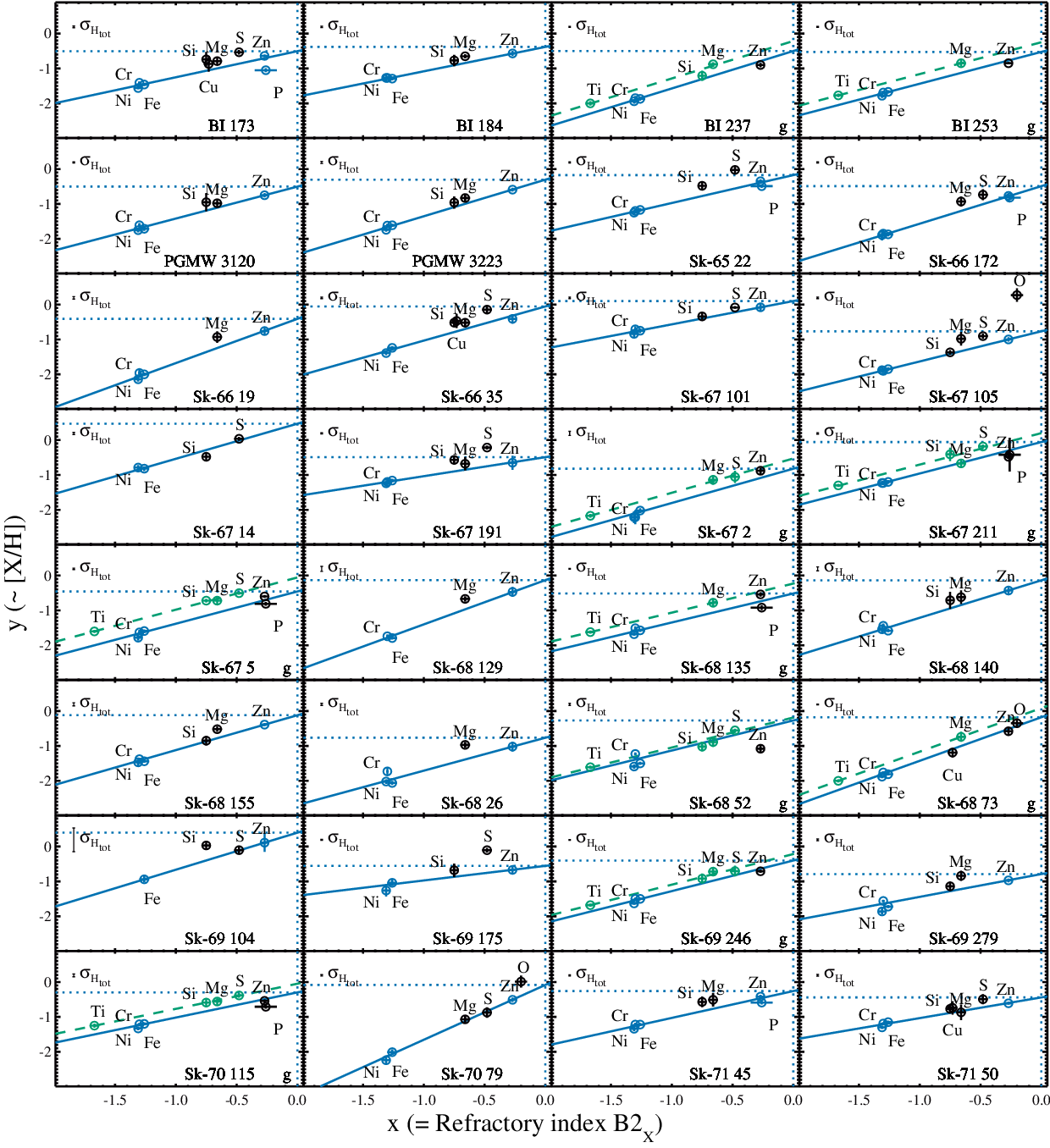}
      \caption{Analysis of the abundance patterns in the neutral ISM in the LMC and determination of the total metallicity and overall strength of depletion. The variables and coefficients of the linear relation are defined in Eqs. \ref{eq xy} to \ref{eq b}, where the $y$-intercept gives the [M/H]$_{\rm tot}$ and the slope of the relation gives the strength of depletion [Zn/Fe]$_{\rm fit}$. In case Ti and another $\alpha$ element are constrained i.e. for the golden sample, labelled with 'g', the slope of the linear fit is determined only from the $\alpha$ elements, which are marked in green, and excluding O. The resulting fits are shown by the green dashed curves, and their slopes are then fixed. The fit for the metallicity uses this slope and is determined by the Fe-group elements Fe, Ni, and Cr (blue solid curves), and excluding Zn and P from the fit. For the non-golden sample we fit linearly all metals together, but excluding the $\alpha$ elements and Mn. The error bars show the 1$\sigma$ uncertainties on the metal columns. The deviations from the linear fit are mostly due to $\alpha$-element enhancement (see Sect. \ref{sec alpha}). The uncertainties shown on the $y$-axis are the formal statistical errors on the column densities. The uncertainty on $N(\mbox{H})_{\rm tot}$ is shown by the error bar on the top left, it affects all points systematically and it is included in the uncertainty of the  [M/H]$_{\rm tot}$.}
         \label{fig xy LMC}
   \end{figure*}

   \begin{figure*}
   \centering
   \includegraphics[width=0.9\textwidth]{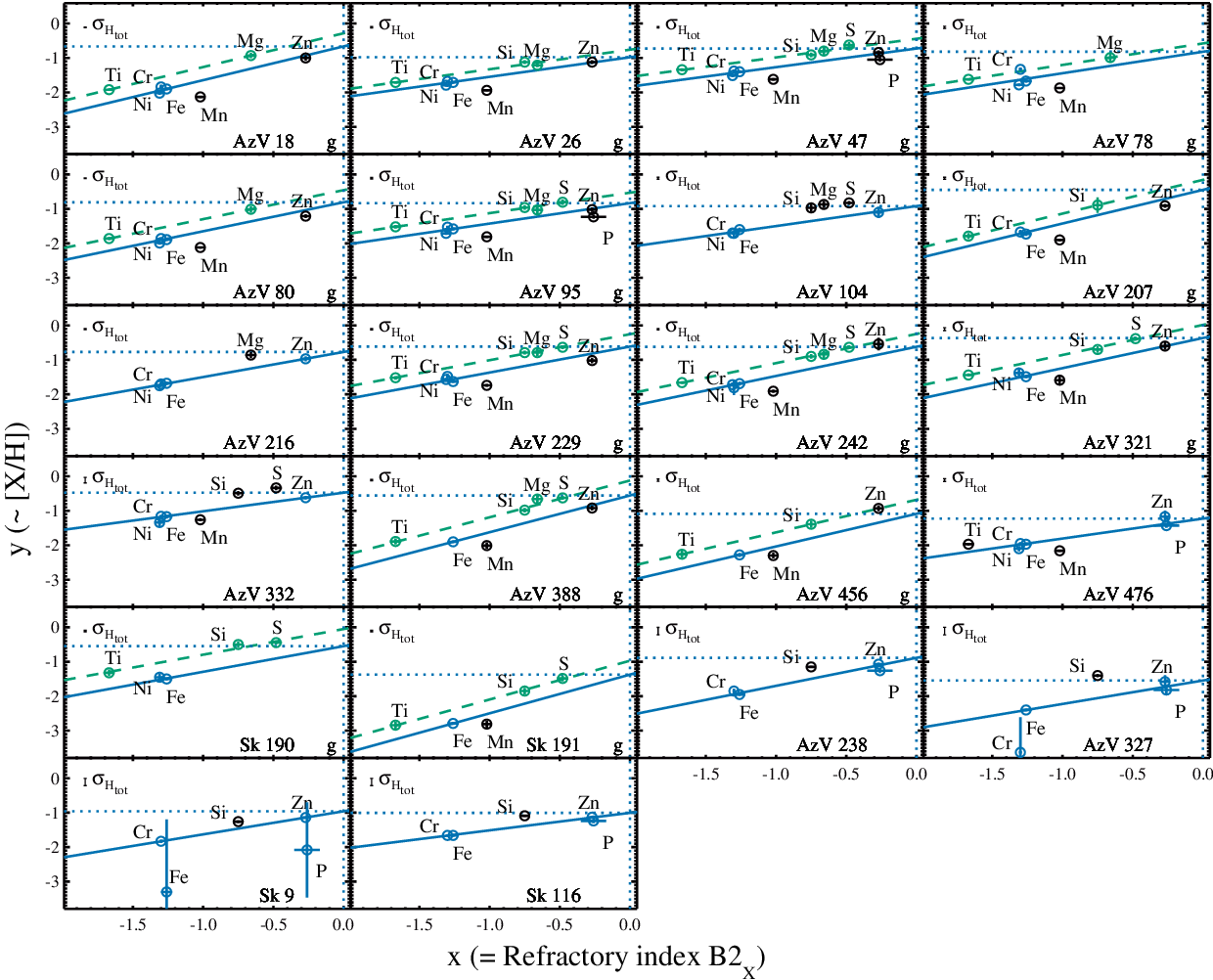}
      \caption{Same as Fig. \ref{fig xy LMC}, but for the SMC.}
         \label{fig xy SMC}
   \end{figure*}
   
  \section{Sample}
\label{sec sample}
We collect a literature sample of metal column densities in the neutral ISM in the SMC and LMC, for a total of 54 lines of sight. Tables \ref{tab LMC} and \ref{tab SMC} present the targets of our LMC and SMC samples, respectively. The locations of the targets in the LMC and SMC are shown in Figs. \ref{fig LMC map} and \ref{fig SMC map}. The measurements of all the 32 LMC lines of sight in our sample are adopted from \citet{Roman-Duval21} as part of the  METAL large HST program \citep{Roman-Duval19}. \citet{Roman-Duval21} measured metal column densities using observations of hot stars taken with the Hubble Space Telescope (HST) Cosmic Origins Spectrograph \citep[COS;][]{Green12} and the Space Telescope Imaging Spectrograph (STIS). \citet{Roman-Duval21} uses data collected with COS medium-resolution gratings (G130M and G160M, with $R\sim15,000$--$20,000$) and STIS (E140M with $R\sim45,000$ and E230M with $R\sim30,000$), with the exceptions of Sk-70~115 and Sk-68~73, for which archival STIS E230H ($R\sim114,000$) was used. The sample of \citet{Roman-Duval21} includes lines of sight which were previously studied by \citet{Tchernyshyov15}, based on data from the Far Ultraviolet Spectroscopic Explorer \citep[FUSE;][]{Sahnow00} from the FUSE MC Legacy Project archive \citep{Blair09}. We adopt all the LMC column density measurements from \citet{Roman-Duval21}. These authors adopt the technique of \citet{Jenkins96} to correct for potential saturation of the absorption lines. Additional information on the observations and column density measurements for the LMC systems can be found in \citet{Roman-Duval21}. Potential systematic differences in the measurements of column densities from data with $R\sim15,000$--$20,000$ or $R>30,000$ is studied by \citet{Roman-Duval21}, who find a potential systematic underestimation of $N$(Zn) and $N$(Si) by up to 0.3~dex in the lowest-resolution data (see their figure 23). Any potential underestimation of $N$(Zn) does not affect our results, because Zn is not used in this paper to estimate neither metallicities for the golden sample nor $\alpha$-element enhancements (see Sect. \ref{sec method}). Any potential underestimation of Si would also not affect our results, other than making the Si overabundances that we observe stronger (see Sect.\ref{sec alpha}). \citet{Roman-Duval21} measure the column densities of \mgii{}  from medium resolution COS data for BI~184, BI~253, Sk-66~129, Sk-68~140, Sk-68~155, Sk-68~73, as well as BI~273, Sk-66~19, Sk-68~26, Sk-69~279, with the last four systems having also \niii{} measured from medium resolution COS data. In most cases our analysis does not depend strongly on these \niii{} and \mgii{} measurements, because \feii{} and \siii{} are well measured from higher resolution STIS data. In two cases, namely BI~253 and Sk-68~73, of our most reliable golden sample (see below) the \mgii{} column density measured with lower-resolution data is the only element representing the less-refractory $\alpha$-elements, adding some uncertainty to our ability to constrain them. These two systems represent $\sim8$\% of our golden sample and thus the potential saturation effects will not affect our results.

For the SMC, we use the column density measurements for $22$ lines of sight of \citet{Jenkins17} and \citet{Tchernyshyov15}. We adopt the column densities for 18 out of 22 systems of \citet{Jenkins17}, who measure them using data from STIS medium-resolution grating E140M and E230M (with $R\sim45,000$ and $R\sim30,000$), similar to the LMC dataset that we adopt from \citet{Roman-Duval21}. \citet{Jenkins17} apply corrections for potential saturation of the absorption lines adopting the technique of \citet{Jenkins96}. In addition, we include the column densities for four systems in the SMC (namely, AzV~327, AzV~238, Sk~116, and Sk~9) from \citet{Tchernyshyov15}, who measured them based on data taken with the COS G185M grating ($R\sim18,000$). \citet{Tchernyshyov15} limit potential saturation of the absorption lines by performing a simultaneous Voigt-profile fit of multiple lines of the same ion. In case of strong saturation, this technique may underestimate the column densities. However, these four lines of sight are not part of our golden sample (see below), and thus will not influence the results of our paper. 

\citet{Konstantopoulou22} collected column density measurements for the MCs for the same sample\footnote{The difference in our SMC sample and that of \citet{Konstantopoulou22} is only for Sk~9, which is not in their sample.} and correct them for the most recent values of oscillator strength, which we adopt here. For \mgii{}, \siii{}, \mnii{}, \cuii{}, \crii{}, and \tiii{} we adopt the oscillator strengths of \citet{Cashman17}, for \niii{} we adopt those of \citet{Boisse19}, for \znii{} those of \citet{Kisielius15}, for \suii{} those of \citet{Kisielius14}, for \pii{} those of \citet{Kurucz17}, and for the other ions with no recent changes of the oscillator strengths we adopt the values of \citet{Morton03}. For the calculation of the (relative) abundances we refer to the solar abundances of \citet{Asplund21}, and choosing the photospheric or meteoritic abundances following the recommendations of \citet{Lodders09}. A table of the solar abundances is reported in \citet{Konstantopoulou22}.

We update the column density of \crii{}  toward Sk-67~2 from $13.91 \pm 0.06$  \citep{Roman-Duval21} to $13.0\pm 0.2$ from measurements of the Apparent Optical Depth of the \crii{} 2062 and 2056 lines (Jenkins, private communication), which are found in disagreement with the original measurement based only on the \crii{} 2066 line, which is likely contaminated by other features. The full updated analysis of the data along this line of sight will be part of future publications from the ULLYSES project \citep{Roman-Duval20}. This substantial change in \crii{} column density for the Sk-67~2 makes the Cr abundance perfectly agree with the abundances of Fe and Ni (see the study of its abundance pattern in Sect. \ref{sec results}).

Overall, we collect column density information for \tiii{}, \niii{}, \crii{}, \feii{}, \mnii{}, \siii{}, \cuii{}, \mgii{},  \suii{}, \pii{}, \znii{}, and \oi{}. For both MC systems, we adopt the column density measurements of \tiii{} from \citet{Welty10}. We define and build a golden sample from the systems with constrained observations of \tiii{} and at least another $\alpha$ element.

\section{Determination of the total metallicity, overall strength of depletion, and deviations}
\label{sec method}

We determine the total (gas $+$ dust) metallicity in the LMC neutral ISM along the different lines of sight using the ``relative method'' described in \citet{DeCia21}. This method is based on the two basic facts: first, that the total abundance $[X/{\rm H}]_{\rm tot}$ can be derived from the observed gas-phase abundance of element $X$, $[X/{\rm H}]= \log N(X) -  \log N({\rm H}) - X_\odot + 12.$, and correcting it for the the depletion of element $X$ in dust, $\delta_X$, as following: $[X/{\rm H}]_{\rm tot} = [X/{\rm H}]- \delta_X$. Note that no attempt of including corrections for nucleosynthesis effects (such as $\alpha$-element enhancement or Mn under-abundance) is made here; however, these may become visible as deviations from the abundance patterns that we expect from depletion effects. Second, $\delta_X$ is a linear function of the overall strength of depletion, which is described by the parameter [Zn/Fe]$_{\rm fit}$ in the relative method. This parameter is related to the observed [Zn/Fe], but is derived from the abundances of several metals, e.g. Ti and other $\alpha$ elements when possible, as we discuss below.

The total metallicity $[{\rm M/H}]_{\rm tot}$ and overall strength of depletion $[{\rm Zn/Fe}]_{\rm fit}$ can then be derived with a linear fit to the data, which are the observed gas-phase abundances of different metals and the information on the tendency to deplete into dust for each metal ($B2_X$ coefficients, or refractory index), as follows: 
\begin{equation}
y = a + bx   
\label{eq xy}
\end{equation}
\begin{equation}
x = B2_X\mbox{,}
\label{eq x}
\end{equation}
\begin{eqnarray}
y &=& \log N(X) -  \log N({\rm H}) - X_\odot + 12 - A2_X  \\
  &\sim& [X/\mbox{H}] \mbox{,} \nonumber
\label{eq y}
\end{eqnarray}
\begin{equation}
a = [{\rm M/H}]_{\rm tot}\mbox{,}
\label{eq a}
\end{equation}
\begin{equation}
b = [{\rm Zn/Fe}]_{\rm fit}\mbox{.}
\label{eq b}
\end{equation}
We adopt the $A2_X$ and $B2_X$ coefficients of \citet{Konstantopoulou22}, which are consistent with those of \citet{DeCia16}, but they include 18 metals, and are based on a large collection of relative abundance measurements for local and distant galaxies and on the most recent set of oscillator strengths. We adopt $B2_{\rm P}=-0.26 \pm 0.08$ from \citet{Konstantopoulou23}, which is updated for the most recent oscillator strengths for P transitions \citep{Brown18}. The coefficients $A2_X$ could in fact be omitted and assumed to be zero (which would imply zero depletion when $[{\rm Zn/Fe}] =0$). The empirically measured values are mostly very close to zero. In this work we include more metals for the analysis of the abundance patterns than in \citet{DeCia21}. Uncertainties in both the $x$ and $y$ axes are included in the calculation of the best linear fit to the abundance patterns\footnote{For this task we make use of the IDL routine MPFITEXY, \citet{Williams10}, which is based on the MPFIT package \citep{Markwardt09}.}. The uncertainties on $N(\mbox{H})_{\rm tot}$ systematically affect all the $y$-values in a uniform way. Therefore, and unlike in \cite{DeCia21}, we do not formally include them in the linear fit to the abundance patterns, because the depletion (slope) does not depend on H. We include the uncertainties on $N(\mbox{H})_{\rm tot}$ to the uncertainty of $[{\rm M/H}]_{\rm tot}$ a posteriori, after the fit. \citet{Konstantopoulou22} find two different series of $B2_X$ values for the MCs for Ti and Mn (see discussion in Sect. \ref{sec caveats}). We adopt the $B2_X$ that are derived globally for local and distant galaxies.\footnote{\citet{Konstantopoulou22} find two series of $B2_X$ values, using or not the assumption of a constant $\alpha$-element abundance in the MCs. The uncertainties we adopt on $B2_X$ do not include the systematic uncertainty due to the two different assumptions. The global values of $B2_{Ti}$ for the two cases are very similar. We test and discuss the possibility of a significantly different $B2_{Ti}$ in Sect. \ref{sec caveats}.}

Figure \ref{fig abu} visualizes how the abundance pattern in the neutral ISM is shaped by different processes. The overall metallicity affects its normalization, i.e it increases or decreases the abundance of all elements. The dust depletion affects its slope, i.e. the refractory elements show lower abundances than the volatile ones. In addition, deviations from a linear fit to the abundance patterns can be caused by nucleosynthesis effects, such as  $\alpha$-element enhancement and Mn under-abundance. We consider this possibility by choosing different sets of metals for the linear fits to the abundance patterns, which we describe in more details below. Here we consider Ti, Mg, Si, S, and O as full $\alpha$ elements, while we consider Fe, Ni, Cr in the Fe group. We treat Mn separately, because its behaviour is somewhat opposite to the $\alpha$ elements \citep[e.g.][]{Kobayashi20}. The possibility of Zn and P having intrinsic enhancement (somewhat following $\alpha$-element processes or massive-star nucleosynthesis) is discussed in Sect. \ref{sec caveats}. Finally, another potential cause of deviations from linear abundance patterns is the presence of a mixture of gases with different metallicities and levels of depletion. This may cause upturns from the linear abundance patterns, because the overall abundance of refractory elements  may arise from gas with low-metallicity and little dust content (empty blue circles in Fig. \ref{fig abu}, panel d), while the abundance of the volatile elements may arise from heavily depleted gas at high metallicities (empty magenta circles in Fig. \ref{fig abu}, panel d). This is just an example, and other configurations are possible. We discuss this in more details in Sect. \ref{sec ISM mix}. The effect of the various processes on the abundance patterns act on distinct elements differently, depending on the refractory and nucleosynthetic properties of each element. Because of this, it is possible to disentangle among the different processes, provided that data for a variety of metals are available.

The choice of which metals to include in the analysis and linear fit to the abundance patterns is vital to the outcome. We test different choices to investigate the robustness of our results and avoid biases due to lack of data or intrinsic variations of some of the metals (see the Appendix). The most solid approach can be applied when several $\alpha$ and non-$\alpha$ elements are constrained. We define our golden sample using systems where at least Ti, another $\alpha$ element, and Fe are constrained, for a total of 10 lines of sight towards the LMC and 14 towards the SMC. In these cases, we determine the amount of dust ([Zn/Fe]$_{\rm fit}$\footnote{Here the slope, or strength of overall depletion, does not depend on Fe nor Zn, but it uses the [Zn/Fe]$_{\rm fit}$ notation because the refractory index $B2_X$ was originally calculated from the depletion sequences observed as a function on [Zn/Fe].}) purely from the slope of the linear fit to the abundance patterns of the $\alpha$ elements only (except O). We then fix this slope and fit linearly the observed abundances of Fe, Ni, and Cr (but not Mn, Zn and P) to find the total metallicity ([M/H]$_{\rm tot}$). For the golden sample, the total metallicities are thus derived from the Fe group abundances (free from potential $\alpha$-element enhancements or Mn underabundance or any potential Zn and P variations), after correcting for dust depletion. The values of [M/H]$_{\rm tot}$ are indicated by the $y$-intercepts of the solid lines at $x = 0$ in the diagrams shown in Figs. \ref{fig xy LMC} and  \ref{fig xy SMC}. Zinc and phosphorus are excluded from the determination of the linear fit to the abundance patterns so that the results do not depend on the abundances of Zn and P and can be used to investigate any eventual deviations of these elements from the abundance patterns. Having constraints on Ti is crucial for this approach, because they provide a sufficiently large dynamical range in $B2_X$ for a solid determination of the slope of the abundance patterns. The calculation of the uncertainties on [M/H]$_{\rm tot}$ include both the uncertainties in the normalization of the fit to the $\alpha$ elements (used for fixing the slope) and the uncertainties on the normalization of the fit to the Fe-group (with a fixed slope).

In case Ti is not constrained (for the non-golden sample), we consider the fit to all metals but excluding the $\alpha$ elements and Mn to avoid the influence of their potential deviations. In this situation, the fits to the abundance patterns are dominated, at the volatile-element end, by Zn and P. This is less optimal, because some intrinsic variations due to nucleosynthesis may be at play, see Sect. \ref{sec caveats}.  Further possibilities on the choice of metals for the analysis of the abundance patterns are discussed in the Appendix. 

In the the case of Sk-67~14, where only a few metals are available, we consider the optimal fit to all metals (a mix of $\alpha$ and non-$\alpha$ elements) and include a posteriori an additional 0.32~dex\footnote{The 0.32~dex is chosen here to be in-between the typical values of $\alpha$-element enhancement in the Milky Way \citep[$\sim0.35$~dex][]{McWilliam97} and DLAs \citep[$\sim0.3$~dex][]{DeCia16}.} in the uncertainty on their total metallicity to account for the paucity of data and potential nucleosynthesis effects causing over- or underestimate of the metallicity.

Figures \ref{fig xy LMC} and  \ref{fig xy SMC} show the determination of the total metallicity and overall strength of depletion from the fit to the abundance patterns. Figures \ref{fig [M/H] comparison} and \ref{fig [Zn/Fe] comparison} present a comparison of the different total metallicity and overall strength of depletion resulting from a different choice of metals in the fit to the abundance patterns. Tables \ref{tab LMC} and \ref{tab SMC} list the total metallicities and strength of dust depletion of the line of sights towards our LMC and SMC samples, respectively.

 \begin{figure}
   \centering
   \includegraphics[width=0.5\textwidth]{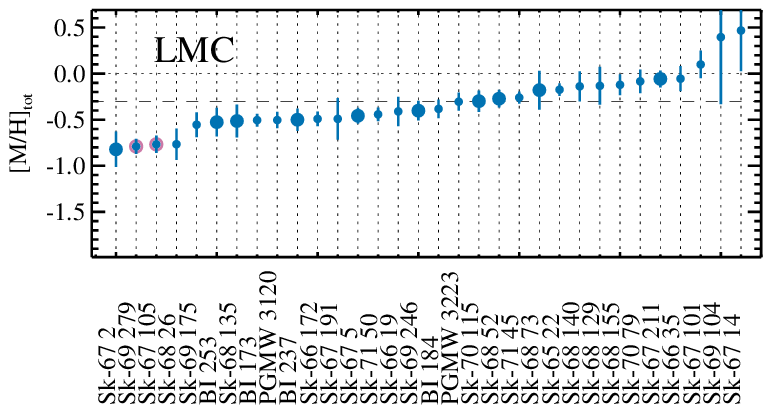}
      \caption{Distribution of the total (dust $+$ gas) metallicities, [$M$/H]$_{\rm tot}$, in the neutral ISM in the LMC. Larger symbols highlight the most solid measurements, for which multiple $\alpha$ (including Ti) and non-$\alpha$ elements are available, which we define as the golden sample). Systems with metallicities that vary significantly from the mean are highlighted with purple circles. The mean stellar metallicity in the LMC is marked by the horizontal dashed line. The total metallicities that we measure are integrated along the line of sight within the LMC. In case of a chemically inhomogeneous ISM, they may be in-between a lower and a higher metallicity, but there are degeneracies in the potential mixes.}
         \label{fig met LMC}
   \end{figure}

\begin{figure}
   \centering
   \includegraphics[width=0.5\textwidth]{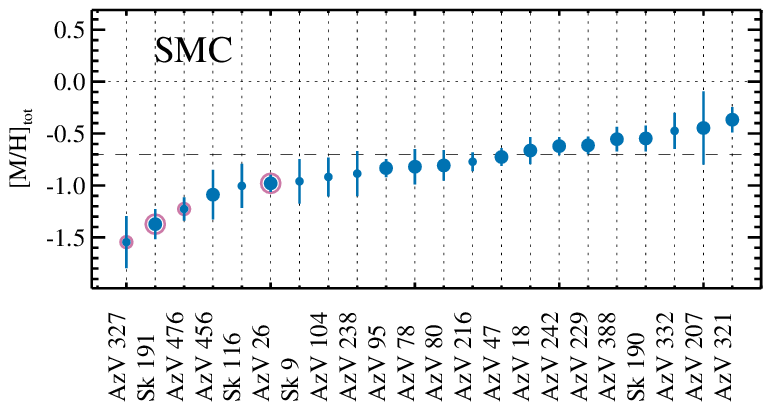}
      \caption{Same as Fig. \ref{fig xy LMC}, but for the SMC.}
         \label{fig met SMC}
   \end{figure}

\begin{figure}
   \centering
   \includegraphics[width=0.5\textwidth]{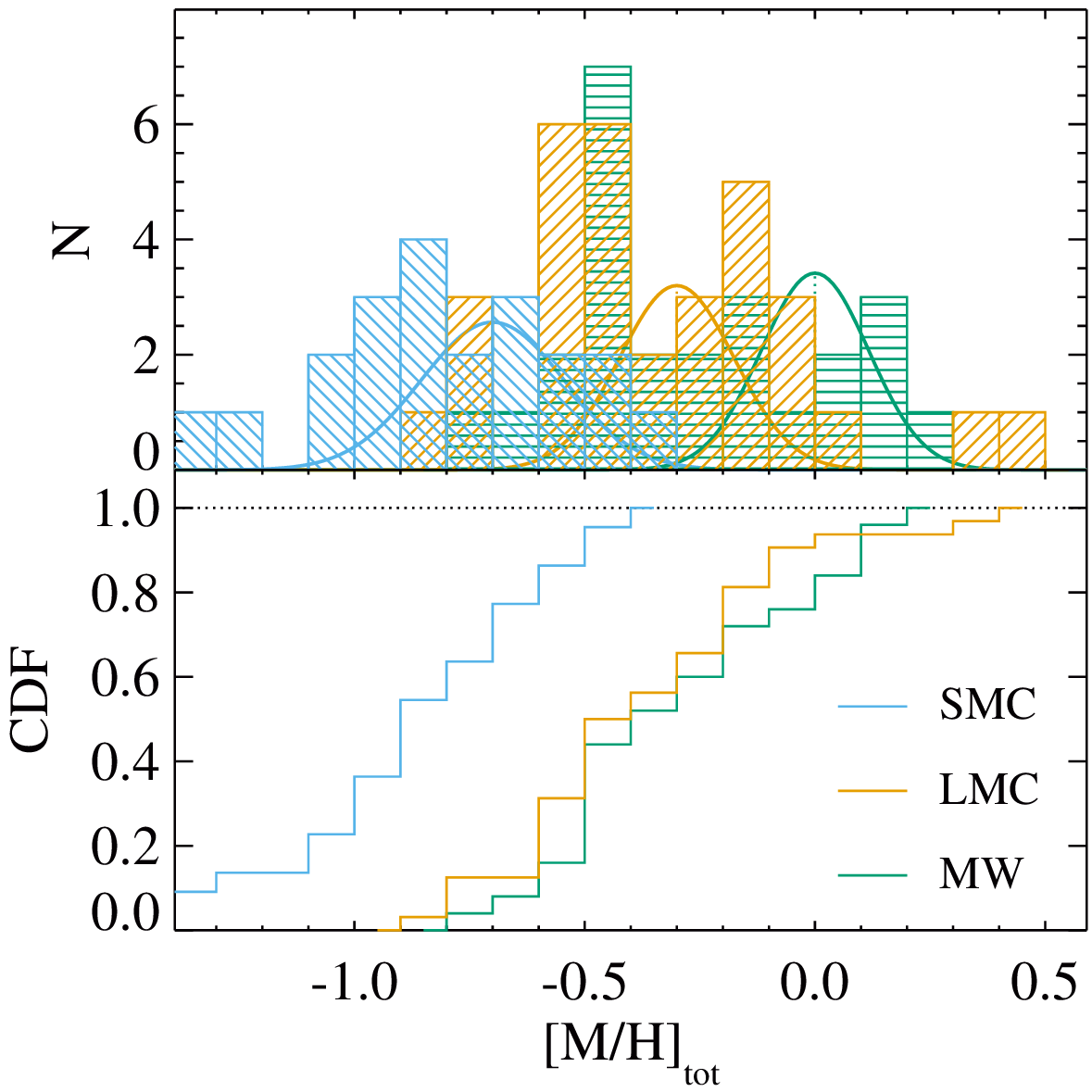}
      \caption{Distribution of the total metallicities in the neutral ISM for the Milky Way \citep[from][]{DeCia21} and Magellanic Clouds measured in this work (top panel, shaded regions). For each of the three environments, the expected distribution if the ISM total metallicities would be the same as in hot stars is shown by a Gaussian curve centred on the hot-stars metallicity and with a $\sigma$ equal to the mean of the uncertainties in $[{\rm M/H}]_{\rm tot}$ that we measure. The values of metallicity in the Milky Way ISM measured by \citet{Ritchey23} are consistent with solar metallicity, with a standard deviation of $\sim0.1$~dex, and are not shown in this figure. The bottom panel shows the cumulative distribution function of the observed total metallicities.}
         \label{fig histo [M/H]}
   \end{figure}

\section{Results and discussion}

\label{sec results}

The neutral ISM is likely complex and possibly chemically inhomogeneous \citep[e.g.][]{Welty20,DeCia21,Ramburuth-Hurt23}. Our understanding of its chemical properties through the study of the abundance patterns is hindered by the several processes that shape them. In this paper we attempt to make sense of this complexity by studying metals with different refractory and nucleosynthetic properties, and thereby separate different effects that play an important role in shaping the abundance patterns: the metallicity, the dust depletion, the $\alpha$-element enhancement, and the presence of ISM mixtures. Here we discuss our results (Sect. \ref{sec met} to \ref{sec ISM mix}) and their caveats (Sect. \ref{sec caveats}).

\subsection{The neutral ISM metallicities}

\label{sec met}

Figures \ref{fig met LMC} and \ref{fig met SMC} show the distribution of total metallicities in neutral ISM in the LMC and SMC towards the lines of sight considered here. The horizontal lines mark the nominal metallicities from measurements of hot stars, $\log Z/Z_{\odot} = -0.3$ and  $-0.7$ for the LMC and SMC, respectively \citep{Rolleston02,Trundle07,Hunter07}. We find that most of the lines of sight have ISM metallicities which are consistent, within the uncertainties, with the nominal values. The mean total metallicity in the neutral ISM in our sample is $[M/\mbox{H}]_{\rm tot} = -0.33$ ($-0.83$) with standard deviation 0.30 (0.30) for the LMC (SMC). In addition, we observe significant variations from the nominal metallicities, with a significance level above $3\sigma$, for two lines of sight in the LMC (namely Sk-69~279 and Sk-67~105, none of which is in the golden sample, with [$M$/H]$_{\rm tot}=-0.77\pm0.09$ and $-0.79\pm0.08$, respectively) and for four lines of sight in the SMC (namely AzV~26 and Sk~191, both in the golden sample, with [$M$/H]$_{\rm tot}=-0.98\pm0.08$ and $-1.37\pm0.14$, respectively, and AzV~327 and AzV~476, both not in the golden sample, with [$M$/H]$_{\rm tot}=-1.55\pm0.25$ and $-1.23\pm0.11$, respectively). These six low-metallicity systems are highlighted in Figs. \ref{fig LMC map}, \ref{fig SMC map}, \ref{fig met LMC} and \ref{fig met SMC}.

Figure \ref{fig histo [M/H]} shows the distribution of the total metallicities in the neutral ISM for the Magellanic Clouds from this work and the Milky Way measurements from \citet{DeCia21}. The distribution of ISM total metallicities in the LMC is overall similar to the Milky Way, while the SMC lines of sight have overall lower total metallicities of the neutral ISM. The Gaussian curves show, for each of the three environments, what the expected distribution of metallicity would be if the ISM total metallicities would be the same as in hot stars and considering the typical uncertainties of our measurements. The center of the Gaussian curves are on the hot stars metallicities and their $\sigma$ is equal to the mean of the uncertainties in $[{\rm M/H}]_{\rm tot}$ that we measure. Two lines of sight toward the LMC have supersolar metallicity but with very large uncertainties (see Fig. \ref{fig met LMC}), and these are excluded from the calculation of the Gaussian width. 

Intriguingly, two of the systems with the lowest neutral ISM metallicities in our sample, Sk~191 and AzV~476, with [$M$/H]$_{\rm tot}=-1.37\pm0.14$ and $-1.23\pm0.11$, respectively, are located in the outskirts of the main body of the SMC, closer to the Magellanic Bridge (or inter-cloud region) and its densest part, the SMC South East Wing \citep[e.g.][]{Westerlund90}. The system with the lowest metallicity that we measure, AzV~327, with [$M$/H]$_{\rm tot}=-1.55\pm0.25$, is located in the northern outskirts of the SMC. The positions of these stars are shown in Figure \ref{fig SMC map}. The metallicity of the gas in the Magellanic Bridge is about 10\% solar (or less), from the observations of volatile elements such as Ar and O \citep{Lehner08}. \citet{Rolleston99,Dufton08} find that B stars in the Magellanic Bridge have $[M/H] \sim -1$, while \citet{Ramachandran21} measure very diverse metallicities (between $-0.3$ and $-1.4$) in O-type stars in the Magellanic Bridge, indicating that the ISM is chemically inhomogeneous and includes low metallicities. On the other hand, \cite{Lee05} find that B-type stars in the SMC South East Wing, not far from Sk~191, show metallicities that are more similar to the main SMC body. The Magellanic Bridge is a tidally stripped region in between the MCs, likely resulting from the interaction between the MCs about $\sim0.2$~Gyr ago \citep{Gardiner96,Besla12}. In the simulations of \citet{Bekki07b} gas had been transferred from the SMC to the LMC, so that the low-metallicities in the Magellanic Bridge are produced by the SMC halo gas, depending on a metallicity gradient. We do not find evidence for a present-day metallicity gradient in the neutral ISM of the MCs (Figs. \ref{fig LMC map} and \ref{fig SMC map}).

The line of sight towards Sk~191 has by far the lowest observed metal abundances with respect to the the other SMC lines of sights observed by \citet{Jenkins17} (see their figure 2). These authors estimate the dust depletion from the observed metal abundances [$X$/H] and conclude that this line of sight may have a large amount of dust depletion, comparable to the most dusty systems in the Milky Way. However, in this interpretation the relative abundances of S, Si, and Ti observed towards Sk~191 do not fit the dust depletion properties known from the Milky Way, and thus \citet{Jenkins17} conclude that the properties of dust depletion in the SMC must be different from the MW, specifically for Mn, Ti and Mg and possibly other elements. To further justify their interpretation of the high depletion towards Sk~191 \citet{Jenkins17} resort to a marginal detection of the weak \oi{} 1356 line. However, as previously stated in this paper, the formalism used by \citet{Jenkins17} to estimate the dust depletion does not take into account any possible variations of the ISM abundances due to different metallicity or nucleosynthesis effects such as $\alpha$-element enhancements. They interpret the low abundances towards Sk~191 as a pure tracer of dust depletion, by construction of their methodology. On the other hand, our analysis of the abundances and relative abundances allows us to independently derive the dust depletion (from the relative abundances, which determine the slope of the fit to the abundance patterns), the metallicity (from the normalization of the fit to the abundance patterns) and $\alpha$-element enhancements or Mn underabundance (from deviations to the fit). This way, we can break the degeneracy and characterize these different aspects. The differences between these methods are further discussed in Sect. \ref{sec caveats}. Finally, we note that the line of sight towards Sk~191 has a low dust extinction ($E(B-V)\sim0.1$~mag), and its environment, significantly separated from the main SMC body, is far more similar to the Magellanic Bridge than the dustiest regions of the Milky Way.

Another case of low observed metal abundances is for the SMC line of sight towards Az V456 \citep[see figure 2 of][]{Jenkins17}, although this system exhibits high molecular content, a relatively large dust extinction, $E(B-V)\sim0.33$~mag, and a Milky-Way-type extinction curve with a 2175 \AA{} bump. In this case, we find marginal indication of low-metallicity (Fig. \ref{fig met SMC}), consistent with the nominal SMC hot-stars metallicity. If some low metallicity gas exists along this line of sight, it may be mixed with a higher metallicity gas, as we discuss in Sect. \ref{sec ISM mix}. The position of AzV~456 is in the direction of the Magellanic Bridge, but closer to the main body of the SMC with respect to Sk~191.

In the LMC the two lines of sight with deviations in neutral ISM metallicity from the nominal value are Sk-69~279 and Sk-67~105, which are located south of the Tarantula Nebula and in the outskirts in the north side of the LMC, respectively, as shown in Fig. \ref{fig LMC map}. In this case, there is no apparent relation with the Magellanic Bridge. Replenishment of gas from the SMC to the LMC \citep[e.g.][]{Bekki07b}, through the Magellanic Bridge, could be a potential explanation for the presence of low-metallicity gas that we observe in these two lines of sight in the LMC.

Overall, previous studies of the abundances in the MCs, such as \citet{Tchernyshyov15,Jenkins17,Roman-Duval21}, do not find variations of metallicity by construction, because they assume no metallicity variations and any deviations of the observed abundances from fixed references are attributed to dust depletion (see Sect. \ref{sec caveats} for more details). \citet{Tchernyshyov15} do not use a single metallicity as one fixed reference, but individual abundances of different metals as fixed references, measured in photospheres of young stars in the MCs, and assume that these are the total (gas+dust) abundances. They study the dust depletion from the deviations of the observed gas-phase abundances [$X$/H] from the reference stellar abundances. \citet{Tchernyshyov15} observe a broad distribution of the observed gas-phase abundances of the volatile elements (see their figure 12), from which they cannot exclude the existence of metal-poor gas.

\citet{Welty97} study the line of sight towards AzV~332 (also known as Sk~108), which is included in our sample, but using high-resolution data. The total column densities along the line of sight within the SMC are quite similar to those we adopt here, with the exception of $N$(Ni), which is 0.4~dex lower in \citet{Welty97}. Regardless of Ni, our analysis of the abundance patterns (Fig. \ref{fig xy SMC}) from the two datasets are equivalent, leading to very similar results for the total metallicity, dust depletion, and $\alpha$-element enhancement.

As an important warning, the total metallicities that we measure along the whole line of sight towards each star (Fig. \ref{fig met LMC}, \ref{fig met SMC}, and \ref{fig histo [M/H]}) are integrated measurements and may in fact be a combination of higher and lower metallicities, if the ISM is not chemically homogeneous along the line of sight. This is discussed further in Sect. \ref{sec ISM mix}. Component-by-component analysis of metal absorption lines from high-resolution spectroscopy are rare towards the MCs, but unveil some of the ISM complexity \citep[e.g.][]{Welty97,Welty99}.

\citet{DeCia21} discover large variations in the metallicity of the neutral ISM in the Milky Way, which was generally assumed to be chemically homogeneous and have solar metallicity. They observe regions of low metallicities, down to 17\% solar and possibly lower, and attribute these to the infall of low-metallicity gas on the Galactic disk. The low-metallicity gas may be distributed in pockets along the line of sight and their mass may be small \citep{DeCia22}. On the other hand, \citet{Ritchey23} find no evidence for low metallicity gas in the Milky Way. This discrepancy arises from the different strategy in the analysis of the abundance patterns. The results of \citet{DeCia21} are based mostly on the study of refractory and mildly refractory elements, which are more sensitive to less dusty and metal-poor gas. The results of \citet{Ritchey23} are based mostly on the study of volatile and mildly volatile elements. The inclusion of the volatile elements in the analysis of the abundance patterns makes it very hard to detect the signatures of any low metallicity gas, if present along the line of sight. This is because the determination of the volatile element abundances is dominated by the gas of higher metallicity, hiding the possible presence of low-metallicity gas. The methodology of \citet{Ritchey23} is not sensitive to the presence of low-metallicity gas because it always includes the volatile elements in the analysis of the abundance patterns and assumes that all the gas along the line of sight has the same metallicity (i.e. performs a linear fit to the abundance patterns of either both volatiles and refractory elements, or excluding the refractory ones). Overall, both \citet{DeCia21} and \citet{Ritchey23} observe deviations from a linear behaviour of the abundance patterns that are due to the complexity of the ISM medium, i.e. differences in depletion and/or metallicity. We discuss this further in Sect. \ref{sec ISM mix}.

A general physical reasons for the presence of gas with lower metallicities in galaxies could be the infall of low-metallicity, possibly near-pristine gas \citep{Fox17}. While gas outflows due to galactic winds are observed in the LMC \citep{Lehner07,Lehner09,Barger16}, it is not clear whether infall of gas is substantial. \citet{Howk02} and \citet{Lehner09} report that infall is rarely observed toward the LMC. In general, gas infall is necessary to explain the chemical evolution of the Milky Way \citep[e.g.][]{Tinsley80}. The time scale of ISM mixing depends on the Milky Way rotation \citep{Edmunds75}, and the survival of pockets of low-metallicity gas depends on the gas masses involved \citep{White83}, and thus the gas accretion rate. \cite{DeCia21} suggest that the observed rate of gas accretion on the Milky Way is largely sufficient for the survival of chemical inhomogeneities in the gas. Galactic fountains can also play an important role for the recycling and mixing of gas and metals, so that potentially some of the inflowing gas could be recycled rather than accreted \citep[e.g.][]{Zech08,Lehner22,Marasco22}. It is possible that the low-metallicity gas that we find towards some lines of sight in the outskirts of the SMC is somehow related to the Magellanic Bridge or a low-metallicity gas halo. The characterization of the origin of the low-metallicity gas in the SMC is beyond the scope of this paper. On the other hand, metallicity variations can be also due to interaction or star formation. Variations of about 0.4~dex in metallicities of young stars (younger than 100 Myr) are observed in the Milky Way, where higher metallicities correlate with the density of the spiral arms, and are likely produced by star formation \citep{Poggio22}. A young stellar population with sub-solar metallicities is observed in the Milky Way and possibly results from a gas infall event \citep{Spitoni22}.

The stellar metallicities in the MCs vary, strongly depending on stellar age, as the metal content of these galaxies build up in time and the galaxies chemically evolve. The mean metallicities of B-Type stars are ${\rm [Fe/H]}= -0.3$ in the LMC and $-0.7$ in the SMC \citep{Rolleston02,Trundle07,Hunter07} and they agree well with the abundances of \hii{} regions \citep{ToribioSanCipriano17}. Other studies found somewhat lower metallicities of B stars \citep[{[}Si/H{]} abundances between $-0.8$ and $-0.4$ for the LMC,][]{Korn00}. The mean metallicity of hot late Type-O stars in the Tarantula Nebula (in clusters NGC 2060 and 2070) in the LMC is ${\rm [Si/H]}=-0.46 \pm 0.04$ \citep{Markova20}. The youngest (1--3 Gyr old) red giant branch (RGB) stars in the LMC have metallicities $\mbox{[Fe/H]}\sim-0.48$ with a small dispersion $\sigma_{\rm [Fe/H]}=0.09$ \citep{Grocholski06}. In the MCs, RGB stars younger than 3--5~Gyr in stellar clusters have a rather flat metallicity gradient and [Fe/H] abundances between $-0.6$ and $-0.8$ for the LMC, and between $-1.2$ and $-0.6$ for the SMC \citep{Parisi09,Cioni09}. Asymptotic giant branch stars, although with probably older ages, show lower abundances and wider spreads, between  $-1.5<\mbox{[Fe/H]}<-1.0$ for the LMC and $-1.5<\mbox{[Fe/H]}<-0.8$ for the SMC \citep{Cioni09}. The metallicities of giant stars in the youngest ($<3$~Gyr) globular clusters spreads between $-1.2<\mbox{[Fe/H]}<0$ \citep{Hill00}. Cepheid stars with an age of 7--8 Gyr have metallicities between $-0.6<\mbox{[Fe/H]}<-0.2$ in the LMC \citep{Romaniello22}.

\subsection{The $\alpha$-element enhancement and Mn under-abundance}
\label{sec alpha}

Type Ia SNe produce most of the Fe-group elements, and including Mn \citep[e.g.][]{Nomoto97}. On the other hand, core-collapse SNe produce many more $\alpha$ elements with respect to Fe, and much less Mn  \citep{Nomoto06}. As a result, $\alpha$-element enhancement and Mn under-abundance are expected in the gas in the early phases following a burst of star formation, before SNe Ia have the time to explode and contribute with the Fe-group elements and consequently cancel out the $\alpha$-element enhancement and Mn under-abundance \citep[e.g.][]{Tinsley79,Kobayashi20,Matteucci21}. Consistently, $\alpha$-element enhancement and Mn under-abundance is observed in stars in the Milky Way \citep[e.g][]{McWilliam97} as well as in nearby dwarf galaxies \citep{deBoer14,Tolstoy09,Hill19}.

We measure $\alpha$-element enhancement in the Magellanic gas that are overall higher than in stars, compared at similar metallicities. The $\alpha$-element enhancement in the gas was likely produced by massive stars that have exploded as core-collapse SNe. In the golden sample we observe deviations from the linear fits to the abundance patterns for both MCs, in the form of $\alpha$-element enhancement. In addition, in the SMC we observe Mn under-abundance, while for the LMC no Mn measurements are available in our dataset. We determine the overall $\alpha$-element enhancement from the difference in metallicity derived from the linear fit to the abundance patterns based on the $\alpha$ elements only and the linear fit (with the same fixed slope) to the Fe, Ni, and Cr abundances (see Figs. \ref{fig xy LMC} and \ref{fig xy SMC}, and top panel of Fig. \ref{fig [M/H] comparison}). The $\alpha$-element enhancements are reported in Tables \ref{tab LMC} and \ref{tab SMC} for the golden sample. Their values are up to 0.40~dex  (0.49~dex) for the LMC (SMC), and on average 0.26~dex (0.35), with a standard deviation of $0.08$~dex. In the SMC golden sample we measure a mean Mn under-abundance of $-0.35$~dex, with a standard deviation of 0.07~dex.

\begin{figure}
   \includegraphics[width=0.9\columnwidth]{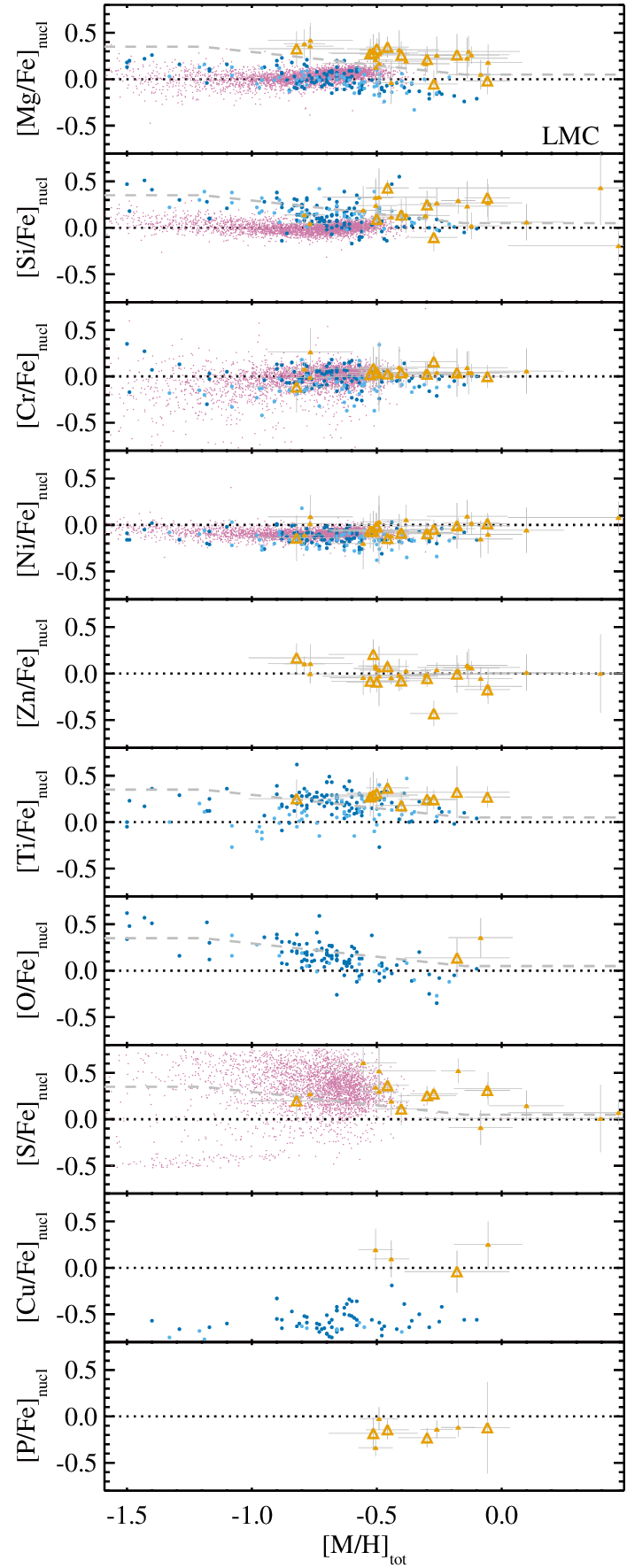}
      \caption{The relative abundances with respect to Fe versus the total metallicity in the LMC ISM (orange triangles), after correcting for dust depletion. Larger symbols highlight the golden sample. The blue circles show the stellar measurements of field red giant stars older than 1 Gyr in the LMC bar (dark blue) and disc (light blue) from \citet{VanderSwaelmen13} and \citet{Pompeia08}. The magenta points are the stellar measurements from APOGEE in the LMC, with the selection of \citet{Hasselquist21}. The APOGEE stellar measurements for S and Cr are less reliable. The grey dashed line shows the average [$\alpha$/Fe] for stars in the Milky Way from \citet{McWilliam97}. Potential variations of [P/Fe] and [Zn/Fe] due to nucleosynthesis are observable only in the golden sample.}
         \label{fig [X/Fe]nucl LMC APOGEE}
   \end{figure}

\begin{figure}
   \includegraphics[width=\columnwidth]{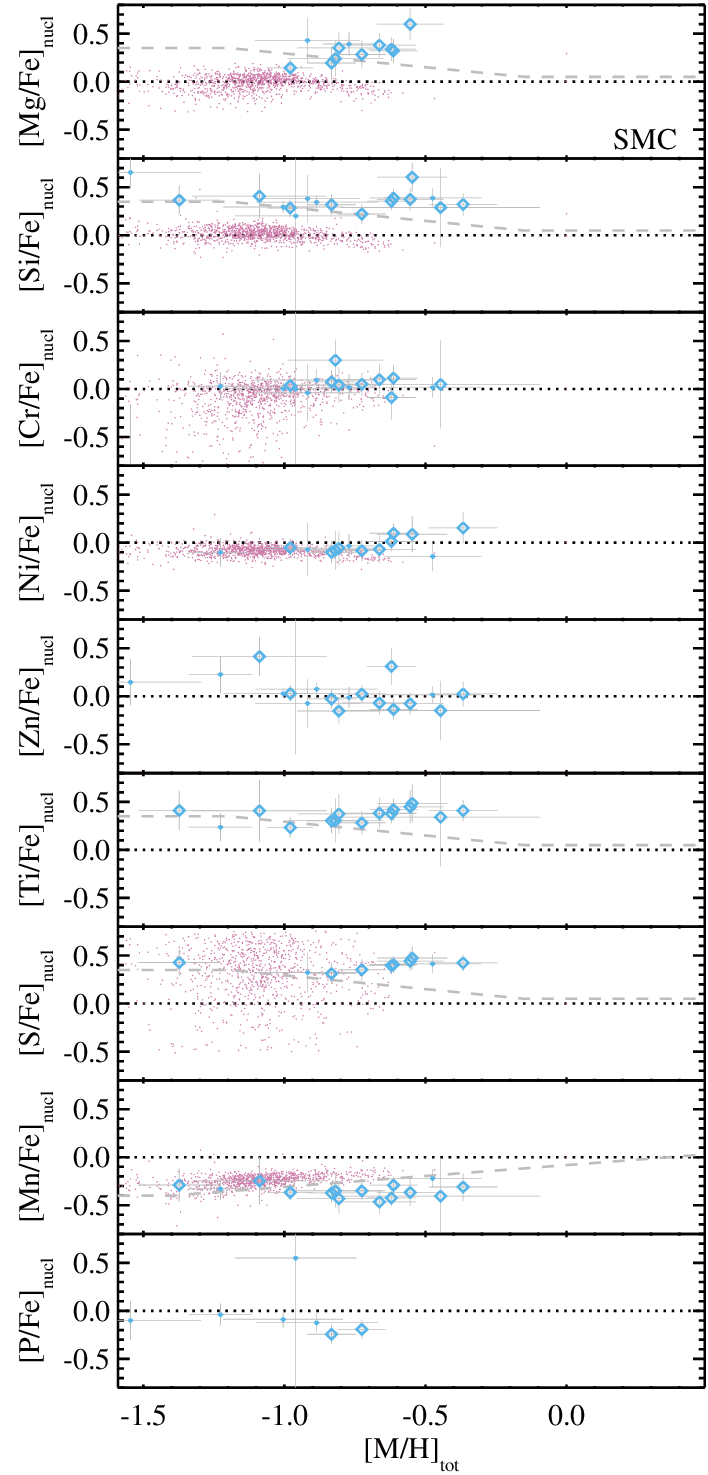}
      \caption{Same as Fig. \ref{fig [X/Fe]nucl LMC APOGEE}, but for the SMC (blue diamonds). The light-grey dashed line for [Mn/Fe] shows the average stellar measurements from \citet{Mishenina15}.}
         \label{fig [X/Fe]nucl SMC APOGEE}
   \end{figure}

In Figs. \ref{fig [X/Fe]nucl LMC APOGEE} and \ref{fig [X/Fe]nucl SMC APOGEE} we present the relative abundances with respect to Fe in the ISM in the LMC and SMC, respectively, after correcting for dust depletion, [$X$/Fe]$_{\rm nucl}$, which are likely due to SN nucleosynthetic effects. These values are also listed in Tables \ref{tab LMC XFe_nucl} and \ref{tab SMC XFe_nucl}. These relative abundances are derived from the residuals of the linear fits to the abundance patterns (Figs. \ref{fig xy LMC} and \ref{fig xy SMC}). While the observed gas-phase relative abundances, [$X$/Fe], are typically distorted (and often dominated) by dust depletion, the values of [$X$/Fe]$_{\rm nucl}$ can be directly compared to stellar measurements. Figure \ref{fig [X/Fe]nucl LMC APOGEE} shows $\alpha$-element enhancement and Mn under-abundance in the ISM of the MCs up to high metallicities, at much higher metallicities than what is observed from stellar observations of the MCs, and to some extent even at higher metallicities than in the Milky Way. Figure \ref{fig [X/Fe]nucl LMC APOGEE} shows stellar [$X$/Fe] measurements for the LMC bar and disc from \citet{VanderSwaelmen13} and \citet{Pompeia08}, based on high-resolution VLT/FLAMES spectra of field red giant stars older than 1 Gyr. Stellar measurements from the SDSS-IV Apache Point Observatory Galactic Evolution Experiment \citep[APOGEE,][]{Majewski17} survey for the LMC and SMC are also shown in  Figs. \ref{fig [X/Fe]nucl LMC APOGEE} and \ref{fig [X/Fe]nucl SMC APOGEE} for comparison.  \citet{Nidever20} and \citet{Hasselquist21} measure [$\alpha$/Fe] and [Fe/H] from high resolution spectra of a few thousands red giant stars in the LMC and SMC, as part of APOGEE. We adopt the same target selection of \citet{Hasselquist21}. Measurements for S and Cr from APOGEE are less reliable.\footnote{https://www.sdss.org/dr17/irspec/abundances}

The extent and occurrence of the $\alpha$-element enhancements in our samples can be assessed from Figs. \ref{fig xy LMC}, \ref{fig xy SMC}, \ref{fig [X/Fe]nucl LMC APOGEE} and \ref{fig [X/Fe]nucl SMC APOGEE}. When observed, Mn appears to be always underabundant in our SMC sample (Fig. \ref{fig [X/Fe]nucl SMC APOGEE}). We observe $\alpha$-element enhancements is in all 14 systems in our SMC golden sample, and in 9 out of 10 systems in our LMC golden sample. Measuring the $\alpha$-element enhancements outside the golden sample is not as reliable, because the fit to the abundance pattern is determined from most elements, washing away deviations due to potential $\alpha$-element enhancements. Overall, all the 18 SMC systems with measurements of $\alpha$-elements show some evidence for enhancements of these elements with respect to Fe (see Fig. \ref{fig xy SMC}). For the LMC, in two out of 32 systems $\alpha$-element enhancements can convincingly be excluded, namely Sk-70~79 and Sk-67~101 (see Fig. \ref{fig xy LMC}). In the system toward Sk-68~52, which is the golden sample, the absence of $\alpha$-element enhancement is hard to establish, because it may arise from a discrepancy between the observed abundance of Cr, which is significantly higher than the abundances of Fe and Ni and may cause an underestimation of the $\alpha$-element enhancement. The low [Zn/H] in this system makes this a viable explanation.

Enhancement of $\alpha$ elements and under-abundance of Mn in the MCs are also observed by \citet{Konstantopoulou22} from an overall study of the depletion sequences (overall relations between the depletions and [Zn/Fe]) in different environments. These authors consider the properties of their MC samples as a whole and find some deviations from the general behaviour of the depletion sequences for $\alpha$ elements and Mn of the MCs, compared to other environments, which they interpret as evidence for $\alpha$-element enhancement and Mn under-abundance. They do not analyse the abundance patterns of individual systems. Here we find that $\alpha$-element enhancement and Mn under-abundance are evident for the individual systems in the Magellanic Clouds. \cite{Jenkins17} also notice a deviation in depletion on Mn for the SMC (i.e. a potentially steeper Mn depletion), which they interpret as due to a potential affinity of Mn with different dust species. Exotic properties of the dust depletion in the MCs that coincidentally mimic $\alpha$-element enhancements and Mn under-abundance is an alternative possibility to explain the deviations from the abundance patterns that we observe. However, in that scenario the total metallicities in the gas would have to be highly supersolar and with amount of dust higher than what is observed even in the dustiest parts of the Milky Way (see Sect. \ref{sec caveats}). The possibility that recent star formation (which is observed in the MCs) produces the systematic $\alpha$-element enhancement and Mn under-abundance that we observe in the gas is far more likely.

Overall, the stellar [$\alpha$/Fe] in the MCs are lower than what is measured in Milky Way red giants \citep{Pompeia08,Lapenna12,Chekhonadskikh12,VanderSwaelmen13}.  The [$\alpha$/Fe] values decrease slowly with increasing metallicity \citep[][see Fig. \ref{fig [X/Fe]nucl LMC APOGEE}]{VanderSwaelmen13,Hasselquist21}. The [$\alpha$/Fe] values measured in B-Type stars in the MCs are sub-solar, around $-0.12$ \citep{Korn00,Rolleston02}. The knee of the $\alpha$-element distribution with metallicity is at $Z\sim0.1\,Z_\odot$ for the Milky Way, as shown by the dashed curves in Figs. \ref{fig [X/Fe]nucl LMC APOGEE} and \ref{fig [X/Fe]nucl SMC APOGEE}. For less massive galaxies the stellar $\alpha$-element knee is expected to be at lower metallicites \citep[e.g.][]{deBoer14}. For the MCs the stellar $\alpha$-element knee is at very low metallicity, consistently from both the APOGEE and the \citet{VanderSwaelmen13} data, and not visible in Figs. \ref{fig [X/Fe]nucl LMC APOGEE} and \ref{fig [X/Fe]nucl SMC APOGEE}. \citep{Nidever20} estimated that the $\alpha$-element knee in stars in the LMC is at [Fe/H] $\sim-2.2$,  indicating a low star formation efficiency.

In the LMC, \citet{Nidever20} and \citet{Hasselquist21} find a change of trend and the stellar [$\alpha$/Fe] increases by $\sim0.1$~dex at the highest metallicities. These data are displayed in Figs. \ref{fig [X/Fe]nucl LMC APOGEE} and \ref{fig [X/Fe]nucl SMC APOGEE}. Chemical evolution models suggest that strong bursts of recent star formation in both the LMC and SMC are needed to explain the observed [$\alpha$/Fe] trends with [Fe/H] \citep{Nidever20,Hasselquist21}. We discuss the star formation history of the MCs in Sec. \ref{sec SFH}.

\citet{Asad22} measure $\alpha$-element enhancement and Mn under-abundance in LMC stellar clusters, which are broadly consistent with the values of \citet{VanderSwaelmen13}. Enhancement of $\alpha$ elements in stellar clusters has also been observed in some extragalactic young clusters \citep[e.g.][]{Larsen06,Larsen08,Hernandez17}.

In the LMC, stellar [Ti/Fe] and [Si/Fe] are somewhat higher than [Mg/Fe] and [O/Fe], likely because of a lower contribution of massive core-collapse SNe \citep{VanderSwaelmen13}. While O and Mg are produced by quiescent He burning in very massive core-collapse SNe progenitors \citep{Woosley95}, Ti and Si are largely produced in the core-collapse SNe explosion of intermediate-mass progenitors \citep[e.g.][]{Nomoto06}. Eventually, most of the [Ti/Fe]$_{\rm nucl}$ values that we measure in the ISM are similar to the values of stellar [Ti/Fe] measured by \citet{VanderSwaelmen13}, as showed in Fig. \ref{fig [X/Fe]nucl LMC APOGEE}. However, in the gas we observe a flat [Ti/Fe]$_{\rm nucl}$ distribution with $[{\rm M/H}]_{\rm tot}$, with enhanced [Ti/Fe]$_{\rm nucl}$ at the highest metallicities as well. Finally, the enhancements in [Ti/Fe] and [Si/Fe] that we observe exclude the possibility that the $\alpha$-element enhancements in the ISM could be produced by winds of massive stars before their core-collapse explosions.

If there would be any intrinsic variations of Zn or P due to nucleosynthesis, these can be observed in our golden sample, when Zn and P are excluded from the fit to the abundance patterns. In the golden sample, we do not observe any enhancement in [P/Fe] in the LMC ISM (Fig. \ref{fig [X/Fe]nucl LMC APOGEE}), unlike in the stellar measurements of \citet{Caffau11}. Enhancement of P was also observed by \citet{Masseron20} for stars with peculiarly high enhancement of $\alpha$ elements. We do not confirm the relation of P nucleosynthesis with massive stars \citep{Cescutti12} based on the ISM abundances measured in the MC. We discuss the case of P and Zn further in Sect. \ref{sec caveats}. 

Strong deviations in [Zn/Fe]$_{\rm nucl}$ are not observed in the golden sample, as shown in Figs. \ref{fig [X/Fe]nucl LMC APOGEE} and \ref{fig [X/Fe]nucl SMC APOGEE}. We observe a potential trend of a deviation of Zn of up to 0.2~dex at lower metallicities, and down to $-0.2$~dex at higher metallicities. This could be consistent with the finding on stellar measurements of \citet{Sitnova22}, but needs to be investigated further with more data. We discuss this possibility further in Sect. \ref{sec caveats}.

Our results suggest that a burst of recent star formation may have increased the $\alpha$ elements in the ISM. We discuss the recent bursts of star formation observed in the MCs in Sec. \ref{sec SFH}. Following the most recent burst of star formation, it is possible that the ISM could show more $\alpha$-element enhancement than the stars produced in the same burst of star formation, while the next generation of stars should reflect the present-day ISM (relative) abundances. The lines of sight in our sample are selected towards hot stars, and thus preferentially cross star-forming regions. Whether and on which time scales the $\alpha$ elements can mix throughout the ISM, and what is the expected chemical composition of the next generation of stars forming from this gas depends, among other things, on the gas and metals masses involved. This needs further investigations, which are beyond the scope of this paper.

\subsection{Recent star formation in the Magellanic Clouds}
\label{sec SFH}

Gas-rich dwarf irregular galaxies tend to have bursty star formation \citep{Weisz14,Atek22}. Both the LMC and SMC experienced recent bursts of star formation. Stars in the Tarantula Nebula in the LMC have ages up to $\sim30$ Myr \citep{DeMarchi11}, with  the youngest being 1 to 2 Myr old, in the R136 group   \citep{Massey98,Crowther16,Bestenlehner20}. The progenitor of the core-collapse SN 1987A in the LMC is likely $\sim12$ Myr old \citep{Panagia20}. In both LMC and SMC there is an increased and common population of Cepheid with an age of $\sim200$ Myr \citep{Joshi19}. The LMC likely experienced a most intense burst of star formation in the last 2~Gyr \citep{Harris09,Nidever21,Hasselquist21}, and possibly more recently $\sim10$ Myr and $\sim100$ Myr ago \citep{Indu11}. The SMC had an increase in star-formation rate in the last 3--5 Gyr \citep{Harris04,Cignoni13,Rubele18,Hasselquist21}, with bursts $\sim2.5$, $\sim0.4$ Gyr, $\sim60$ Myr and $\sim10$ Myr ago \citep{Harris04,Indu11}. Some of this star-formation activity is probably related to the interaction between the LMC, SMC, and the Milky Way, which played a role in the creation of the Magellanic Stream $\sim1.7$~Gyr ago \citep{Nidever08}. A close encounter or interaction between the MCs could have happened over the past 1-2 Gyr \citep{Besla12,Pardy18,Lucchini21}. Gas outflows are observed around the LMC \citep{Barger16}, also suggesting a recent burst of star formation.

While there is evidence for very recent star formation in the MCs, it is not straightforward to estimate which stellar population could be responsible for the $\alpha$-element enhancement in the ISM that we observe, including the timescale for the diffusion of the $\alpha$ elements in the ISM. The bursts of star formation that produced $\alpha$-element enhancements in the Magellanic ISM are likely more recent than $\sim1$ Gyr, before SNe Type Ia could contribute in rising the Fe-group elements. Further modelling is required to assess which stellar population and masses could reproduce the current observations of the ISM.

\subsection{Dust, metals, and \hh{} molecules.}
\label{sec H2}

\begin{figure}
   \centering
   \includegraphics[width=0.5\textwidth]{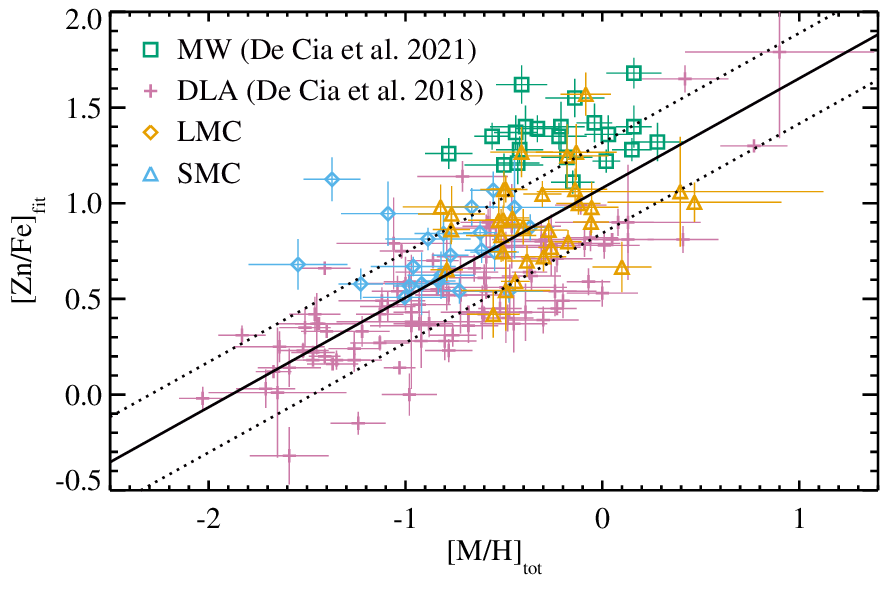}
      \caption{Relation between the metallicity, [M/H]$_{\rm tot}$, and the strength of dust depletion, [Zn/Fe]$_{\rm fit}$, in the neutral ISM in the Milky Way, LMC, SMC, and DLAs. The solid line shows a linear fit to the data, including uncertainties on both axis, $[{\rm Zn/Fe}]_{\rm fit} =  1.08 + 0.57 \times [{\rm M/H}]_{\rm tot}$, with uncertainties of 0.03 and 0.04 for the intercept and slope, respectively. The dotted lines mark the region within the resulting 0.26 internal scatter of the relation. The Milky Way data are from \citet{DeCia21}. The DLA data are from \citep{DeCia18b} and they show the observed [Zn/Fe].}
      
         \label{fig ZnFe met}
   \end{figure}

Dust is made of metals and the surface of dust grains facilitates the formation of molecules. Here we compare the metallicity, dust depletion, and molecular components of the ISM in local galaxies. Figure \ref{fig ZnFe met} shows that the total metallicity broadly correlates with the strength of dust depletion, albeit with a large scatter. Thus, the amount of dust in the ISM overall depends on the metallicity of the gas. This is in consistent with a significant amount of dust being built through grain-growth in the ISM, as suggested for example by \citep{DeCia13,Dwek16,Mattsson14}, depending on the gas metallicity, density, and temperature. A similar relation between [Zn/Fe] and metallicity has previously been observed in DLAs \citep[e.g.][]{Ledoux02,Noterdaeme08,DeCia16}.\footnote{The correlation observed in Fig. \ref{fig ZnFe met} cannot be produced by covariance between the metallicity and the dust depletion, which are the resulting intercept and slope of the linear fit to the abundance pattern. Indeed, there is a similar correlation (with a different slope) between the observed [Zn/Fe] and [Zn/H] in the ISM in the MCs, with a similar internal scatter.} The data for the Milky Way do not strongly follow the dust-metallicity relation, and show higher levels of depletion than DLAs, even at comparable metallicities (including at a fixed solar metallicity). This is possibly the result of the fact that not only metallicity, but also density, temperature, and pressure, have an important role in the growth of dust grains in the ISM. Dust is expected to grow more easily in the Milky Way disk, thanks to the higher pressure and higher density of colder gas \citep[as also observed for molecules, e.g.][]{Blitz06}, than in the more diffuse warm neutral medium probed by DLAs.

The properties of dust depletion that we find in the neutral ISM of the MCs do not appear to be strongly correlated with the dust extinction curves. For example, the line of sight towards AzV~456 has a Milky-Way dust extinction curve with a 2015 \AA{} bump, but we find no special properties of the abundance patterns in this system, as already remarked by \citet{Sofia06}, with the exception of a potential presence of chemically inhomogeneous gas (possibly including a metal- and dust-rich component) along this line of sight (see Sect. \ref{sec ISM mix}).

\begin{figure}
   \centering
   \includegraphics[width=0.5\textwidth]{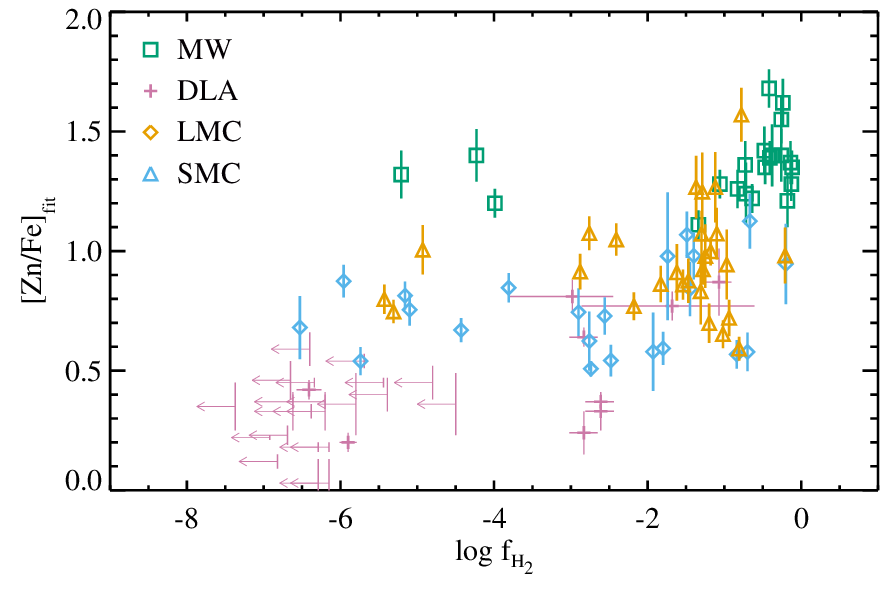}
      \caption{The relation between the strength of dust depletion in the neutral ISM and the fraction of molecular \hh{}. The Milky Way data are from \citet{DeCia21}. The DLA data are from \citet{Noterdaeme08} and \citep{DeCia18b} and they show the observed [Zn/Fe]. }
         \label{fig H2 [Zn/Fe]}
   \end{figure}

\begin{figure}
   \centering
   \includegraphics[width=0.5\textwidth]{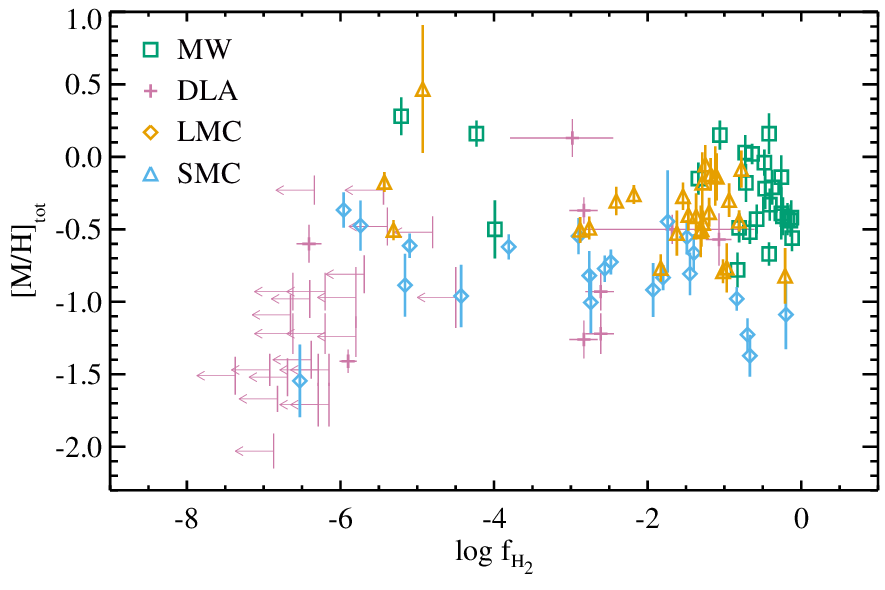}
      \caption{The relation between the total metallicity of the neutral ISM and the fraction of molecular \hh{}. Symbols are like in Fig. \ref{fig ZnFe met}. The Milky Way data are from \citet{DeCia21}. The DLA data are from \citet{Noterdaeme08} and \citep{DeCia18b}.}
         \label{fig H2 [M/H]}
   \end{figure}

The amount of molecules could in principle be related to the metallicity of the ISM, or its dust content. The rate of \hh{} formation is thought to be proportional to the metallicity and the square of the density of the gas \citep{Glover11}. Among DLAs, higher levels of dust depletion are observed in systems bearing \hh{} \citep{Noterdaeme08,Balashev19,Bolmer19}. Figures \ref{fig H2 [Zn/Fe]} and \ref{fig H2 [M/H]} show that the Milky Way harbours overall higher metallicities, dust depletion, and molecular fractions, followed by the LMC, and then the SMC.  At the same level of metallicity, the Milky Way shows a higher molecular fraction (Fig. \ref{fig H2 [M/H]}) and a higher level of dust depletion (Fig. \ref{fig ZnFe met}) with respect to the MCs and DLAs. This is likely because not only metallicity plays an important role for the production of dust and molecules, but also because of the presence of cold dense gas. In turn, the presence of cold dense gas and the molecular fraction strongly depends on pressure more than metallicity \citep{Blitz06,Balashev17}, making the formation of \hh{}  more favourable in the Milky Way disk than in DLAs \citep{Noterdaeme15}.

In our observations, the \hh{} molecular fraction $f_{\mbox{\hh{}}}$ is not clearly correlated with the metallicity (Fig. \ref{fig H2 [M/H]}) nor the amount of dust depletion (Fig. \ref{fig H2 [Zn/Fe]}). Interestingly, there are systems in the SMC that show a high molecular fraction, but low metallicity and mild levels of dust depletion, for example Sk~143 (in the SMC) and Sk-67~105 (in the LMC). In the case of Sk~143, both [Zn/Fe] and [Si/Ti] give strong indication that there is only mild dust depletion in the system, despite the high fraction of \hh{}. In the case of Sk-67~105, the oxygen abundance strongly deviates from the otherwise linear abundance pattern shown by Ni, Cr, Fe, Si, Mg, S, and Zn. Both the high molecular content and the deviation to high abundances of the most volatile elements suggest that there is actually a mix of ISM gases along this line of sight, with potentially different metallicities and levels of dust depletion. We discuss this possibility in more details in Sec. \ref{sec ISM mix}. Overall, while singly ionized species such as \znii{} and \feii{} trace the warm neutral medium, it is very likely that \hh{} does not form or survive in this medium, but rather in a colder and denser phase. This would naturally explain the lack of clear correlations in Figs. \ref{fig H2 [Zn/Fe]} and \ref{fig H2 [M/H]}.

\subsection{A mixture of different ISM gas types}
\label{sec ISM mix} 
 
The ISM is a complex medium, and likely chemically inhomogeneous in its metal and dust content. The ISM in the Milky Way has regions with different metallicities \citep{DeCia21} and strengths of dust depletion \citep{Jenkins09,Welty20}. Variations of metallicities \citep{SanchezAlmeida15,Kreckel19,Wang22,Chartab22} and dust depletion \citep[][]{Prochaska03b,Dessauges-Zavadsky06,Rodriguez06,Wiseman17,deUgartePostigo18,Noterdaeme17,Guber18,Ramburuth-Hurt23} have been observed within external galaxies as well. It is well possible that the total line of sight towards a star intercepts gases with different chemical properties. Indeed, \citet{DeCia21} suggest that this is often the case in the Milky Way, and that the ISM may be a mix of gases with different metallicities and depletion properties, which are averaged out when studying the overall metal absorption-line profiles. The actual mass of the different gas components is not straightforward to estimate \citep{DeCia22}.

\begin{figure}
   \centering
   \includegraphics[width=0.5\textwidth]{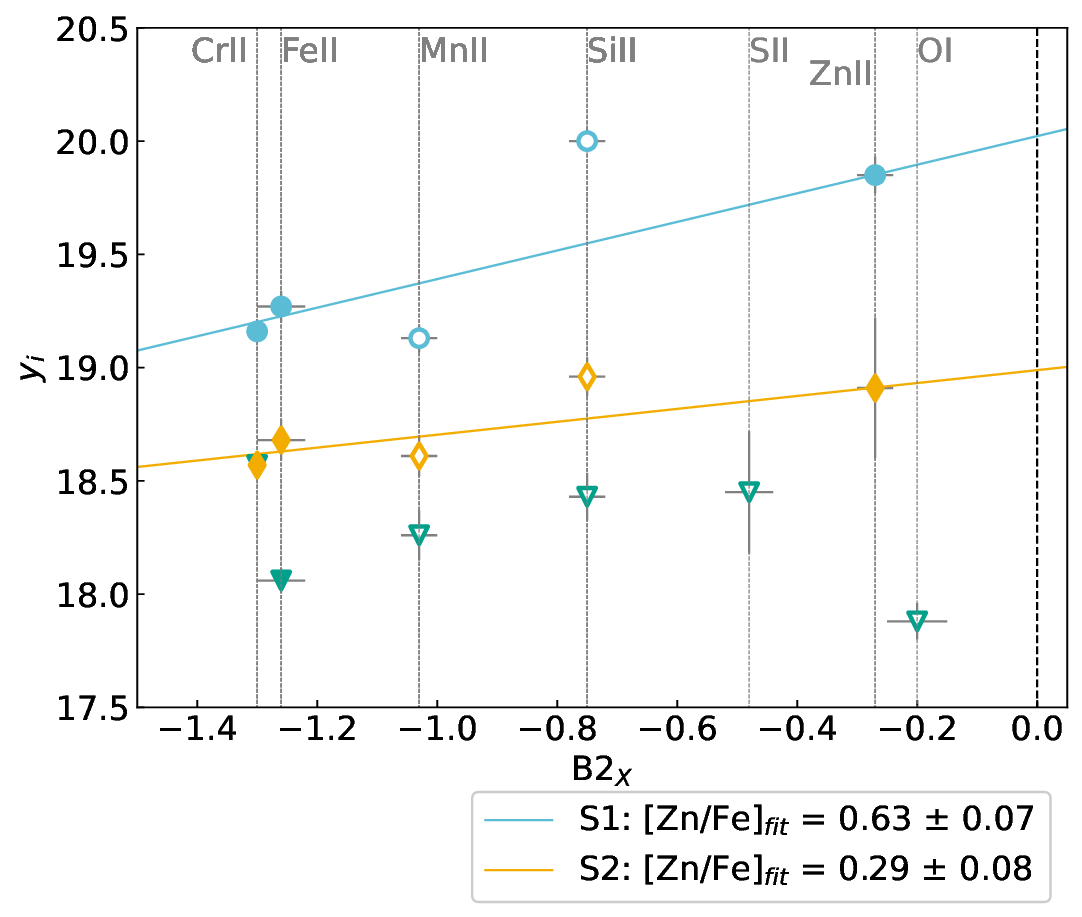}
      \caption{The metal pattern of the three main individual components observed towards AzV~332, based on the technique of \citet{Ramburuth-Hurt23} and using the metal column densities measured by \citet{Welty97}. The $y$-axis represents the equivalent metal column densities.}
         \label{fig AzV332}
   \end{figure}

Differences in the amount of depletion and type of depletion patterns have been observed in different groups of components in the line profiles observed at high spectral resolution towards AzV~332 (i.e. Sk~108) within the SMC \citep{Welty97}, a target which is also in our sample, and SN 1987A within the LMC \citep{Welty99}. Variations of depletion within absorbing systems have been observed for distant galaxies by \citet{Ramburuth-Hurt23}. Figure \ref{fig AzV332} shows our analysis of the metal patterns of the groups of individual component using the measurements of \citet{Welty97}, and with the same technique of \citet{Ramburuth-Hurt23}. This technique analyses the metal patterns in a very similar way we analyse the abundance patterns, but without including H, so that it cannot give information on metallicities, but it characterises the dust depletion and any deviations due to nucleosynthesis, in particular for individual components. Component S1 carries about 90\% of the metals \citep[as shown from the normalization of the metal patterns in Fig. \ref{fig AzV332}, and as remarked by][]{Welty97} and shows the strongest strength of depletion ([Zn/Fe]$_{\rm fit}$, i.e. the slope of the metal patterns), as well as the stronger $\alpha$-element enhancement and Mn underabundance than the other components.

The presence of multiple ISM components adds one more level of complexity to the abundance patterns that we observe. For example, there may be the contribution of high-metallicity and highly-depleted gas superimposed to the contribution of metal-poor, dust-free gas. This is illustrated in Fig. \ref{fig abu}, panel d). In this example the metal-rich gas dominates the abundances of the volatile elements, while the metal- and dust-poor gas dominates the refractory ones. This effect can cause deviations of the most volatile elements from the overall linear abundance patterns in the Milky Way ISM \citep{DeCia21,DeCia22}, although this is not a unique explanation. Different mixture of ISM gas types with different properties are possible, including gas with low dust depletion but high metallicity, as possibly the result of dust destruction.

The observations of elements more volatile than Zn is limited in our sample. A strong deviation of the observed oxygen abundance from the abundance patterns is observed towards Sk-67 105 in the LMC (Fig. \ref{fig xy LMC}), which may be a sign of a chemically inhomogeneous ISM mix along this line of sight. However, this \oi{} measurement is based on the weak 1355 \AA{} line and it is uncertain. Oxygen is constrained also for Sk-70~79 and Sk-68~73 and in both cases is consistent with [$M$/H]$_{\rm tot}$ within the uncertainties.

For AzV~456 in the SMC, an alternative interpretation of the Ti over-abundance could be a potential bend of the abundance pattern due to ISM mixes. While the presence of ISM mixes might be the case also for other lines of sight, it is not clear from the available data.

In this work, by analysing the column densities of the whole line profiles, we have assumed that the ISM along the line of sight is chemically homogeneous. Nevertheless, the $\alpha$-element enhancement and Mn under-abundance that we measure often towards both the LMC and SMC cannot be possibly explained with different ISM components, so these results hold even if the ISM turns out to be chemically inhomogeneous.

\section{Caveats}

\label{sec caveats}

It is challenging to understand the observed ISM abundance patterns, especially when fewer metal species are observed, and measure reliable slopes to the linear fits to the data. For the golden sample it is possible to establish the slope purely based on the $\alpha$ elements (Ti and at least another more volatile $\alpha$ element, either Mg, Si, or S), thereby removing the uncertainties on the determination of the dust depletion due to different nucleosynthesis. Moreover, for these systems, the determination of the metallicity is based on the Fe-group elements only, and independently from P and Zn, elements with a potentially more uncertain nucleosynthetic origin. We discuss the main uncertainties and caveats below.

\subsection{On the relative and $F*$ methods}

Figure 2 of \citet{Jenkins17} shows the quantity $D(X)$, which is used by the $F*$ method to measure dust depletion. The quantity $D(X)$ is determined from the fixed metallicity assumed for the gas and the observed [$X$/H], which traces not only dust depletion, but potentially also metallicity variations and alpha-element enhancements. Variations in $D(X)$ may be caused also by variations in metallicity and $\alpha$-element enhancement. We find that $\alpha$-element enhancements, variations in ISM metallicity, and a mild dependence of depletion on metallicity can produce a change of slope with respect to the Milky Way of the trend of $D({\rm Ti})$ with $D({\rm Fe})$, and not for $D({\rm Si})$, an overall effect which is mildly observed also in Fig. 2 of \citet{Jenkins17}. 

The relative method \citep{DeCia21} determines the neutral ISM metallicity and dust depletion as separate parameters, i.e. the normalization and slope, where the dust depletion (slope) is purely determined from the relative abundances. The deviations of specific elements from the linear fit to the abundance patterns, for example $\alpha$-element overabundances and Mn underabundance, can reveal additional nucleosynthesis effects. This is illustrated in Fig. \ref{fig abu}.

The difference in assumptions of the relative and $F*$ methods may in principle lead to different interpretations. The two methods agree fairly well in their estimations of the metallicities in the neutral ISM of the Milky Way \citet{DeCia21}. There is an overall correlation between the refractory indexes of the two methods, $B2_X$ for the relative method and $A_X$ for the $F*$ method. In particular, in the formalism of the relative method Si and Mg are found to be more volatile, and Mn more refractory than what is estimated in the $F*$ formalism for the Milky Way. In this work we measure with the relative method an over-abundance of Si and Mg by up to 0.4~dex, and an under-abundance of Mn in the neutral ISM towards the MCs. If we would assume that Si and Mg would be more refractory (more negative $B2_X$) and that Mn would be more volatile (less negative $B2_X$) like in the $F*$ formalism for the Milky Way, than we would observe a larger enhancement of Si and Mn, and a larger under-abundance of Mn. The values of the $F*$-method refractory indexes $A_{\rm Mg}$ and $A_{\rm Si}$ for the SMC are higher (and lower for $A_{\rm Mn}$) than in the Milky Way, however \citep{Jenkins17}. Because the $F*$ method measures the depletion of metals from the observed [$X$/H], any variation in abundances of Mg, Si, Mn, etc. are by construction interpreted as an effect of dust depletion in the $F*$ formalism and affect the values of $A_X$. Thus, it is not possible to estimate potential $\alpha$-element enhancement in the MCs with the $F*$ method itself and compare it with our results.

The determination of the $B2_X$ values in \citet{DeCia16} and \citet{Konstantopoulou22} uses data from different environments, including DLAs and the Milky Way. In DLAs $\alpha$-element enhancements (and Mn under-abundance) are observed for systems with no dust depletion ([Zn/Fe] $\sim0$), and these are corrected for by assuming an $\alpha$-element distribution, using the observed $\alpha$-element enhancements and assuming that the position of the so-called $\alpha$-element knee is at the same metallicity as in the Milky Way, around [M/H] $\sim -1$. The determination of the $B2_X$ values are little sensitive to this assumption, and instead are dominated from the observations of dust-free DLAs and the Milky Way. Overall, the correction for $\alpha$-element enhancements is relevant for systems with [M/H] $< -1$  \citep{DeCia16,Konstantopoulou22}. In the MCs, the gas metallicities are well above this value, and the determination of total metallicity and $\alpha$-element enhancements is independent from the assumptions on nucleosynthesis of \citet{DeCia16} and \citet{Konstantopoulou22}.

\subsection{Zinc}

Our work uses the refractory index $B2_X$, which relies on the observation of [Zn/Fe] as a tracer of dust. The main issue with this is whether Zn and Fe follow each other nucleosynthetically. To some extent, Zn could be considered as behaving somewhat similarly to $\alpha$ elements, albeit with a small amplitude \citep[e.g.][]{Duffau17,Sitnova22}. The DLA systems are observed down to [Zn/Fe] $\sim0$, i.e. with no dust depletion, and these show $\alpha$-element enhancement and Mn under-abundance of about 0.3~dex. Similar enhancements of Zn can be excluded, but smaller effects ($<0.2$~dex) could be at play and we discuss it further below. We further discuss the reliability of [Zn/Fe] as a dust tracer in in more details the Appendix. 

In Sect. \ref{sec alpha} we mentioned a potential trend in the deviation of Zn of up to 0.2~dex at lower metallicities, and down to $-0.2$~dex at higher metallicities, potentially due to nucleosynthesis, which is tentatively observed in Fig. \ref{fig [X/Fe]nucl LMC APOGEE} for the golden sample. For the non-golden sample, the fit to the abundance patterns and the determination of the total metallicities relies on Zn. In case such deviations of Zn would be present in the non-golden sample, the estimates of the total metallicities could be about 0.2~dex higher at the high-metallicities end, and about 0.2~dex lower at the low-metallicities end, thereby increasing the variations of total metallicities in the neutral ISM. The total metallicities for the non-golden sample are shown in Fig. \ref{fig met LMC} and \ref{fig met SMC}.

In the case that [Zn/Fe] would have an intrinsic trend varying up to $+0.2$~dex in low-metallicity systems and down to $-0.2$~dex for high metallicity systems, the initial determination of $B2_X$ \citep{DeCia16,Konstantopoulou22} could also be affected, so that all depletions could be slightly underestimated, i.e. all $B2_X$ could be somewhat lower (steeper depletion sequences). This could lead to somewhat higher total metallicities in the analysis of the abundance patterns. On the other hand, the estimates of depletion of the most refractory elements (e.g. Ti) towards the most dusty lines of sight in the MW are similar between the determinations based on [Zn/Fe] \citep{DeCia16,Konstantopoulou22} and the determinations based on the $F*$ factor \citep{Jenkins09}. This ensures that if any intrinsic [Zn/Fe] trend with metallicity is present, it does not heavily affect the absolute values of the depletions. Moreover, a global shift of all the $B2_X$ values would not affect the relative abundances of the metals in the analysis of the abundance pattern, and thus our results on the $\alpha$ element enhancements are robust.

The determination of $B2_X$ is based on the assumption that the depletion of Zn, $\delta_{\rm Zn}$, is a linear function of [Zn/Fe] \citep{DeCia16}. The effects of varying the slope of $\delta_{\rm Zn}$ with [Zn/Fe], $B2_{\rm Zn}$, are investigated in \citet{DeCia21}. A $B2_{\rm Zn}$ twice as steep than the optimal value is highly unlikely, and it would produce metallicities that are 0.3--0.4~dex higher. A slope $B2_{\rm Zn}$ that is twice as shallow would produce metallicities 0.15--0.20~dex lower. A relation of $\delta_{\rm Zn}$ with [Zn/Fe] that is not linear, but keeps at zero for higher values of [Zn/Fe] and then turns steeper, like in Figure 5 of \citet{DeCia16} would be similar to the latter scenario, so possibly produce slightly lower dust-corrected metallicities. \citet{Roman-Duval22} explored this non-linear possibility, using a different methodology.

Zinc is not considered when calculating dust depletion, metallicity, and $\alpha$-element enhancements in the golden sample. Thus, our results do not depend on Zn measurements and their uncertainties. For the non-golden sample, in case of measurements based on data at lower spectral resolution, for example for the four SMC systems from \citet{Tchernyshyov15}, we cannot exclude hidden saturation of \znii{} lines. The abundances of Zn in these systems are similar to the abundances of P, which is as volatile as Zn. This possibly hints at no strong saturation effects of \znii{}, but may be a coincidence.

\subsection{Phosphorus}

\citet{Konstantopoulou22} find that the depletion of P is significantly shallower than what estimated by \citet{Jenkins09}, who find that P has a higher tendency of depleting than Zn.

The ISM abundances that we observe in the LMC do not show any enhancement of P. \citet{Caffau11} observe high values (up to $\sim0.4$~dex) of [P/Fe] in a dozen of Galactic stars, and interpret them as P being potentially produced through $\alpha$ capture. To reproduce these observations, \citet{Cescutti12} suggest that P is formed mainly in massive stars. Phosphorus is likely not produced in core-collapse SNe \citep{Woosley95}, but rather in the late evolutionary stages of massive stars \citep{Arnett96}. For DLAs with [Zn/Fe] <~ 0.4, for which the effects of dust depletion are minimal and the $\alpha$-element enhancements are most noticeable, \citet{Konstantopoulou22} observe a branch of DLAs (not all) with increased P(see their Figure 1). However, these measurements are mostly based on only one \pii{} absorption line in the Ly-$\alpha$ forest, and thus potentially contaminated \citet{DeCia16}. 

\subsection{Sulphur}

\citet{Sofia06} suggested that sulphur is a better reference than zinc for the estimate of the depletion of silicon. Indeed [S/Si] mildly traces dust depletion, because S and Si are both $\alpha$ elements produced by oxygen burning in core-collapse SNe, and their depletion properties are mildly different. We raise nevertheless a first caveat, that S has a non-negligible depletion \citep[e.g.][]{Jenkins09,DeCia16}. Furthermore, S has been considered a troublesome element \citep{Jenkins09,Jenkins17} because measurements of \suii{} in the local group may be affected by ionization, i.e. the contribution to the line-of-sight  N(\suii{}) from \hii{} regions. While this effect is to some extent also observed in \citet{Konstantopoulou22}, the determination of $B2_S$ is determined largely by QSO DLAs, for which such troublesome ionization effects are not at play, because the lines of sight are not preferentially crossing through \hii{} regions. While we still include S in our analysis, any strong variations due to ionization would be observable. \citet{Jenkins17} found that three SMC systems showed S over-abundance, namely AzV~321, AzV~332, and Sk~190. We interpret these as a combination of $\alpha$-element enhancement and a relatively high total metallicity (see Fig. \ref{fig xy SMC} and Table \ref{tab SMC}), while the likely ionization effects do not seem to cause strong deviations.

\subsection{Manganese, titanium, and peculiar depletion}

\citet{Jenkins17} find steeper depletion sequences of Mn in the SMC than in the Milky Way. \citet{Konstantopoulou22} find that the apparently steeper Mn depletion sequence for the SMC may be instead the effect of a Mn under-abundance. They also find the opposite trend for Ti (and possibly Si). Overall, they interpret these deviations as $\alpha$-element enhancement and Mn under-abundance, possibly due to recent core-collapse SN nucleosynthesis. The coincidence of the Mn under-abundance and the enhancement of Ti, Si, Mg, and S that we observe in the ISM of the MCs make the hypothesis of recent core-collapse SNe very appealing \citep[e.g.][]{Kobayashi20}. 

\citet{Jenkins17} also find that the depletion of Ti in the SMC is steeper than in the Milky Way, while the depletion of Si is rather similar in the two environments, which is shown in their Figure 2. In that work, the measurements of depletions are in fact the difference between the observed abundances and the fixed reference metallicity (assumed from the metallicity of hot stars), and these values depends as well on potential variations in intrinsic metallicity and other effects such as $\alpha$-element enhancements. The different behaviours of the depletion of Ti and Si observed in the SMC by \citet{Jenkins17} casts some doubts on whether the variations that we observe in the abundance patterns in the MCs are actually due to nucleosynthesis of core-collapse SNe, or whether they are due to different dust-depletion properties in the MCs. \citet{Konstantopoulou22} finds that the depletion of Ti in the MCs is significantly shallower than for the Milky Way and distant galaxies, an effect which is removed if one assumes that there are $\alpha$-element enhancements in the gas in the MCs. We test the effects of a different Ti depletion in the MCs on our results by assuming a shallower depletion of Ti \citep[$B2_{Ti} =-1.01\pm0.13$ and  $-1.11\pm0.06$ for LMC and SMC, respectively,][]{Konstantopoulou22}. In this case, for our golden sample we would estimate total metallicities in the gas that are super-solar for most systems (between 0.3 and 3~dex for the LMC and between $-0.26$ and 1.4~dex for the SMC) as well as extremely high values of the overall strength of dust depletion (with $1.3<{\rm[Zn/Fe]}_{\rm fit}<3.7$ for the LMC and  $1.1<{\rm[Zn/Fe]}_{\rm fit}<2.4$), much higher than what is observed in the dustiest lines of sight in the Milky Way. In addition, this alternative solution would imply also strong deficiencies (by more than 1~dex), of Zn and O. The simpler solution, where the deviations from the abundance patterns are regular $\alpha$-element enhancements and Mn under-abundance caused by recent star formation, is far more likely.

There is a non-negligible possibility that the [$X$/Fe]$_{\rm nucl}$ variations that we observe could not be due to recent core-collapse SNe, but to a different make-up of dust. Our targets are all hot stars, so they preferentially select star-forming regions and molecular clouds, and almost all of them show $\alpha$-element enhancements. The fraction of polycyclic aromatic hydrocarbons (PAHs, composed of C and H atoms) is lower in the SMC than in the Milky Way, and it is not uniformly distributed, being higher in molecular clouds \citep{Sandstrom10}. To explain this, the PAHs could be formed in molecular clouds, or alternatively recent star formation (younger than $\sim25$ Myr) may have destroyed the PAH in the diffuse ISM. Overall, we observe coherent enhancement of [Si/Fe]$_{\rm nucl}$, [Ti/Fe]$_{\rm nucl}$, [S/Fe]$_{\rm nucl}$, [Mg/Fe]$_{\rm nucl}$, and under-abundance of [Mn/Fe]$_{\rm nucl}$ in the neutral ISM towards star-forming regions the MCs, which can be all explained by the contribution from recent core-collapse SNe. The alternative explanation of a different dust grain composition in molecular clouds being the cause of the [$X$/Fe]$_{\rm nucl}$ variations cannot be completely excluded but is unlikely, given the consistency in these variations. Whether not only the physical conditions, but also $\alpha$-element enhancements or in general the ISM composition in molecular clouds have an influence on the PAH fraction will need to be further investigated.

\subsection{Caveats due to possible ISM heterogeneity}

Several processes shape the observed abundance patterns, as illustrated in Fig. \ref{fig abu}. It is necessary to observe several metals and with different nucleosynthetic and refractory properties to be able to properly disentangle these. The level of uncertainty is higher whenever only a few metals are observed. It is possible to discriminate $\alpha$-element enhancement and Mn under-abundance from the effects of ISM mixing. This is because ISM mixing can mimic abundance deviations of the most refractory (e.g. Ti) or the most volatile elements (e.g. O). But ISM mixing cannot selectively reproduce deviations of specific elements in the middle range of the refractory index, such as Si, Mg, and Mn, like $\alpha$-element enhancements instead do.

\section{Conclusions}
\label{sec conclusions}

In this paper we study the abundance patterns in the neutral ISM towards 32 and $22$ hot stars in the LMC and SMC, respectively, and characterise (and correct for) the depletion of metals into dust grains in these systems. For the first time, we measure the total (gas + metals) metallicities of the neutral ISM, as well as dust-corrected [$X$/Fe]$_{\rm nucl}$ in the gas that can be compared with stellar values, in particular $\alpha$-element enhancements.

We find enhancement of $\alpha$ elements (Ti, Si, Mg, and S) in the systems of our golden sample in both MCs and Mn under-abundance in the golden sample for the SMC (where Mn measurements are available), based on the deviations from the linear fits to their abundance patterns, i.e. after correcting for dust depletion. We observe these $\alpha$-element enhancements and Mn under-abundance at all metallicities, including at the high-metallicity end. The average $\alpha$-element enhancements in the ISM are of 0.26~dex (0.35~dex) for the LMC (SMC), with a standard deviation of 0.08 for both MCs. The average Mn under-abundance for the SMC is $-0.35$~dex, with a standard deviation of 0.07. This suggests that a recent burst of star formation has enriched in $\alpha$ elements (and impoverished in Mn relatively to Fe) the neutral ISM of these elements, i.e. from the nucleosynthesis contribution of core-collapse SNe, before the onset of Type Ia SNe ($\sim 1$~Gyr). This effect is partly observed in the stars in the MCs as well, although to a lesser extent, hinting to a delay in chemical enrichment between the gas and the stars. Evidence for recent star formation is also found from independent studies on gas outflows, stellar ages, and stellar evolution observations and models, and possibly triggered by the interaction between the SMC, LMC, and the Milky Way.

We measure the total metallicities in the neutral ISM in the LMC and SMC, and find that most values are consistent, within the uncertainties, with the nominal reference metallicities from OB stars and \hii{} regions ($\log Z/Z_\odot =-0.3$ and $-0.7$ for the LMC and SMC, respectively). The average dust-corrected metallicity in the neutral ISM is $[M/\mbox{H}]_{\rm tot} = -0.33$ ($-0.83$) with standard deviation 0.30 (0.30) for the LMC (SMC). 

In six systems we measure a significantly lower metallicity than nominal, two in the LMC (Sk-67~105 and Sk-69~279) with $\sim16$\% solar and four in the SMC (AzV~327, Sk~191, AzV~476, and AzV~26) with metallicities spanning between $3$ and $10$\% solar. Three of the SMC lines of sight with low-metallicity gas are located in the SMC outskirts, two of which in the direction of the Magellanic Bridge, a region previously known to have low-metallicity gas and stars \citep[e.g.][]{Lehner08,Ramachandran21}. Thus, with the exception of lines of sight towards the Magellanic Bridge, the neutral gas in the LMC and SMC appears fairly well mixed in terms of metallicity. The neutral ISM metallicity distribution in the LMC is similar to the Milky Way, while the SMC lines of sight have overall lower metallicities.

To probe the existence of low-metallicity gas in the ISM it is necessary to either exclude the volatile metals from the analysis of the abundance patterns \citep[e.g.][]{DeCia21} or model the abundance patterns with at least two components, one driven by the refractory metals that can be sensitive to the metal-poor gas, and one driven by the volatile metals that are mostly sensitive to the metal-rich gas. Indeed, the abundances of refractory elements are very low in dusty and metal-rich gas, but are higher dust-poor and/or metal-poor gas. On the other hand, the abundances of volatile metals are always higher for gas that is more-metal-rich. Studies of abundance patterns that systematically include the volatile metals in a single fit to the abundance patterns, even when non-linear deviations are evident \citep[e.g][]{Ritchey23} are not sensitive to the presence of low-metallicity gas, but are instead dominated by metal-rich gas.

We find that the overall strength of dust depletion, [Zn/Fe]$_{\rm fit}$, varies within each MC, and overall correlates with the total metallicity. Comparing different environments, at similar metallicities, the Milky Way shows higher levels of dust depletion and molecular fractions than the LMC, and then progressively the SMC and DLAs. This suggests that not only metallicity plays an important role for the formation of dust and molecules \cite[e.g.,][]{Petitjean06}, but probably also density, temperature and pressure \citep[][]{Noterdaeme15}, so a higher fraction of cold dense gas, like in the case of the Milky Way disk. We find no obvious trend of the molecular fraction with either the total metallicity or the overall strength of dust depletion. The lack of this correlation is probably due to the complexity of the ISM, having a mix of warm, cold, and molecular medium, as well as chemical inhomogeneities along the line of sight. Indeed, the presence of some low-metallicity and low-depletion gas along the line of sight can lower the total metallicity and dust depletion that we measure in the warm ISM, but does not affect the molecular gas which is mostly confined in a colder and denser gas phase.

Studying the chemical properties of the neutral ISM based on the relative abundances of metals (i.e. the relative method) allows the simultaneous characterization of dust depletion, metallicity and nucleosynthesis effects such as $\alpha$-element enhancements and Mn underabundance. Characterising the chemical properties of the neutral ISM in the SMC and LMC and eventually linking them to the chemical properties of stars that form from this gas is important for a more complete understanding of their chemical evolution. This study makes one important step forward in this direction.

\begin{acknowledgements}
      We thank the anonymous referee for an insightful and constructive report, which sharpened and further enriched this paper. We thank Edward Jenkins for the stimulating and profound discussions which helped improving the paper substantially. We thank Eline Tolstoy, Corinne Charbonnel, Nikos Prantzos, Yves Revaz, Pascale Jablonka, Nicolas Lehner, Guido De Marchi, S{\o}ren Larsen, Sviatoslav Borisov, and Kirill Tchernyshyov for insightful conversations and useful suggestions, and Mathieu Van der Swaelmen for sharing the LMC stellar data with us. A.D.C., A.V., C.K., I.J., J.K. K., and T.R.H. acknowledge support by the Swiss National Science Foundation under grant 185692 funding the "Interstellar One" project. This research has made use of ESASky \citep{Baines17,Giordano18}, developed by the ESAC Science Data Centre (ESDC) team and maintained alongside other ESA science mission's archives at ESA's European Space Astronomy Centre (ESAC, Madrid, Spain).
\end{acknowledgements}

%
%

   \bibliographystyle{aa} 
   \bibliography{biblio.bib} 

\begin{thebibliography}{166}
\expandafter\ifx\csname natexlab\endcsname\relax\def\natexlab#1{#1}\fi

\bibitem[{{Arnett}(1996)}]{Arnett96}
{Arnett}, D. 1996, {Supernovae and Nucleosynthesis: An Investigation of the
  History of Matter from the Big Bang to the Present}

\bibitem[{{Asa'd} {et~al.}(2022){Asa'd}, {Hernandez}, {As'ad}, {Molero},
  {Matteucci}, {Larsen}, \& {Chilingarian}}]{Asad22}
{Asa'd}, R., {Hernandez}, S., {As'ad}, A., {et~al.} 2022, \apj, 929, 174

\bibitem[{{Asplund} {et~al.}(2021){Asplund}, {Amarsi}, \&
  {Grevesse}}]{Asplund21}
{Asplund}, M., {Amarsi}, A.~M., \& {Grevesse}, N. 2021, \aap, 653, A141

\bibitem[{{Atek} {et~al.}(2022){Atek}, {Furtak}, {Oesch}, {van Dokkum},
  {Reddy}, {Contini}, {Illingworth}, \& {Wilkins}}]{Atek22}
{Atek}, H., {Furtak}, L.~J., {Oesch}, P., {et~al.} 2022, \mnras, 511, 4464

\bibitem[{{Baines} {et~al.}(2017){Baines}, {Giordano}, {Racero}, {Salgado},
  {L{\'o}pez Mart{\'\i}}, {Mer{\'\i}n}, {Sarmiento}, {Guti{\'e}rrez}, {Ortiz de
  Landaluce}, {Le{\'o}n}, {de Teodoro}, {Gonz{\'a}lez}, {Nieto}, {Segovia},
  {Pollock}, {Rosa}, {Arviset}, {Lennon}, {O'Mullane}, \& {de
  Marchi}}]{Baines17}
{Baines}, D., {Giordano}, F., {Racero}, E., {et~al.} 2017, \pasp, 129, 028001

\bibitem[{{Balashev} {et~al.}(2019){Balashev}, {Klimenko}, {Noterdaeme},
  {Krogager}, {Varshalovich}, {Ivanchik}, {Petitjean}, {Srianand}, \&
  {Ledoux}}]{Balashev19}
{Balashev}, S.~A., {Klimenko}, V.~V., {Noterdaeme}, P., {et~al.} 2019, \mnras,
  490, 2668

\bibitem[{{Balashev} {et~al.}(2017){Balashev}, {Noterdaeme}, {Rahmani},
  {Klimenko}, {Ledoux}, {Petitjean}, {Srianand}, {Ivanchik}, \&
  {Varshalovich}}]{Balashev17}
{Balashev}, S.~A., {Noterdaeme}, P., {Rahmani}, H., {et~al.} 2017, \mnras, 470,
  2890

\bibitem[{{Barbuy} {et~al.}(2015){Barbuy}, {Fria{\c c}a}, {da Silveira},
  {Hill}, {Zoccali}, {Minniti}, {Renzini}, {Ortolani}, \&
  {G{\'o}mez}}]{Barbuy15}
{Barbuy}, B., {Fria{\c c}a}, A.~C.~S., {da Silveira}, C.~R., {et~al.} 2015,
  \aap, 580, A40

\bibitem[{{Barger} {et~al.}(2016){Barger}, {Lehner}, \& {Howk}}]{Barger16}
{Barger}, K.~A., {Lehner}, N., \& {Howk}, J.~C. 2016, \apj, 817, 91

\bibitem[{{Becker} {et~al.}(2012){Becker}, {Sargent}, {Rauch}, \&
  {Carswell}}]{Becker12}
{Becker}, G.~D., {Sargent}, W.~L.~W., {Rauch}, M., \& {Carswell}, R.~F. 2012,
  \apj, 744, 91

\bibitem[{{Bekki} \& {Chiba}(2007)}]{Bekki07b}
{Bekki}, K. \& {Chiba}, M. 2007, \mnras, 381, L16

\bibitem[{{Bekki} {et~al.}(2004){Bekki}, {Couch}, {Beasley}, {Forbes}, {Chiba},
  \& {Da Costa}}]{Bekki04}
{Bekki}, K., {Couch}, W.~J., {Beasley}, M.~A., {et~al.} 2004, \apjl, 610, L93

\bibitem[{{Bekki} \& {Stanimirovi{\'c}}(2009)}]{Bekki09}
{Bekki}, K. \& {Stanimirovi{\'c}}, S. 2009, \mnras, 395, 342

\bibitem[{{Besla} {et~al.}(2012){Besla}, {Kallivayalil}, {Hernquist}, {van der
  Marel}, {Cox}, \& {Kere{\v{s}}}}]{Besla12}
{Besla}, G., {Kallivayalil}, N., {Hernquist}, L., {et~al.} 2012, \mnras, 421,
  2109

\bibitem[{{Bestenlehner} {et~al.}(2020){Bestenlehner}, {Crowther},
  {Caballero-Nieves}, {Schneider}, {Sim{\'o}n-D{\'\i}az}, {Brands}, {de Koter},
  {Gr{\"a}fener}, {Herrero}, {Langer}, {Lennon}, {Ma{\'\i}z Apell{\'a}niz},
  {Puls}, \& {Vink}}]{Bestenlehner20}
{Bestenlehner}, J.~M., {Crowther}, P.~A., {Caballero-Nieves}, S.~M., {et~al.}
  2020, \mnras, 499, 1918

\bibitem[{{Blair} {et~al.}(2009){Blair}, {Oliveira}, {LaMassa}, {Gutman},
  {Danforth}, {Fullerton}, {Sankrit}, \& {Gruendl}}]{Blair09}
{Blair}, W.~P., {Oliveira}, C., {LaMassa}, S., {et~al.} 2009, \pasp, 121, 634

\bibitem[{{Blitz} \& {Rosolowsky}(2006)}]{Blitz06}
{Blitz}, L. \& {Rosolowsky}, E. 2006, \apj, 650, 933

\bibitem[{{Boiss{\'e}} \& {Bergeron}(2019)}]{Boisse19}
{Boiss{\'e}}, P. \& {Bergeron}, J. 2019, \aap, 622, A140

\bibitem[{{Bolmer} {et~al.}(2019){Bolmer}, {Ledoux}, {Wiseman}, {De Cia},
  {Selsing}, {Schady}, {Greiner}, {Savaglio}, {Burgess}, {D'Elia}, {Fynbo},
  {Goldoni}, {Hartmann}, {Heintz}, {Jakobsson}, {Japelj}, {Kaper}, {Tanvir},
  {Vreeswijk}, \& {Zafar}}]{Bolmer19}
{Bolmer}, J., {Ledoux}, C., {Wiseman}, P., {et~al.} 2019, \aap, 623, A43

\bibitem[{{Brown} {et~al.}(2018){Brown}, {Alkhayat}, {Irving}, {Heidarian},
  {Bancroft Brown}, {Federman}, {Cheng}, \& {Curtis}}]{Brown18}
{Brown}, M.~S., {Alkhayat}, R.~B., {Irving}, R.~E., {et~al.} 2018, \apj, 868,
  42

\bibitem[{{Caffau} {et~al.}(2011){Caffau}, {Bonifacio}, {Fran{\c c}ois},
  {Sbordone}, {Monaco}, {Spite}, {Spite}, {Ludwig}, {Cayrel}, {Zaggia},
  {Hammer}, {Randich}, {Molaro}, \& {Hill}}]{Caffau11}
{Caffau}, E., {Bonifacio}, P., {Fran{\c c}ois}, P., {et~al.} 2011, \nat, 477,
  67

\bibitem[{{Cashman} {et~al.}(2017){Cashman}, {Kulkarni}, {Kisielius},
  {Ferland}, \& {Bogdanovich}}]{Cashman17}
{Cashman}, F.~H., {Kulkarni}, V.~P., {Kisielius}, R., {Ferland}, G.~J., \&
  {Bogdanovich}, P. 2017, \apjs, 230, 8

\bibitem[{{Cescutti} {et~al.}(2012){Cescutti}, {Matteucci}, {Caffau}, \&
  {Fran{\c{c}}ois}}]{Cescutti12}
{Cescutti}, G., {Matteucci}, F., {Caffau}, E., \& {Fran{\c{c}}ois}, P. 2012,
  \aap, 540, A33

\bibitem[{{Chartab} {et~al.}(2022){Chartab}, {Cooray}, {Ma}, {Nayyeri},
  {Zilliot}, {Lopez}, {Fadda}, {Herrera-Camus}, {Malkan}, {Rigopoulou},
  {Sheth}, \& {Wardlow}}]{Chartab22}
{Chartab}, N., {Cooray}, A., {Ma}, J., {et~al.} 2022, Nature Astronomy, 6, 844

\bibitem[{{Chekhonadskikh}(2012)}]{Chekhonadskikh12}
{Chekhonadskikh}, F.~A. 2012, Kinematics and Physics of Celestial Bodies, 28,
  128

\bibitem[{{Cignoni} {et~al.}(2013){Cignoni}, {Cole}, {Tosi}, {Gallagher},
  {Sabbi}, {Anderson}, {Grebel}, \& {Nota}}]{Cignoni13}
{Cignoni}, M., {Cole}, A.~A., {Tosi}, M., {et~al.} 2013, \apj, 775, 83

\bibitem[{{Cioni}(2009)}]{Cioni09}
{Cioni}, M. R.~L. 2009, \aap, 506, 1137

\bibitem[{{Cooke} {et~al.}(2011){Cooke}, {Pettini}, {Steidel}, {Rudie}, \&
  {Nissen}}]{Cooke11}
{Cooke}, R., {Pettini}, M., {Steidel}, C.~C., {Rudie}, G.~C., \& {Nissen},
  P.~E. 2011, \mnras, 417, 1534

\bibitem[{{Crowther} {et~al.}(2016){Crowther}, {Caballero-Nieves}, {Bostroem},
  {Ma{\'\i}z Apell{\'a}niz}, {Schneider}, {Walborn}, {Angus}, {Brott},
  {Bonanos}, {de Koter}, {de Mink}, {Evans}, {Gr{\"a}fener}, {Herrero},
  {Howarth}, {Langer}, {Lennon}, {Puls}, {Sana}, \& {Vink}}]{Crowther16}
{Crowther}, P.~A., {Caballero-Nieves}, S.~M., {Bostroem}, K.~A., {et~al.} 2016,
  \mnras, 458, 624

\bibitem[{{Cullen} {et~al.}(2021){Cullen}, {Shapley}, {McLure}, {Dunlop},
  {Sanders}, {Topping}, {Reddy}, {Amor{\'\i}n}, {Begley}, {Bolzonella},
  {Calabr{\`o}}, {Carnall}, {Castellano}, {Cimatti}, {Cirasuolo}, {Cresci},
  {Fontana}, {Fontanot}, {Garilli}, {Guaita}, {Hamadouche}, {Hathi},
  {Mannucci}, {McLeod}, {Pentericci}, {Saxena}, {Talia}, \&
  {Zamorani}}]{Cullen21}
{Cullen}, F., {Shapley}, A.~E., {McLure}, R.~J., {et~al.} 2021, \mnras, 505,
  903

\bibitem[{{da Silveira} {et~al.}(2018){da Silveira}, {Barbuy}, {Fria{\c{c}}a},
  {Hill}, {Zoccali}, {Rafelski}, {Gonzalez}, {Minniti}, {Renzini}, \&
  {Ortolani}}]{daSilveira18}
{da Silveira}, C.~R., {Barbuy}, B., {Fria{\c{c}}a}, A.~C.~S., {et~al.} 2018,
  \aap, 614, A149

\bibitem[{{de Boer} {et~al.}(2014){de Boer}, {Belokurov}, {Beers}, \&
  {Lee}}]{deBoer14}
{de Boer}, T.~J.~L., {Belokurov}, V., {Beers}, T.~C., \& {Lee}, Y.~S. 2014,
  \mnras, 443, 658

\bibitem[{{De Cia}(2018)}]{DeCia18a}
{De Cia}, A. 2018, \aap, 613, L2

\bibitem[{{De Cia} {et~al.}(2021){De Cia}, {Jenkins}, {Fox}, {Ledoux},
  {Ramburuth-Hurt}, {Konstantopoulou}, {Petitjean}, \& {Krogager}}]{DeCia21}
{De Cia}, A., {Jenkins}, E.~B., {Fox}, A.~J., {et~al.} 2021, \nat, 597, 206

\bibitem[{{De Cia} {et~al.}(2022){De Cia}, {Jenkins}, {Fox}, {Ledoux},
  {Ramburuth-Hurt}, {Konstantopoulou}, {Petitjean}, \& {Krogager}}]{DeCia22}
{De Cia}, A., {Jenkins}, E.~B., {Fox}, A.~J., {et~al.} 2022, \nat, 605, E8

\bibitem[{{De Cia} {et~al.}(2016){De Cia}, {Ledoux}, {Mattsson}, {Petitjean},
  {Srianand}, {Gavignaud}, \& {Jenkins}}]{DeCia16}
{De Cia}, A., {Ledoux}, C., {Mattsson}, L., {et~al.} 2016, \aap, 596, A97

\bibitem[{{De Cia} {et~al.}(2018){De Cia}, {Ledoux}, {Petitjean}, \&
  {Savaglio}}]{DeCia18b}
{De Cia}, A., {Ledoux}, C., {Petitjean}, P., \& {Savaglio}, S. 2018, \aap, 611,
  A76

\bibitem[{{De Cia} {et~al.}(2013){De Cia}, {Ledoux}, {Savaglio}, {Schady}, \&
  {Vreeswijk}}]{DeCia13}
{De Cia}, A., {Ledoux}, C., {Savaglio}, S., {Schady}, P., \& {Vreeswijk}, P.~M.
  2013, \aap, 560, A88

\bibitem[{{De Marchi} {et~al.}(2011){De Marchi}, {Paresce}, {Panagia},
  {Beccari}, {Spezzi}, {Sirianni}, {Andersen}, {Mutchler}, {Balick}, {Dopita},
  {Frogel}, {Whitmore}, {Bond}, {Calzetti}, {Carollo}, {Disney}, {Hall},
  {Holtzman}, {Kimble}, {McCarthy}, {O'Connell}, {Saha}, {Silk}, {Trauger},
  {Walker}, {Windhorst}, \& {Young}}]{DeMarchi11}
{De Marchi}, G., {Paresce}, F., {Panagia}, N., {et~al.} 2011, \apj, 739, 27

\bibitem[{{de Ugarte Postigo} {et~al.}(2018){de Ugarte Postigo}, {Th{\"o}ne},
  {Bolmer}, {Schulze}, {Mart{\'\i}n}, {Kann}, {D'Elia}, {Selsing},
  {Martin-Carrillo}, {Perley}, {Kim}, {Izzo}, {S{\'a}nchez-Ram{\'\i}rez},
  {Guidorzi}, {Klotz}, {Wiersema}, {Bauer}, {Bensch}, {Campana}, {Cano},
  {Covino}, {Coward}, {De Cia}, {de Gregorio-Monsalvo}, {De Pasquale}, {Fynbo},
  {Greiner}, {Gomboc}, {Hanlon}, {Hansen}, {Hartmann}, {Heintz}, {Jakobsson},
  {Kobayashi}, {Malesani}, {Martone}, {Meintjes}, {Micha{\l}owski}, {Mundell},
  {Murphy}, {Oates}, {Salmon}, {van Soelen}, {Tanvir}, {Turpin}, {Xu}, \&
  {Zafar}}]{deUgartePostigo18}
{de Ugarte Postigo}, A., {Th{\"o}ne}, C.~C., {Bolmer}, J., {et~al.} 2018, \aap,
  620, A119

\bibitem[{{Delgado Mena} {et~al.}(2019){Delgado Mena}, {Moya}, {Adibekyan},
  {Tsantaki}, {Gonz{\'a}lez Hern{\'a}ndez}, {Israelian}, {Davies}, {Chaplin},
  {Sousa}, {Ferreira}, \& {Santos}}]{DelgadoMena19}
{Delgado Mena}, E., {Moya}, A., {Adibekyan}, V., {et~al.} 2019, \aap, 624, A78

\bibitem[{{Dessauges-Zavadsky} {et~al.}(2002){Dessauges-Zavadsky}, {Prochaska},
  \& {D'Odorico}}]{Dessauges-Zavadsky02}
{Dessauges-Zavadsky}, M., {Prochaska}, J.~X., \& {D'Odorico}, S. 2002, \aap,
  391, 801

\bibitem[{{Dessauges-Zavadsky} {et~al.}(2006){Dessauges-Zavadsky}, {Prochaska},
  {D'Odorico}, {Calura}, \& {Matteucci}}]{Dessauges-Zavadsky06}
{Dessauges-Zavadsky}, M., {Prochaska}, J.~X., {D'Odorico}, S., {Calura}, F., \&
  {Matteucci}, F. 2006, \aap, 445, 93

\bibitem[{{D'Onghia} \& {Fox}(2016)}]{Donghia16}
{D'Onghia}, E. \& {Fox}, A.~J. 2016, \araa, 54, 363

\bibitem[{{Duffau} {et~al.}(2017){Duffau}, {Caffau}, {Sbordone}, {Bonifacio},
  {Andrievsky}, {Korotin}, {Babusiaux}, {Salvadori}, {Monaco},
  {Fran{\c{c}}ois}, {Sk{\'u}lad{\'o}ttir}, {Bragaglia}, {Donati}, {Spina},
  {Gallagher}, {Ludwig}, {Christlieb}, {Hansen}, {Mott}, {Steffen}, {Zaggia},
  {Blanco-Cuaresma}, {Calura}, {Friel}, {Jim{\'e}nez-Esteban}, {Koch},
  {Magrini}, {Pancino}, {Tang}, {Tautvai{\v{s}}ien{\.{e}}}, {Vallenari},
  {Hawkins}, {Gilmore}, {Randich}, {Feltzing}, {Bensby}, {Flaccomio},
  {Smiljanic}, {Bayo}, {Carraro}, {Casey}, {Costado}, {Damiani}, {Franciosini},
  {Hourihane}, {Jofr{\'e}}, {Lardo}, {Lewis}, {Morbidelli}, {Sousa}, \&
  {Worley}}]{Duffau17}
{Duffau}, S., {Caffau}, E., {Sbordone}, L., {et~al.} 2017, \aap, 604, A128

\bibitem[{{Dufton} {et~al.}(2008){Dufton}, {Ryans}, {Thompson}, \&
  {Street}}]{Dufton08}
{Dufton}, P.~L., {Ryans}, R.~S.~I., {Thompson}, H.~M.~A., \& {Street}, R.~A.
  2008, \mnras, 385, 2261

\bibitem[{{Dwek}(2016)}]{Dwek16}
{Dwek}, E. 2016, \apj, 825, 136

\bibitem[{{Edmunds}(1975)}]{Edmunds75}
{Edmunds}, M.~G. 1975, \apss, 32, 483

\bibitem[{{Field}(1974)}]{Field74}
{Field}, G.~B. 1974, \apj, 187, 453

\bibitem[{{Fox} \& {Dav{\'e}}(2017)}]{Fox17}
{Fox}, A.~J. \& {Dav{\'e}}, R., eds. 2017, Astrophysics and Space Science
  Library, Vol. 430, {Gas Accretion onto Galaxies} (Springer International
  Publishing AG)

\bibitem[{{Fox} {et~al.}(2010){Fox}, {Wakker}, {Smoker}, {Richter}, {Savage},
  \& {Sembach}}]{Fox10}
{Fox}, A.~J., {Wakker}, B.~P., {Smoker}, J.~V., {et~al.} 2010, \apj, 718, 1046

\bibitem[{{Gardiner} \& {Noguchi}(1996)}]{Gardiner96}
{Gardiner}, L.~T. \& {Noguchi}, M. 1996, \mnras, 278, 191

\bibitem[{{Giordano} {et~al.}(2018){Giordano}, {Racero}, {Norman},
  {Vall{\'e}s}, {Mer{\'\i}n}, {Baines}, {L{\'o}pez-Caniego}, {Mart{\'\i}}, {de
  Teodoro}, {Salgado}, {Sarmiento}, {Guti{\'e}rrez-S{\'a}nchez}, {Prieto},
  {Lorca}, {Alberola}, {Valtchanov}, {de Marchi}, {{\'A}lvarez}, \&
  {Arviset}}]{Giordano18}
{Giordano}, F., {Racero}, E., {Norman}, H., {et~al.} 2018, Astronomy and
  Computing, 24, 97

\bibitem[{{Glover} \& {Mac Low}(2011)}]{Glover11}
{Glover}, S.~C.~O. \& {Mac Low}, M.~M. 2011, \mnras, 412, 337

\bibitem[{{Green} {et~al.}(2012){Green}, {Froning}, {Osterman}, {Ebbets},
  {Heap}, {Leitherer}, {Linsky}, {Savage}, {Sembach}, {Shull}, {Siegmund},
  {Snow}, {Spencer}, {Stern}, {Stocke}, {Welsh}, {B{\'e}land}, {Burgh},
  {Danforth}, {France}, {Keeney}, {McPhate}, {Penton}, {Andrews},
  {Brownsberger}, {Morse}, \& {Wilkinson}}]{Green12}
{Green}, J.~C., {Froning}, C.~S., {Osterman}, S., {et~al.} 2012, \apj, 744, 60

\bibitem[{{Grocholski} {et~al.}(2006){Grocholski}, {Cole}, {Sarajedini},
  {Geisler}, \& {Smith}}]{Grocholski06}
{Grocholski}, A.~J., {Cole}, A.~A., {Sarajedini}, A., {Geisler}, D., \&
  {Smith}, V.~V. 2006, \aj, 132, 1630

\bibitem[{{Guber} {et~al.}(2018){Guber}, {Richter}, \& {Wendt}}]{Guber18}
{Guber}, C.~R., {Richter}, P., \& {Wendt}, M. 2018, \aap, 609, A85

\bibitem[{{Harris} \& {Zaritsky}(2004)}]{Harris04}
{Harris}, J. \& {Zaritsky}, D. 2004, \aj, 127, 1531

\bibitem[{{Harris} \& {Zaritsky}(2009)}]{Harris09}
{Harris}, J. \& {Zaritsky}, D. 2009, \aj, 138, 1243

\bibitem[{{Hasselquist} {et~al.}(2021){Hasselquist}, {Hayes}, {Lian},
  {Weinberg}, {Zasowski}, {Horta}, {Beaton}, {Feuillet}, {Garro}, {Gallart},
  {Smith}, {Holtzman}, {Minniti}, {Lacerna}, {Shetrone}, {J{\"o}nsson},
  {Cioni}, {Fillingham}, {Cunha}, {O'Connell}, {Fern{\'a}ndez-Trincado},
  {Mu{\~n}oz}, {Schiavon}, {Almeida}, {Anguiano}, {Beers}, {Bizyaev},
  {Brownstein}, {Cohen}, {Frinchaboy}, {Garc{\'\i}a-Hern{\'a}ndez}, {Geisler},
  {Lane}, {Majewski}, {Nidever}, {Nitschelm}, {Povick}, {Price-Whelan},
  {Roman-Lopes}, {Rosado}, {Sobeck}, {Stringfellow}, {Valenzuela}, {Villanova},
  \& {Vincenzo}}]{Hasselquist21}
{Hasselquist}, S., {Hayes}, C.~R., {Lian}, J., {et~al.} 2021, \apj, 923, 172

\bibitem[{{Hernandez} {et~al.}(2017){Hernandez}, {Larsen}, {Trager}, {Groot},
  \& {Kaper}}]{Hernandez17}
{Hernandez}, S., {Larsen}, S., {Trager}, S., {Groot}, P., \& {Kaper}, L. 2017,
  \aap, 603, A119

\bibitem[{{Hill} {et~al.}(2000){Hill}, {Fran{\c{c}}ois}, {Spite}, {Primas}, \&
  {Spite}}]{Hill00}
{Hill}, V., {Fran{\c{c}}ois}, P., {Spite}, M., {Primas}, F., \& {Spite}, F.
  2000, \aap, 364, L19

\bibitem[{{Hill} {et~al.}(2019){Hill}, {Sk{\'u}lad{\'o}ttir}, {Tolstoy},
  {Venn}, {Shetrone}, {Jablonka}, {Primas}, {Battaglia}, {de Boer},
  {Fran{\c{c}}ois}, {Helmi}, {Kaufer}, {Letarte}, {Starkenburg}, \&
  {Spite}}]{Hill19}
{Hill}, V., {Sk{\'u}lad{\'o}ttir}, {\'A}., {Tolstoy}, E., {et~al.} 2019, \aap,
  626, A15

\bibitem[{{Howk} {et~al.}(2002){Howk}, {Sembach}, {Savage}, {Massa},
  {Friedman}, \& {Fullerton}}]{Howk02}
{Howk}, J.~C., {Sembach}, K.~R., {Savage}, B.~D., {et~al.} 2002, \apj, 569, 214

\bibitem[{{Hunter} {et~al.}(2007){Hunter}, {Dufton}, {Smartt}, {Ryans},
  {Evans}, {Lennon}, {Trundle}, {Hubeny}, \& {Lanz}}]{Hunter07}
{Hunter}, I., {Dufton}, P.~L., {Smartt}, S.~J., {et~al.} 2007, \aap, 466, 277

\bibitem[{{Indu} \& {Subramaniam}(2011)}]{Indu11}
{Indu}, G. \& {Subramaniam}, A. 2011, \aap, 535, A115

\bibitem[{{Jenkins}(1996)}]{Jenkins96}
{Jenkins}, E.~B. 1996, \apj, 471, 292

\bibitem[{{Jenkins}(2009)}]{Jenkins09}
{Jenkins}, E.~B. 2009, \apj, 700, 1299

\bibitem[{{Jenkins} {et~al.}(1986){Jenkins}, {Savage}, \&
  {Spitzer}}]{Jenkins86}
{Jenkins}, E.~B., {Savage}, B.~D., \& {Spitzer}, Jr., L. 1986, \apj, 301, 355

\bibitem[{{Jenkins} \& {Wallerstein}(2017)}]{Jenkins17}
{Jenkins}, E.~B. \& {Wallerstein}, G. 2017, \apj, 838, 85

\bibitem[{{Joshi} \& {Panchal}(2019)}]{Joshi19}
{Joshi}, Y.~C. \& {Panchal}, A. 2019, \aap, 628, A51

\bibitem[{{Kisielius} {et~al.}(2014){Kisielius}, {Kulkarni}, {Ferland},
  {Bogdanovich}, \& {Lykins}}]{Kisielius14}
{Kisielius}, R., {Kulkarni}, V.~P., {Ferland}, G.~J., {Bogdanovich}, P., \&
  {Lykins}, M.~L. 2014, \apj, 780, 76

\bibitem[{{Kisielius} {et~al.}(2015){Kisielius}, {Kulkarni}, {Ferland},
  {Bogdanovich}, {Som}, \& {Lykins}}]{Kisielius15}
{Kisielius}, R., {Kulkarni}, V.~P., {Ferland}, G.~J., {et~al.} 2015, \apj, 804,
  76

\bibitem[{{Kobayashi} {et~al.}(2020){Kobayashi}, {Karakas}, \&
  {Lugaro}}]{Kobayashi20}
{Kobayashi}, C., {Karakas}, A.~I., \& {Lugaro}, M. 2020, \apj, 900, 179

\bibitem[{{Konstantopoulou} {et~al.}(2022){Konstantopoulou}, {De Cia},
  {Krogager}, {Ledoux}, {Noterdaeme}, {Fynbo}, {Heintz}, {Watson}, {Andersen},
  {Ramburuth-Hurt}, \& {Jermann}}]{Konstantopoulou22}
{Konstantopoulou}, C., {De Cia}, A., {Krogager}, J.-K., {et~al.} 2022, arXiv
  e-prints, arXiv:2207.08804

\bibitem[{{Konstantopoulou} {et~al.}(2023){Konstantopoulou}, {De Cia},
  {Krogager}, {Ledoux}, {Noterdaeme}, {Fynbo}, {Heintz}, {Watson}, {Andersen},
  {Ramburuth-Hurt}, \& {Jermann}}]{Konstantopoulou23}
{Konstantopoulou}, C., {De Cia}, A., {Krogager}, J.-K., {et~al.} 2023, \aap,
  674, C1

\bibitem[{{Korn} {et~al.}(2000){Korn}, {Becker}, {Gummersbach}, \&
  {Wolf}}]{Korn00}
{Korn}, A.~J., {Becker}, S.~R., {Gummersbach}, C.~A., \& {Wolf}, B. 2000, \aap,
  353, 655

\bibitem[{{Kreckel} {et~al.}(2019){Kreckel}, {Ho}, {Blanc}, {Groves},
  {Santoro}, {Schinnerer}, {Bigiel}, {Chevance}, {Congiu}, {Emsellem}, {Faesi},
  {Glover}, {Grasha}, {Kruijssen}, {Lang}, {Leroy}, {Meidt}, {McElroy}, {Pety},
  {Rosolowsky}, {Saito}, {Sandstrom}, {Sanchez-Blazquez}, \&
  {Schruba}}]{Kreckel19}
{Kreckel}, K., {Ho}, I.~T., {Blanc}, G.~A., {et~al.} 2019, \apj, 887, 80

\bibitem[{{Kurucz}(2017)}]{Kurucz17}
{Kurucz}, R.~L. 2017, Canadian Journal of Physics, 95, 825

\bibitem[{{Lambert}(1987)}]{Lambert87}
{Lambert}, D.~L. 1987, Journal of Astrophysics and Astronomy, 8, 103

\bibitem[{{Lapenna} {et~al.}(2012){Lapenna}, {Mucciarelli}, {Origlia}, \&
  {Ferraro}}]{Lapenna12}
{Lapenna}, E., {Mucciarelli}, A., {Origlia}, L., \& {Ferraro}, F.~R. 2012,
  \apj, 761, 33

\bibitem[{{Larsen} {et~al.}(2008){Larsen}, {Origlia}, {Brodie}, \&
  {Gallagher}}]{Larsen08}
{Larsen}, S.~S., {Origlia}, L., {Brodie}, J., \& {Gallagher}, J.~S. 2008,
  \mnras, 383, 263

\bibitem[{{Larsen} {et~al.}(2006){Larsen}, {Origlia}, {Brodie}, \&
  {Gallagher}}]{Larsen06}
{Larsen}, S.~S., {Origlia}, L., {Brodie}, J.~P., \& {Gallagher}, J.~S. 2006,
  \mnras, 368, L10

\bibitem[{{Ledoux} {et~al.}(2002){Ledoux}, {Bergeron}, \&
  {Petitjean}}]{Ledoux02}
{Ledoux}, C., {Bergeron}, J., \& {Petitjean}, P. 2002, \aap, 385, 802

\bibitem[{{Lee} {et~al.}(2005){Lee}, {Rolleston}, {Dufton}, \& {Ryans}}]{Lee05}
{Lee}, J.~K., {Rolleston}, W.~R.~J., {Dufton}, P.~L., \& {Ryans}, R.~S.~I.
  2005, \aap, 429, 1025

\bibitem[{{Lehner} \& {Howk}(2007)}]{Lehner07}
{Lehner}, N. \& {Howk}, J.~C. 2007, \mnras, 377, 687

\bibitem[{{Lehner} {et~al.}(2008){Lehner}, {Howk}, {Keenan}, \&
  {Smoker}}]{Lehner08}
{Lehner}, N., {Howk}, J.~C., {Keenan}, F.~P., \& {Smoker}, J.~V. 2008, \apj,
  678, 219

\bibitem[{{Lehner} {et~al.}(2022){Lehner}, {Howk}, {Marasco}, \&
  {Fraternali}}]{Lehner22}
{Lehner}, N., {Howk}, J.~C., {Marasco}, A., \& {Fraternali}, F. 2022, \mnras,
  513, 3228

\bibitem[{{Lehner} {et~al.}(2009){Lehner}, {Staveley-Smith}, \&
  {Howk}}]{Lehner09}
{Lehner}, N., {Staveley-Smith}, L., \& {Howk}, J.~C. 2009, \apj, 702, 940

\bibitem[{{Lodders} {et~al.}(2009){Lodders}, {Palme}, \& {Gail}}]{Lodders09}
{Lodders}, K., {Palme}, H., \& {Gail}, H.-P. 2009, in ''Landolt-B{\"o}rnstein -
  Group VI Astronomy and Astrophysics Numerical Data, ed. {J.~E.~Tr{\"u}mper},
  44

\bibitem[{{Lucchini} {et~al.}(2021){Lucchini}, {D'Onghia}, \&
  {Fox}}]{Lucchini21}
{Lucchini}, S., {D'Onghia}, E., \& {Fox}, A.~J. 2021, \apjl, 921, L36

\bibitem[{{Lucchini} {et~al.}(2020){Lucchini}, {D'Onghia}, {Fox}, {Bustard},
  {Bland-Hawthorn}, \& {Zweibel}}]{Lucchini20}
{Lucchini}, S., {D'Onghia}, E., {Fox}, A.~J., {et~al.} 2020, \nat, 585, 203

\bibitem[{{Majewski} {et~al.}(2017){Majewski}, {Schiavon}, {Frinchaboy},
  {Allende Prieto}, {Barkhouser}, {Bizyaev}, {Blank}, {Brunner}, {Burton},
  {Carrera}, {Chojnowski}, {Cunha}, {Epstein}, {Fitzgerald}, {Garc{\'\i}a
  P{\'e}rez}, {Hearty}, {Henderson}, {Holtzman}, {Johnson}, {Lam}, {Lawler},
  {Maseman}, {M{\'e}sz{\'a}ros}, {Nelson}, {Nguyen}, {Nidever}, {Pinsonneault},
  {Shetrone}, {Smee}, {Smith}, {Stolberg}, {Skrutskie}, {Walker}, {Wilson},
  {Zasowski}, {Anders}, {Basu}, {Beland}, {Blanton}, {Bovy}, {Brownstein},
  {Carlberg}, {Chaplin}, {Chiappini}, {Eisenstein}, {Elsworth}, {Feuillet},
  {Fleming}, {Galbraith-Frew}, {Garc{\'\i}a}, {Garc{\'\i}a-Hern{\'a}ndez},
  {Gillespie}, {Girardi}, {Gunn}, {Hasselquist}, {Hayden}, {Hekker}, {Ivans},
  {Kinemuchi}, {Klaene}, {Mahadevan}, {Mathur}, {Mosser}, {Muna}, {Munn},
  {Nichol}, {O'Connell}, {Parejko}, {Robin}, {Rocha-Pinto}, {Schultheis},
  {Serenelli}, {Shane}, {Silva Aguirre}, {Sobeck}, {Thompson}, {Troup},
  {Weinberg}, \& {Zamora}}]{Majewski17}
{Majewski}, S.~R., {Schiavon}, R.~P., {Frinchaboy}, P.~M., {et~al.} 2017, \aj,
  154, 94

\bibitem[{{Marasco} {et~al.}(2022){Marasco}, {Fraternali}, {Lehner}, \&
  {Howk}}]{Marasco22}
{Marasco}, A., {Fraternali}, F., {Lehner}, N., \& {Howk}, J.~C. 2022, \mnras,
  515, 4176

\bibitem[{{Markova} {et~al.}(2020){Markova}, {Puls}, {Dufton}, {Lennon},
  {Evans}, {de Koter}, {Ram{\'\i}rez-Agudelo}, {Sana}, \& {Vink}}]{Markova20}
{Markova}, N., {Puls}, J., {Dufton}, P.~L., {et~al.} 2020, \aap, 634, A16

\bibitem[{{Markwardt}(2009)}]{Markwardt09}
{Markwardt}, C.~B. 2009, in Astronomical Society of the Pacific Conference
  Series, Vol. 411, Astronomical Data Analysis Software and Systems XVIII, ed.
  D.~A. {Bohlender}, D.~{Durand}, \& P.~{Dowler}, 251

\bibitem[{{Masseron} {et~al.}(2020){Masseron}, {Garc{\'\i}a-Hern{\'a}ndez},
  {Santove{\~n}a}, {Manchado}, {Zamora}, {Manteiga}, \& {Dafonte}}]{Masseron20}
{Masseron}, T., {Garc{\'\i}a-Hern{\'a}ndez}, D.~A., {Santove{\~n}a}, R.,
  {et~al.} 2020, Nature Communications, 11, 3759

\bibitem[{{Massey} \& {Hunter}(1998)}]{Massey98}
{Massey}, P. \& {Hunter}, D.~A. 1998, \apj, 493, 180

\bibitem[{{Matteucci}(2021)}]{Matteucci21}
{Matteucci}, F. 2021, \aapr, 29, 5

\bibitem[{{Mattsson} {et~al.}(2014){Mattsson}, {De Cia}, {Andersen}, \&
  {Zafar}}]{Mattsson14}
{Mattsson}, L., {De Cia}, A., {Andersen}, A.~C., \& {Zafar}, T. 2014, \mnras,
  440, 1562

\bibitem[{{McWilliam}(1997)}]{McWilliam97}
{McWilliam}, A. 1997, \araa, 35, 503

\bibitem[{{Mishenina} {et~al.}(2015){Mishenina}, {Gorbaneva}, {Pignatari},
  {Thielemann}, \& {Korotin}}]{Mishenina15}
{Mishenina}, T., {Gorbaneva}, T., {Pignatari}, M., {Thielemann}, F.-K., \&
  {Korotin}, S.~A. 2015, \mnras, 454, 1585

\bibitem[{{Morton}(2003)}]{Morton03}
{Morton}, D.~C. 2003, \apjs, 149, 205

\bibitem[{{Nidever} {et~al.}(2020){Nidever}, {Hasselquist}, {Hayes}, {Hawkins},
  {Povick}, {Majewski}, {Smith}, {Anguiano}, {Stringfellow}, {Sobeck}, {Cunha},
  {Beers}, {Bestenlehner}, {Cohen}, {Garcia-Hernandez}, {J{\"o}nsson},
  {Nitschelm}, {Shetrone}, {Lacerna}, {Allende Prieto}, {Beaton}, {Dell'Agli},
  {Fern{\'a}ndez-Trincado}, {Feuillet}, {Gallart}, {Hearty}, {Holtzman},
  {Manchado}, {Mu{\~n}oz}, {O'Connell}, \& {Rosado}}]{Nidever20}
{Nidever}, D.~L., {Hasselquist}, S., {Hayes}, C.~R., {et~al.} 2020, \apj, 895,
  88

\bibitem[{{Nidever} {et~al.}(2008){Nidever}, {Majewski}, \& {Butler
  Burton}}]{Nidever08}
{Nidever}, D.~L., {Majewski}, S.~R., \& {Butler Burton}, W. 2008, \apj, 679,
  432

\bibitem[{{Nidever} {et~al.}(2021){Nidever}, {Olsen}, {Choi}, {Ruiz-Lara},
  {Miller}, {Johnson}, {Bell}, {Blum}, {Cioni}, {Gallart}, {Majewski},
  {Martin}, {Massana}, {Monachesi}, {No{\"e}l}, {Sakowska}, {van der Marel},
  {Walker}, {Zaritsky}, {Bell}, {Conn}, {de Boer}, {Gruendl}, {Monelli},
  {Mu{\~n}oz}, {Saha}, {Vivas}, {Bernard}, {Besla}, {Carballo-Bello}, {Dorta},
  {Martinez-Delgado}, {Goater}, {Rusakov}, \& {Stringfellow}}]{Nidever21}
{Nidever}, D.~L., {Olsen}, K., {Choi}, Y., {et~al.} 2021, \aj, 161, 74

\bibitem[{{Nissen} {et~al.}(2007){Nissen}, {Akerman}, {Asplund}, {Fabbian},
  {Kerber}, {Kaufl}, \& {Pettini}}]{Nissen07}
{Nissen}, P.~E., {Akerman}, C., {Asplund}, M., {et~al.} 2007, \aap, 469, 319

\bibitem[{{Nomoto} {et~al.}(1997){Nomoto}, {Iwamoto}, {Nakasato}, {Thielemann},
  {Brachwitz}, {Tsujimoto}, {Kubo}, \& {Kishimoto}}]{Nomoto97}
{Nomoto}, K., {Iwamoto}, K., {Nakasato}, N., {et~al.} 1997, Nuclear Physics A,
  621, 467

\bibitem[{{Nomoto} {et~al.}(2006){Nomoto}, {Tominaga}, {Umeda}, {Kobayashi}, \&
  {Maeda}}]{Nomoto06}
{Nomoto}, K., {Tominaga}, N., {Umeda}, H., {Kobayashi}, C., \& {Maeda}, K.
  2006, Nuclear Physics A, 777, 424

\bibitem[{{Noterdaeme} {et~al.}(2017){Noterdaeme}, {Krogager}, {Balashev},
  {Ge}, {Gupta}, {Kr{\"u}hler}, {Ledoux}, {Murphy}, {P{\^a}ris}, {Petitjean},
  {Rahmani}, {Srianand}, \& {Ubachs}}]{Noterdaeme17}
{Noterdaeme}, P., {Krogager}, J.-K., {Balashev}, S., {et~al.} 2017, \aap, 597,
  A82

\bibitem[{{Noterdaeme} {et~al.}(2008){Noterdaeme}, {Ledoux}, {Petitjean}, \&
  {Srianand}}]{Noterdaeme08}
{Noterdaeme}, P., {Ledoux}, C., {Petitjean}, P., \& {Srianand}, R. 2008, \aap,
  481, 327

\bibitem[{{Noterdaeme} {et~al.}(2015){Noterdaeme}, {Petitjean}, \&
  {Srianand}}]{Noterdaeme15}
{Noterdaeme}, P., {Petitjean}, P., \& {Srianand}, R. 2015, \aap, 578, L5

\bibitem[{{Panagia} {et~al.}(2000){Panagia}, {Romaniello}, {Scuderi}, \&
  {Kirshner}}]{Panagia20}
{Panagia}, N., {Romaniello}, M., {Scuderi}, S., \& {Kirshner}, R.~P. 2000,
  \apj, 539, 197

\bibitem[{{Pardy} {et~al.}(2018){Pardy}, {D'Onghia}, \& {Fox}}]{Pardy18}
{Pardy}, S.~A., {D'Onghia}, E., \& {Fox}, A.~J. 2018, \apj, 857, 101

\bibitem[{{Parisi} {et~al.}(2009){Parisi}, {Grocholski}, {Geisler},
  {Sarajedini}, \& {Clari{\'a}}}]{Parisi09}
{Parisi}, M.~C., {Grocholski}, A.~J., {Geisler}, D., {Sarajedini}, A., \&
  {Clari{\'a}}, J.~J. 2009, \aj, 138, 517

\bibitem[{{Petitjean} {et~al.}(2006){Petitjean}, {Ledoux}, {Noterdaeme}, \&
  {Srianand}}]{Petitjean06}
{Petitjean}, P., {Ledoux}, C., {Noterdaeme}, P., \& {Srianand}, R. 2006, \aap,
  456, L9

\bibitem[{{Phillips} {et~al.}(1982){Phillips}, {Gondhalekar}, \&
  {Pettini}}]{Phillips82}
{Phillips}, A.~P., {Gondhalekar}, P.~M., \& {Pettini}, M. 1982, \mnras, 200,
  687

\bibitem[{{Poggio} {et~al.}(2022){Poggio}, {Recio-Blanco}, {Palicio}, {Re
  Fiorentin}, {de Laverny}, {Drimmel}, {Kordopatis}, {Lattanzi}, {Schultheis},
  {Spagna}, \& {Spitoni}}]{Poggio22}
{Poggio}, E., {Recio-Blanco}, A., {Palicio}, P.~A., {et~al.} 2022, arXiv
  e-prints, arXiv:2206.14849

\bibitem[{{Pomp{\'e}ia} {et~al.}(2008){Pomp{\'e}ia}, {Hill}, {Spite}, {Cole},
  {Primas}, {Romaniello}, {Pasquini}, {Cioni}, \& {Smecker Hane}}]{Pompeia08}
{Pomp{\'e}ia}, L., {Hill}, V., {Spite}, M., {et~al.} 2008, \aap, 480, 379

\bibitem[{{Primas} {et~al.}(2000){Primas}, {Brugamyer}, {Sneden}, {King},
  {Beers}, {Boesgaard}, \& {Deliyannis}}]{Primas00}
{Primas}, F., {Brugamyer}, E., {Sneden}, C., {et~al.} 2000, in The First Stars,
  ed. A.~{Weiss}, T.~G. {Abel}, \& V.~{Hill}, 51

\bibitem[{{Prochaska}(2003)}]{Prochaska03b}
{Prochaska}, J.~X. 2003, \apj, 582, 49

\bibitem[{{Ramachandran} {et~al.}(2021){Ramachandran}, {Oskinova}, \&
  {Hamann}}]{Ramachandran21}
{Ramachandran}, V., {Oskinova}, L.~M., \& {Hamann}, W.~R. 2021, \aap, 646, A16

\bibitem[{{Ramburuth-Hurt} {et~al.}(2023){Ramburuth-Hurt}, {De Cia},
  {Krogager}, {Ledoux}, {Petitjean}, {P{\'e}roux}, {Dessauges-Zavadsky},
  {Fynbo}, {Wendt}, {Bouch{\'e}}, {Konstantopoulou}, \&
  {Jermann}}]{Ramburuth-Hurt23}
{Ramburuth-Hurt}, T., {De Cia}, A., {Krogager}, J.~K., {et~al.} 2023, \aap,
  672, A68

\bibitem[{{Ritchey} {et~al.}(2023){Ritchey}, {Jenkins}, {Shull}, {Savage},
  {Federman}, \& {Lambert}}]{Ritchey23}
{Ritchey}, A.~M., {Jenkins}, E.~B., {Shull}, J.~M., {et~al.} 2023, arXiv
  e-prints, arXiv:2301.09743

\bibitem[{{Rodr{\'{\i}}guez} {et~al.}(2006){Rodr{\'{\i}}guez}, {Petitjean},
  {Aracil}, {Ledoux}, \& {Srianand}}]{Rodriguez06}
{Rodr{\'{\i}}guez}, E., {Petitjean}, P., {Aracil}, B., {Ledoux}, C., \&
  {Srianand}, R. 2006, \aap, 446, 791

\bibitem[{{Rolleston} {et~al.}(1999){Rolleston}, {Dufton}, {McErlean}, \&
  {Venn}}]{Rolleston99}
{Rolleston}, W.~R.~J., {Dufton}, P.~L., {McErlean}, N.~D., \& {Venn}, K.~A.
  1999, \aap, 348, 728

\bibitem[{{Rolleston} {et~al.}(2002){Rolleston}, {Trundle}, \&
  {Dufton}}]{Rolleston02}
{Rolleston}, W.~R.~J., {Trundle}, C., \& {Dufton}, P.~L. 2002, \aap, 396, 53

\bibitem[{{Roman-Duval} {et~al.}(2022){Roman-Duval}, {Jenkins}, {Tchernyshyov},
  {Clark}, {De Cia}, {Gordon}, {Hamanowicz}, {Lebouteiller}, {Rafelski},
  {Sandstrom}, {Werk}, \& {Yanchulova Merica-Jones}}]{Roman-Duval22}
{Roman-Duval}, J., {Jenkins}, E.~B., {Tchernyshyov}, K., {et~al.} 2022, arXiv
  e-prints, arXiv:2206.03639

\bibitem[{{Roman-Duval} {et~al.}(2021){Roman-Duval}, {Jenkins}, {Tchernyshyov},
  {Williams}, {Clark}, {Gordon}, {Meixner}, {Hagen}, {Peek}, {Sandstrom},
  {Werk}, \& {Yanchulova Merica-Jones}}]{Roman-Duval21}
{Roman-Duval}, J., {Jenkins}, E.~B., {Tchernyshyov}, K., {et~al.} 2021, \apj,
  910, 95

\bibitem[{{Roman-Duval} {et~al.}(2019){Roman-Duval}, {Jenkins}, {Williams},
  {Tchernyshyov}, {Gordon}, {Meixner}, {Hagen}, {Peek}, {Sandstrom}, {Werk}, \&
  {Yanchulova Merica-Jones}}]{Roman-Duval19}
{Roman-Duval}, J., {Jenkins}, E.~B., {Williams}, B., {et~al.} 2019, \apj, 871,
  151

\bibitem[{{Roman-Duval} {et~al.}(2020){Roman-Duval}, {De Rosa}, {Proffitt},
  {Reid}, {Brown}, {Fullerton}, {Fischer}, {Aloisi}, {Britt}, {Busko},
  {Carlberg}, {Fox}, {Frazer}, {James}, {Jedrzejewski}, {Lockwood}, {Monroe},
  {Oliveira}, {Plesha}, {Riedel}, {Riley}, {Shaw}, {Smith}, {Sohn}, {Taylor},
  {Ubeda}, \& {Welty}}]{Roman-Duval20}
{Roman-Duval}, J.~C., {De Rosa}, G., {Proffitt}, C., {et~al.} 2020, in American
  Astronomical Society Meeting Abstracts, Vol. 235, American Astronomical
  Society Meeting Abstracts \#235, 232.03

\bibitem[{{Romaniello} {et~al.}(2022){Romaniello}, {Riess}, {Mancino},
  {Anderson}, {Freudling}, {Kudritzki}, {Macr{\`\i}}, {Mucciarelli}, \&
  {Yuan}}]{Romaniello22}
{Romaniello}, M., {Riess}, A., {Mancino}, S., {et~al.} 2022, \aap, 658, A29

\bibitem[{{Rubele} {et~al.}(2018){Rubele}, {Pastorelli}, {Girardi}, {Cioni},
  {Zaggia}, {Marigo}, {Bekki}, {Bressan}, {Clementini}, {de Grijs}, {Emerson},
  {Groenewegen}, {Ivanov}, {Muraveva}, {Nanni}, {Oliveira}, {Ripepi}, {Sun}, \&
  {van Loon}}]{Rubele18}
{Rubele}, S., {Pastorelli}, G., {Girardi}, L., {et~al.} 2018, \mnras, 478, 5017

\bibitem[{{Sahnow} {et~al.}(2000){Sahnow}, {Moos}, {Ake}, {Andersen},
  {Andersson}, {Andre}, {Artis}, {Berman}, {Blair}, {Brownsberger}, {Calvani},
  {Chayer}, {Conard}, {Feldman}, {Friedman}, {Fullerton}, {Gaines}, {Gawne},
  {Green}, {Gummin}, {Jennings}, {Joyce}, {Kaiser}, {Kruk}, {Lindler}, {Massa},
  {Murphy}, {Oegerle}, {Ohl}, {Roberts}, {Romelfanger}, {Roth}, {Sankrit},
  {Sembach}, {Shelton}, {Siegmund}, {Silva}, {Sonneborn}, {Vaclavik}, {Weaver},
  \& {Wilkinson}}]{Sahnow00}
{Sahnow}, D.~J., {Moos}, H.~W., {Ake}, T.~B., {et~al.} 2000, \apjl, 538, L7

\bibitem[{{Saito} {et~al.}(2009){Saito}, {Takada-Hidai}, {Honda}, \&
  {Takeda}}]{Saito09}
{Saito}, Y.-J., {Takada-Hidai}, M., {Honda}, S., \& {Takeda}, Y. 2009, \pasj,
  61, 549

\bibitem[{{Salvadori} \& {Ferrara}(2012)}]{Salvadori12}
{Salvadori}, S. \& {Ferrara}, A. 2012, \mnras, 421, L29

\bibitem[{{S{\'a}nchez Almeida} {et~al.}(2015){S{\'a}nchez Almeida},
  {Elmegreen}, {Mu{\~n}oz-Tu{\~n}{\'o}n}, {Elmegreen}, {P{\'e}rez-Montero},
  {Amor{\'\i}n}, {Filho}, {Ascasibar}, {Papaderos}, \&
  {V{\'\i}lchez}}]{SanchezAlmeida15}
{S{\'a}nchez Almeida}, J., {Elmegreen}, B.~G., {Mu{\~n}oz-Tu{\~n}{\'o}n}, C.,
  {et~al.} 2015, \apjl, 810, L15

\bibitem[{{Sandstrom} {et~al.}(2010){Sandstrom}, {Bolatto}, {Draine}, {Bot}, \&
  {Stanimirovi{\'c}}}]{Sandstrom10}
{Sandstrom}, K.~M., {Bolatto}, A.~D., {Draine}, B.~T., {Bot}, C., \&
  {Stanimirovi{\'c}}, S. 2010, \apj, 715, 701

\bibitem[{{Savage} \& {Sembach}(1996)}]{Savage96}
{Savage}, B.~D. \& {Sembach}, K.~R. 1996, \araa, 34, 279

\bibitem[{{Shipp} {et~al.}(2021){Shipp}, {Erkal}, {Drlica-Wagner}, {Li},
  {Pace}, {Koposov}, {Cullinane}, {Da Costa}, {Ji}, {Kuehn}, {Lewis}, {Mackey},
  {Simpson}, {Wan}, {Zucker}, {Bland-Hawthorn}, {Ferguson}, {Lilleengen}, \&
  {Lilleengen}}]{Shipp21}
{Shipp}, N., {Erkal}, D., {Drlica-Wagner}, A., {et~al.} 2021, \apj, 923, 149

\bibitem[{{Sitnova} {et~al.}(2022){Sitnova}, {Yakovleva}, {Belyaev}, \&
  {Mashonkina}}]{Sitnova22}
{Sitnova}, T.~M., {Yakovleva}, S.~A., {Belyaev}, A.~K., \& {Mashonkina}, L.~I.
  2022, \mnras [\eprint[arXiv]{2205.05819}]

\bibitem[{{Sk{\'u}lad{\'o}ttir} {et~al.}(2017){Sk{\'u}lad{\'o}ttir}, {Tolstoy},
  {Salvadori}, {Hill}, \& {Pettini}}]{Skuladottir17}
{Sk{\'u}lad{\'o}ttir}, {\'A}., {Tolstoy}, E., {Salvadori}, S., {Hill}, V., \&
  {Pettini}, M. 2017, \aap, 606, A71

\bibitem[{{Sneden} {et~al.}(1991){Sneden}, {Gratton}, \& {Crocker}}]{Sneden91}
{Sneden}, C., {Gratton}, R.~G., \& {Crocker}, D.~A. 1991, \aap, 246, 354

\bibitem[{{Sofia} {et~al.}(2006){Sofia}, {Gordon}, {Clayton}, {Misselt},
  {Wolff}, {Cox}, \& {Ehrenfreund}}]{Sofia06}
{Sofia}, U.~J., {Gordon}, K.~D., {Clayton}, G.~C., {et~al.} 2006, \apj, 636,
  753

\bibitem[{{Spitoni} {et~al.}(2022){Spitoni}, {Recio-Blanco}, {de Laverny},
  {Palicio}, {Kordopatis}, {Schultheis}, {Contursi}, {Poggio}, {Romano}, \&
  {Matteucci}}]{Spitoni22}
{Spitoni}, E., {Recio-Blanco}, A., {de Laverny}, P., {et~al.} 2022, arXiv
  e-prints, arXiv:2206.12436

\bibitem[{{Tchernyshyov} {et~al.}(2015){Tchernyshyov}, {Meixner}, {Seale},
  {Fox}, {Friedman}, {Dwek}, \& {Galliano}}]{Tchernyshyov15}
{Tchernyshyov}, K., {Meixner}, M., {Seale}, J., {et~al.} 2015, \apj, 811, 78

\bibitem[{{Tinsley}(1979)}]{Tinsley79}
{Tinsley}, B.~M. 1979, \apj, 229, 1046

\bibitem[{{Tinsley}(1980)}]{Tinsley80}
{Tinsley}, B.~M. 1980, \fcp, 5, 287

\bibitem[{{Tolstoy} {et~al.}(2009){Tolstoy}, {Hill}, \& {Tosi}}]{Tolstoy09}
{Tolstoy}, E., {Hill}, V., \& {Tosi}, M. 2009, \araa, 47, 371

\bibitem[{{Toribio San Cipriano} {et~al.}(2017){Toribio San Cipriano},
  {Dom{\'\i}nguez-Guzm{\'a}n}, {Esteban}, {Garc{\'\i}a-Rojas}, {Mesa-Delgado},
  {Bresolin}, {Rodr{\'\i}guez}, \&
  {Sim{\'o}n-D{\'\i}az}}]{ToribioSanCipriano17}
{Toribio San Cipriano}, L., {Dom{\'\i}nguez-Guzm{\'a}n}, G., {Esteban}, C.,
  {et~al.} 2017, \mnras, 467, 3759

\bibitem[{{Trundle} {et~al.}(2007){Trundle}, {Dufton}, {Hunter}, {Evans},
  {Lennon}, {Smartt}, \& {Ryans}}]{Trundle07}
{Trundle}, C., {Dufton}, P.~L., {Hunter}, I., {et~al.} 2007, {The VLT-FLAMES
  survey of massive stars: evolution of surface N abundances and effective
  temperature scales in the Galaxy and Magellanic Clouds}, Astronomy and
  Astrophysics, Volume 471, Issue 2, August IV 2007, pp.625-643

\bibitem[{{Van der Swaelmen} {et~al.}(2013){Van der Swaelmen}, {Hill},
  {Primas}, \& {Cole}}]{VanderSwaelmen13}
{Van der Swaelmen}, M., {Hill}, V., {Primas}, F., \& {Cole}, A.~A. 2013, \aap,
  560, A44

\bibitem[{{Viegas}(1995)}]{Viegas95}
{Viegas}, S.~M. 1995, \mnras, 276, 268

\bibitem[{{Wang} {et~al.}(2022){Wang}, {Li}, {Cai}, {Shi}, {Fan}, {Zheng},
  {Bian}, {Teplitz}, {Alavi}, {Colbert}, {Henry}, \& {Malkan}}]{Wang22}
{Wang}, X., {Li}, Z., {Cai}, Z., {et~al.} 2022, \apj, 926, 70

\bibitem[{{Weisz} {et~al.}(2014){Weisz}, {Dolphin}, {Skillman}, {Holtzman},
  {Gilbert}, {Dalcanton}, \& {Williams}}]{Weisz14}
{Weisz}, D.~R., {Dolphin}, A.~E., {Skillman}, E.~D., {et~al.} 2014, \apj, 789,
  147

\bibitem[{{Welty} \& {Crowther}(2010)}]{Welty10}
{Welty}, D.~E. \& {Crowther}, P.~A. 2010, \mnras, 404, 1321

\bibitem[{{Welty} {et~al.}(1999){Welty}, {Frisch}, {Sonneborn}, \&
  {York}}]{Welty99}
{Welty}, D.~E., {Frisch}, P.~C., {Sonneborn}, G., \& {York}, D.~G. 1999, \apj,
  512, 636

\bibitem[{{Welty} {et~al.}(1997){Welty}, {Lauroesch}, {Blades}, {Hobbs}, \&
  {York}}]{Welty97}
{Welty}, D.~E., {Lauroesch}, J.~T., {Blades}, J.~C., {Hobbs}, L.~M., \& {York},
  D.~G. 1997, \apj, 489, 672

\bibitem[{{Welty} {et~al.}(2020){Welty}, {Sonnentrucker}, {Snow}, \&
  {York}}]{Welty20}
{Welty}, D.~E., {Sonnentrucker}, P., {Snow}, T.~P., \& {York}, D.~G. 2020,
  \apj, 897, 36

\bibitem[{{Westerlund}(1990)}]{Westerlund90}
{Westerlund}, B.~E. 1990, \aapr, 2, 29

\bibitem[{{White} \& {Audouze}(1983)}]{White83}
{White}, S.~D.~M. \& {Audouze}, J. 1983, \mnras, 203, 603

\bibitem[{{Williams} {et~al.}(2010){Williams}, {Bureau}, \&
  {Cappellari}}]{Williams10}
{Williams}, M.~J., {Bureau}, M., \& {Cappellari}, M. 2010, \mnras, 409, 1330

\bibitem[{{Wiseman} {et~al.}(2017){Wiseman}, {Schady}, {Bolmer}, {Kr{\"u}hler},
  {Yates}, {Greiner}, \& {Fynbo}}]{Wiseman17}
{Wiseman}, P., {Schady}, P., {Bolmer}, J., {et~al.} 2017, \aap, 599, A24

\bibitem[{{Wolfe} {et~al.}(2005){Wolfe}, {Gawiser}, \& {Prochaska}}]{Wolfe05}
{Wolfe}, A.~M., {Gawiser}, E., \& {Prochaska}, J.~X. 2005, \araa, 43, 861

\bibitem[{{Woosley} \& {Weaver}(1995)}]{Woosley95}
{Woosley}, S.~E. \& {Weaver}, T.~A. 1995, \apjs, 101, 181

\bibitem[{{Zech} {et~al.}(2008){Zech}, {Lehner}, {Howk}, {Dixon}, \&
  {Brown}}]{Zech08}
{Zech}, W.~F., {Lehner}, N., {Howk}, J.~C., {Dixon}, W. V.~D., \& {Brown},
  T.~M. 2008, \apj, 679, 460

\end{thebibliography}

\begin{table*}
\centering
\caption{Sample of lines of sight towards the LMC and chemical properties of their neutral ISM. The [$\alpha$/Fe]$_{\rm nucl}$ is the average $\alpha$-element enhancement, reported only for the golden sample.[1] \citet{Roman-Duval21}. The metal column densities have been corrected for the latest oscillator strengths, following \citet{Konstantopoulou22}. $N$(Ti) are taken from \citet{Welty10}.}
\begin{tabular}{l | c c c c c c | c c c}
\hline
\hline
   ID                        & RA                        & Dec & $\log N({\rm H})_{\rm tot}$                    & [Zn/Fe]         & $\log f_{\rm H_2}$ & Ref.           & [Zn/Fe]$_{\rm fit}$             & [M/H]$_{\rm tot}$  & [$\alpha$/Fe]$_{\rm nucl}$\\
\hline
   BI~173 &                05:27:09.94 &               -69:07:56.46 &             $21.25\pm0.05$ &             $ 0.83\pm0.05$ &                    $-5.31$ &  [1] &                $ 0.75\pm0.05$ &                $-0.50\pm0.07$ &                           --  \\
   BI~184 &                05:30:30.66 &               -71:02:31.60 &             $21.15\pm0.04$ &             $ 0.73\pm0.08$ &                    $-1.20$ &  [1] &                $ 0.70\pm0.08$ &                $-0.38\pm0.10$ &                           --  \\
   BI~237 &                05:36:14.63 &               -67:39:19.18 &             $21.65\pm0.03$ &             $ 0.98\pm0.07$ &                    $-1.30$ &  [1] &                $ 1.07\pm0.10$ &                $-0.50\pm0.12$ &                $ 0.29\pm0.16$ \\
   BI~253 &                05:37:34.46 &               -69:01:10.20 &             $21.68\pm0.03$ &             $ 0.83\pm0.04$ &                    $-1.62$ &  [1] &                $ 0.91\pm0.12$ &                $-0.53\pm0.16$ &                $ 0.28\pm0.22$ \\
PGMW~3120 &                04:56:46.81 &               -66:24:46.72 &             $21.48\pm0.03$ &             $ 0.97\pm0.08$ &                    $-2.88$ &  [1] &                $ 0.91\pm0.07$ &                $-0.50\pm0.09$ &                           --  \\
PGMW~3223 &                04:57:00.86 &               -66:24:25.12 &             $21.40\pm0.06$ &             $ 1.03\pm0.06$ &                    $-2.41$ &  [1] &                $ 1.05\pm0.07$ &                $-0.30\pm0.10$ &                           --  \\
 Sk-65~22 &                05:01:23.07 &               -65:52:33.40 &             $20.66\pm0.03$ &             $ 0.83\pm0.06$ &                    $-5.43$ &  [1] &                $ 0.80\pm0.06$ &                $-0.17\pm0.07$ &                           --  \\
Sk-66~172 &                05:37:05.39 &               -66:21:35.18 &             $21.27\pm0.03$ &             $ 1.11\pm0.06$ &                    $-2.76$ &  [1] &                $ 1.07\pm0.07$ &                $-0.49\pm0.08$ &                           --  \\
 Sk-66~19 &                04:55:53.95 &               -66:24:59.35 &             $21.87\pm0.07$ &             $ 1.25\pm0.13$ &                    $-1.37$ &  [1] &                $ 1.27\pm0.13$ &                $-0.41\pm0.16$ &                           --  \\
 Sk-66~35 &                04:57:04.44 &               -66:34:38.45 &             $20.85\pm0.04$ &             $ 0.84\pm0.14$ &                    $-1.25$ &  [1] &                $ 0.98\pm0.13$ &                $-0.05\pm0.14$ &                           --  \\
Sk-67~101 &                05:25:56.22 &               -67:30:28.67 &             $20.20\pm0.04$ &             $ 0.68\pm0.13$ &                        --  &  [1] &                $ 0.67\pm0.13$ &                $ 0.10\pm0.15$ &                           --  \\
Sk-67~105 &                05:26:06.19 &               -67:10:56.79 &             $21.26\pm0.04$ &             $ 0.86\pm0.07$ &                    $-1.83$ &  [1] &                $ 0.86\pm0.08$ &                $-0.77\pm0.09$ &                           --  \\
 Sk-67~14 &                04:54:31.89 &               -67:15:24.58 &             $20.24\pm0.06$ &             $ 0.00\pm0.00$ &                    $-4.93$ &  [1] &                $ 1.01\pm0.10$ &                $ 0.47\pm0.44$ &                           --  \\
Sk-67~191 &                05:33:34.03 &               -67:30:19.72 &             $20.78\pm0.03$ &             $ 0.52\pm0.20$ &                        --  &  [1] &                $ 0.54\pm0.21$ &                $-0.49\pm0.23$ &                           --  \\
  Sk-67~2 &                04:47:04.45 &               -67:06:53.12 &             $21.46\pm0.12$ &             $ 1.15\pm0.11$ &                    $-0.21$ &  [1] &                $ 0.98\pm0.12$ &                $-0.82\pm0.19$ &                $ 0.29\pm0.26$ \\
Sk-67~211 &                05:35:13.91 &               -67:33:27.51 &             $20.81\pm0.04$ &             $ 0.73\pm0.06$ &                        --  &  [1] &                $ 0.90\pm0.07$ &                $-0.06\pm0.09$ &                $ 0.25\pm0.12$ \\
  Sk-67~5 &                04:50:18.92 &               -67:39:38.10 &             $21.04\pm0.04$ &             $ 1.00\pm0.05$ &                    $-1.28$ &  [1] &                $ 0.92\pm0.06$ &                $-0.46\pm0.09$ &                $ 0.40\pm0.11$ \\
Sk-68~129 &                05:36:26.77 &               -68:57:31.90 &             $21.62\pm0.14$ &             $ 1.33\pm0.14$ &                    $-1.12$ &  [1] &                $ 1.27\pm0.15$ &                $-0.13\pm0.20$ &                           --  \\
Sk-68~135 &                05:37:49.11 &               -68:55:01.69 &             $21.48\pm0.02$ &             $ 1.04\pm0.07$ &                    $-1.31$ &  [1] &                $ 0.83\pm0.14$ &                $-0.51\pm0.18$ &                $ 0.28\pm0.25$ \\
Sk-68~140 &                05:38:57.18 &               -68:56:53.10 &             $21.51\pm0.11$ &             $ 1.16\pm0.10$ &                    $-1.10$ &  [1] &                $ 1.07\pm0.11$ &                $-0.14\pm0.16$ &                           --  \\
Sk-68~155 &                05:42:54.93 &               -68:56:54.50 &             $21.47\pm0.09$ &             $ 1.06\pm0.05$ &                    $-1.18$ &  [1] &                $ 1.00\pm0.06$ &                $-0.12\pm0.11$ &                           --  \\
 Sk-68~26 &                05:01:32.25 &               -68:10:42.93 &             $21.65\pm0.06$ &             $ 1.05\pm0.15$ &                    $-0.97$ &  [1] &                $ 0.94\pm0.15$ &                $-0.77\pm0.17$ &                           --  \\
 Sk-68~52 &                05:07:20.42 &               -68:32:08.59 &             $21.31\pm0.06$ &             $ 0.43\pm0.12$ &                    $-1.54$ &  [1] &                $ 0.86\pm0.06$ &                $-0.27\pm0.09$ &                $ 0.09\pm0.13$ \\
 Sk-68~73 &                05:22:59.78 &               -68:01:46.62 &             $21.68\pm0.02$ &             $ 1.24\pm0.12$ &                    $-1.29$ &  [1] &                $ 1.25\pm0.16$ &                $-0.18\pm0.21$ &                $ 0.26\pm0.30$ \\
Sk-69~104 &                05:18:59.50 &               -69:12:54.82 &             $19.57\pm0.68$ &             $ 1.06\pm0.27$ &                        --  &  [1] &                $ 1.06\pm0.29$ &                $ 0.40\pm0.73$ &                           --  \\
Sk-69~175 &                05:31:25.52 &               -69:05:38.59 &             $20.64\pm0.03$ &             $ 0.38\pm0.12$ &                        --  &  [1] &                $ 0.42\pm0.12$ &                $-0.55\pm0.13$ &                           --  \\
Sk-69~246 &                05:38:53.38 &               -69:02:00.93 &             $21.48\pm0.02$ &             $ 0.80\pm0.04$ &                    $-1.47$ &  [1] &                $ 0.88\pm0.09$ &                $-0.40\pm0.11$ &                $ 0.18\pm0.15$ \\
Sk-69~279 &                04:54:14.26 &               -69:15:13.35 &             $21.63\pm0.05$ &             $ 0.76\pm0.06$ &                    $-1.02$ &  [1] &                $ 0.65\pm0.06$ &                $-0.79\pm0.08$ &                           --  \\
Sk-70~115 &                05:48:49.65 &               -70:03:57.82 &             $21.18\pm0.08$ &             $ 0.67\pm0.05$ &                    $-0.94$ &  [1] &                $ 0.72\pm0.08$ &                $-0.30\pm0.12$ &                $ 0.25\pm0.16$ \\
 Sk-70~79 &                05:06:37.26 &               -70:29:24.16 &             $21.34\pm0.04$ &             $ 1.51\pm0.09$ &                    $-0.78$ &  [1] &                $ 1.57\pm0.11$ &                $-0.08\pm0.13$ &                           --  \\
 Sk-71~45 &                05:31:15.65 &               -71:04:09.69 &             $21.11\pm0.03$ &             $ 0.81\pm0.06$ &                    $-2.18$ &  [1] &                $ 0.77\pm0.06$ &                $-0.26\pm0.07$ &                           --  \\
 Sk-71~50 &                05:40:43.19 &               -71:29:00.65 &             $21.24\pm0.05$ &             $ 0.55\pm0.05$ &                    $-0.81$ &  [1] &                $ 0.59\pm0.05$ &                $-0.44\pm0.07$ &                           --  \\
\hline
\hline
\end{tabular}
\label{tab LMC}
\end{table*}

\begin{table*}
\centering
\caption{Same as Table \ref{tab LMC}, but for the SMC. [2] \citet{Jenkins17}; [3] \citet{Tchernyshyov15}. The metal column densities have been corrected for the latest oscillator strengths, following \citet{Konstantopoulou22}. $N$(Ti) are taken from \citet{Welty10}.}
\begin{tabular}{l | c c c c c c | c c c}
\hline
\hline
   ID                        & RA                        & Dec & $\log N({\rm H})_{\rm tot}$                    & [Zn/Fe]         & $\log f_{\rm H_2}$ & Ref.           & [Zn/Fe]$_{\rm fit}$             & [M/H]$_{\rm tot}$  & [$\alpha$/Fe]$_{\rm nucl}$\\
\hline
   AzV~18 &                00:47:12.22 &               -73:06:33.11 &             $22.06\pm0.02$ &             $ 0.91\pm0.09$ &                    $-1.40$ &  [2] &                $ 0.98\pm0.10$ &                $-0.66\pm0.13$ &                $ 0.38\pm0.18$ \\
   AzV~26 &                00:47:50.05 &               -73:08:21.05 &             $21.77\pm0.04$ &             $ 0.60\pm0.05$ &                    $-0.84$ &  [2] &                $ 0.57\pm0.06$ &                $-0.98\pm0.08$ &                $ 0.22\pm0.11$ \\
   AzV~47 &                00:48:51.49 &               -73:25:58.54 &             $21.32\pm0.04$ &             $ 0.57\pm0.07$ &                    $-2.48$ &  [2] &                $ 0.54\pm0.07$ &                $-0.73\pm0.08$ &                $ 0.28\pm0.12$ \\
   AzV~78 &                00:50:38.39 &               -73:28:18.26 &             $21.70\pm0.06$ &             $ 0.00\pm0.00$ &                    $-2.76$ &  [2] &                $ 0.62\pm0.12$ &                $-0.82\pm0.17$ &                $ 0.24\pm0.24$ \\
   AzV~80 &                00:50:43.81 &               -72:47:41.55 &             $21.83\pm0.02$ &             $ 0.69\pm0.06$ &                    $-1.45$ &  [2] &                $ 0.84\pm0.11$ &                $-0.81\pm0.15$ &                $ 0.35\pm0.21$ \\
   AzV~95 &                00:51:21.60 &               -72:44:14.88 &             $21.50\pm0.04$ &             $ 0.57\pm0.07$ &                    $-1.80$ &  [2] &                $ 0.59\pm0.07$ &                $-0.83\pm0.09$ &                $ 0.30\pm0.12$ \\
  AzV~104 &                00:51:38.43 &               -72:48:06.08 &             $21.46\pm0.06$ &             $ 0.51\pm0.16$ &                    $-1.93$ &  [2] &                $ 0.58\pm0.16$ &                $-0.92\pm0.19$ &                           --  \\
  AzV~207 &                00:58:33.19 &               -71:55:46.72 &             $21.44\pm0.06$ &             $ 0.83\pm0.06$ &                    $-1.74$ &  [2] &                $ 0.98\pm0.27$ &                $-0.45\pm0.35$ &                $ 0.29\pm0.50$ \\
  AzV~216 &                00:58:59.14 &               -72:44:34.13 &             $21.64\pm0.03$ &             $ 0.72\pm0.08$ &                    $-2.56$ &  [2] &                $ 0.73\pm0.08$ &                $-0.77\pm0.09$ &                           --  \\
  AzV~229 &                00:59:26.58 &               -72:09:53.93 &             $21.06\pm0.04$ &             $ 0.62\pm0.08$ &                    $-5.10$ &  [2] &                $ 0.75\pm0.07$ &                $-0.61\pm0.08$ &                $ 0.36\pm0.12$ \\
  AzV~242 &                01:00:06.88 &               -72:13:57.37 &             $21.32\pm0.04$ &             $ 1.16\pm0.18$ &                    $-3.81$ &  [2] &                $ 0.85\pm0.06$ &                $-0.62\pm0.09$ &                $ 0.37\pm0.11$ \\
  AzV~321 &                01:02:57.08 &               -72:08:09.11 &             $20.70\pm0.08$ &             $ 0.90\pm0.11$ &                    $-5.96$ &  [2] &                $ 0.87\pm0.07$ &                $-0.37\pm0.12$ &                $ 0.38\pm0.16$ \\
  AzV~332 &                01:03:25.23 &               -72:06:43.86 &             $20.54\pm0.16$ &             $ 0.56\pm0.06$ &                    $-5.74$ &  [2] &                $ 0.54\pm0.06$ &                $-0.47\pm0.17$ &                           --  \\
  AzV~388 &                01:05:39.53 &               -72:29:26.94 &             $21.19\pm0.04$ &             $ 0.99\pm0.07$ &                    $-1.49$ &  [2] &                $ 1.07\pm0.10$ &                $-0.55\pm0.12$ &                $ 0.43\pm0.16$ \\
  AzV~456 &                01:10:55.76 &               -72:42:56.22 &             $21.43\pm0.07$ &             $ 1.36\pm0.09$ &                    $-0.20$ &  [2] &                $ 0.95\pm0.17$ &                $-1.09\pm0.24$ &                $ 0.41\pm0.33$ \\
  AzV~476 &                01:13:42.45 &               -73:17:29.52 &             $21.95\pm0.08$ &             $ 0.81\pm0.17$ &                    $-0.70$ &  [2] &                $ 0.58\pm0.08$ &                $-1.23\pm0.11$ &                           --  \\
   Sk~190 &                01:31:27.98 &               -73:22:14.22 &             $20.62\pm0.04$ &             $ 0.00\pm0.00$ &                    $-2.90$ &  [2] &                $ 0.74\pm0.10$ &                $-0.55\pm0.12$ &                $ 0.50\pm0.17$ \\
   Sk~191 &                01:41:42.07 &               -73:50:38.18 &             $21.62\pm0.04$ &             $ 0.00\pm0.00$ &                    $-0.67$ &  [2] &                $ 1.12\pm0.11$ &                $-1.37\pm0.14$ &                $ 0.40\pm0.19$ \\
  AzV~238 &                00:59:55.51 &               -72:13:37.78 &             $21.41\pm0.21$ &             $ 0.89\pm0.04$ &                    $-5.16$ &  [3] &                $ 0.81\pm0.06$ &                $-0.89\pm0.22$ &                           --  \\
  AzV~327 &                01:03:10.55 &               -72:02:14.33 &             $21.62\pm0.22$ &             $ 0.83\pm0.19$ &                    $-6.53$ &  [3] &                $ 0.68\pm0.13$ &                $-1.55\pm0.25$ &                           --  \\
     Sk~9 &                00:46:32.63 &               -73:06:05.56 &             $21.76\pm0.21$ &             $ 2.17\pm2.10$ &                    $-4.43$ &  [3] &                $ 0.67\pm0.05$ &                $-0.96\pm0.21$ &                           --  \\
   Sk~116 &                01:04:55.74 &               -72:46:48.15 &             $21.57\pm0.21$ &             $ 0.54\pm0.04$ &                    $-2.74$ &  [3] &                $ 0.51\pm0.03$ &                $-1.00\pm0.21$ &                           --  \\
\hline
\hline
\end{tabular}
\label{tab SMC}
\end{table*}

\begin{landscape}
\begin{table}
\centering
\caption{The [$X$/Fe]$_{\rm nucl}$ due to nucleosynthesis, after correction for dust-depletion, in the LMC neutral ISM. These values are shown in Fig. \ref{fig [X/Fe]nucl LMC APOGEE}.}
\begin{tabular}{l | c c c c c c c c c c}
\hline
\hline
            ID 
  & [Mg/Fe]$_{\rm nucl}$ 
  & [Si/Fe]$_{\rm nucl}$ 
  & [Cr/Fe]$_{\rm nucl}$ 
  & [Ni/Fe]$_{\rm nucl}$ 
  & [Zn/Fe]$_{\rm nucl}$ 
  & [Ti/Fe]$_{\rm nucl}$ 
   & [O/Fe]$_{\rm nucl}$ 
   & [S/Fe]$_{\rm nucl}$ 
  & [Cu/Fe]$_{\rm nucl}$ 
   & [P/Fe]$_{\rm nucl}$ 
\\ \hline
         BI~173
       & $ 0.21\pm 0.12$ 
       & $ 0.32\pm 0.21$ 
       & $ 0.07\pm 0.08$ 
       & $-0.06\pm 0.08$ 
       & $ 0.08\pm 0.07$ 
                    & -- 
                    & -- 
       & $ 0.34\pm 0.08$ 
       & $ 0.19\pm 0.23$ 
       & $-0.34\pm 0.09$ 
 \\
         BI~184
       & $ 0.19\pm 0.11$ 
       & $ 0.14\pm 0.19$ 
       & $ 0.02\pm 0.15$ 
       & $ 0.05\pm 0.16$ 
       & $ 0.03\pm 0.11$ 
                    & -- 
                    & -- 
                    & -- 
                    & -- 
                    & -- 
 \\
         BI~237
       & $ 0.33\pm 0.11$ 
       & $ 0.09\pm 0.17$ 
       & $ 0.04\pm 0.16$ 
       & $-0.02\pm 0.15$ 
       & $-0.09\pm 0.11$ 
       & $ 0.31\pm 0.16$ 
                    & -- 
                    & -- 
                    & -- 
                    & -- 
 \\
         BI~253
       & $ 0.28\pm 0.16$ 
                    & -- 
       & $ 0.02\pm 0.19$ 
       & $-0.06\pm 0.19$ 
       & $-0.08\pm 0.12$ 
       & $ 0.27\pm 0.21$ 
                    & -- 
                    & -- 
                    & -- 
                    & -- 
 \\
      PGMW~3120
       & $ 0.13\pm 0.15$ 
       & $ 0.24\pm 0.27$ 
       & $ 0.08\pm 0.13$ 
       & $ 0.01\pm 0.12$ 
       & $ 0.06\pm 0.10$ 
                    & -- 
                    & -- 
                    & -- 
                    & -- 
                    & -- 
 \\
      PGMW~3223
       & $ 0.17\pm 0.13$ 
       & $ 0.13\pm 0.18$ 
       & $ 0.05\pm 0.12$ 
       & $-0.08\pm 0.10$ 
       & $-0.02\pm 0.08$ 
                    & -- 
                    & -- 
                    & -- 
                    & -- 
                    & -- 
 \\
       Sk-65~22
                    & -- 
       & $ 0.29\pm 0.14$ 
       & $ 0.01\pm 0.12$ 
       & $-0.04\pm 0.11$ 
       & $ 0.03\pm 0.08$ 
                    & -- 
                    & -- 
       & $ 0.52\pm 0.13$ 
                    & -- 
       & $-0.12\pm 0.10$ 
 \\
      Sk-66~172
       & $ 0.27\pm 0.16$ 
                    & -- 
       & $ 0.04\pm 0.14$ 
       & $ 0.03\pm 0.18$ 
       & $ 0.04\pm 0.09$ 
                    & -- 
                    & -- 
       & $ 0.30\pm 0.17$ 
                    & -- 
       & $-0.02\pm 0.12$ 
 \\
       Sk-66~19
       & $ 0.32\pm 0.21$ 
                    & -- 
       & $ 0.10\pm 0.22$ 
       & $-0.08\pm 0.19$ 
       & $-0.01\pm 0.17$ 
                    & -- 
                    & -- 
                    & -- 
                    & -- 
                    & -- 
 \\
       Sk-66~35
       & $ 0.18\pm 0.19$ 
       & $ 0.27\pm 0.20$ 
                    & -- 
       & $-0.10\pm 0.21$ 
       & $-0.14\pm 0.19$ 
                    & -- 
                    & -- 
       & $ 0.33\pm 0.18$ 
       & $ 0.25\pm 0.25$ 
                    & -- 
 \\
      Sk-67~101
                    & -- 
       & $ 0.06\pm 0.20$ 
       & $ 0.06\pm 0.24$ 
       & $-0.06\pm 0.24$ 
       & $ 0.01\pm 0.20$ 
                    & -- 
                    & -- 
       & $ 0.15\pm 0.16$ 
                    & -- 
                    & -- 
 \\
      Sk-67~105
       & $ 0.36\pm 0.21$ 
       & $ 0.04\pm 0.11$ 
       & $-0.01\pm 0.13$ 
       & $ 0.01\pm 0.14$ 
       & $-0.00\pm 0.10$ 
                    & -- 
       & $ 1.21\pm 0.21$ 
       & $ 0.27\pm 0.11$ 
                    & -- 
                    & -- 
 \\
       Sk-67~14
                    & -- 
       & $-0.19\pm 0.16$ 
                    & -- 
       & $ 0.08\pm 0.17$ 
                    & -- 
                    & -- 
                    & -- 
       & $ 0.07\pm 0.12$ 
                    & -- 
                    & -- 
 \\
      Sk-67~191
       & $ 0.17\pm 0.33$ 
       & $ 0.33\pm 0.31$ 
       & $-0.00\pm 0.38$ 
       & $-0.05\pm 0.37$ 
       & $-0.03\pm 0.32$ 
                    & -- 
                    & -- 
       & $ 0.52\pm 0.27$ 
                    & -- 
                    & -- 
 \\
        Sk-67~2
       & $ 0.33\pm 0.19$ 
                    & -- 
       & $-0.11\pm 0.28$ 
       & $-0.14\pm 0.25$ 
       & $ 0.17\pm 0.16$ 
       & $ 0.25\pm 0.21$ 
                    & -- 
       & $ 0.20\pm 0.21$ 
                    & -- 
                    & -- 
 \\
      Sk-67~211
       & $-0.02\pm 0.14$ 
       & $ 0.32\pm 0.21$ 
       & $-0.00\pm 0.12$ 
       & $ 0.02\pm 0.11$ 
       & $-0.17\pm 0.09$ 
       & $ 0.27\pm 0.11$ 
                    & -- 
       & $ 0.31\pm 0.09$ 
                    & -- 
       & $-0.12\pm 0.49$ 
 \\
        Sk-67~5
       & $ 0.35\pm 0.11$ 
       & $ 0.43\pm 0.08$ 
       & $ 0.03\pm 0.12$ 
       & $-0.14\pm 0.12$ 
       & $ 0.08\pm 0.07$ 
       & $ 0.37\pm 0.10$ 
                    & -- 
       & $ 0.36\pm 0.08$ 
                    & -- 
       & $-0.14\pm 0.10$ 
 \\
      Sk-68~129
       & $ 0.30\pm 0.17$ 
                    & -- 
       & $ 0.04\pm 0.23$ 
                    & -- 
       & $ 0.07\pm 0.20$ 
                    & -- 
                    & -- 
                    & -- 
                    & -- 
                    & -- 
 \\
      Sk-68~135
       & $ 0.28\pm 0.20$ 
                    & -- 
       & $ 0.09\pm 0.23$ 
       & $-0.07\pm 0.22$ 
       & $ 0.21\pm 0.16$ 
       & $ 0.29\pm 0.25$ 
                    & -- 
                    & -- 
                    & -- 
       & $-0.18\pm 0.17$ 
 \\
      Sk-68~140
       & $ 0.23\pm 0.22$ 
       & $ 0.23\pm 0.27$ 
       & $ 0.09\pm 0.18$ 
       & $ 0.09\pm 0.18$ 
       & $ 0.09\pm 0.14$ 
                    & -- 
                    & -- 
                    & -- 
                    & -- 
                    & -- 
 \\
      Sk-68~155
       & $ 0.26\pm 0.09$ 
       & $ 0.02\pm 0.14$ 
       & $ 0.04\pm 0.09$ 
       & $ 0.02\pm 0.12$ 
       & $ 0.06\pm 0.07$ 
                    & -- 
                    & -- 
                    & -- 
                    & -- 
                    & -- 
 \\
       Sk-68~26
       & $ 0.42\pm 0.19$ 
                    & -- 
       & $ 0.26\pm 0.26$ 
       & $ 0.09\pm 0.23$ 
       & $ 0.11\pm 0.21$ 
                    & -- 
                    & -- 
                    & -- 
                    & -- 
                    & -- 
 \\
       Sk-68~52
       & $-0.05\pm 0.15$ 
       & $-0.10\pm 0.15$ 
       & $ 0.16\pm 0.10$ 
       & $-0.05\pm 0.11$ 
       & $-0.43\pm 0.14$ 
       & $ 0.24\pm 0.10$ 
                    & -- 
       & $ 0.27\pm 0.08$ 
                    & -- 
                    & -- 
 \\
       Sk-68~73
       & $ 0.26\pm 0.22$ 
                    & -- 
       & $ 0.04\pm 0.26$ 
       & $-0.01\pm 0.24$ 
       & $-0.01\pm 0.20$ 
       & $ 0.32\pm 0.28$ 
       & $ 0.14\pm 0.21$ 
                    & -- 
       & $-0.04\pm 0.22$ 
                    & -- 
 \\
      Sk-69~104
                    & -- 
       & $ 0.43\pm 0.40$ 
                    & -- 
                    & -- 
       & $ 0.00\pm 0.42$ 
                    & -- 
                    & -- 
       & $ 0.01\pm 0.36$ 
                    & -- 
                    & -- 
 \\
      Sk-69~175
                    & -- 
       & $ 0.19\pm 0.26$ 
                    & -- 
       & $-0.20\pm 0.28$ 
       & $-0.05\pm 0.19$ 
                    & -- 
                    & -- 
       & $ 0.61\pm 0.18$ 
                    & -- 
                    & -- 
 \\
      Sk-69~246
       & $ 0.26\pm 0.15$ 
       & $ 0.14\pm 0.17$ 
       & $ 0.04\pm 0.15$ 
       & $-0.09\pm 0.14$ 
       & $-0.08\pm 0.10$ 
       & $ 0.18\pm 0.16$ 
                    & -- 
       & $ 0.11\pm 0.18$ 
                    & -- 
                    & -- 
 \\
      Sk-69~279
       & $ 0.38\pm 0.16$ 
       & $ 0.14\pm 0.15$ 
       & $ 0.08\pm 0.10$ 
       & $-0.11\pm 0.12$ 
       & $ 0.10\pm 0.08$ 
                    & -- 
                    & -- 
                    & -- 
                    & -- 
                    & -- 
 \\
      Sk-70~115
       & $ 0.21\pm 0.14$ 
       & $ 0.25\pm 0.15$ 
       & $ 0.02\pm 0.12$ 
       & $-0.09\pm 0.12$ 
       & $-0.05\pm 0.09$ 
       & $ 0.25\pm 0.14$ 
                    & -- 
       & $ 0.25\pm 0.10$ 
                    & -- 
       & $-0.23\pm 0.11$ 
 \\
       Sk-70~79
       & $ 0.05\pm 0.16$ 
                    & -- 
                    & -- 
       & $-0.15\pm 0.20$ 
       & $-0.05\pm 0.13$ 
                    & -- 
       & $ 0.36\pm 0.21$ 
       & $-0.09\pm 0.19$ 
                    & -- 
                    & -- 
 \\
       Sk-71~45
       & $ 0.26\pm 0.20$ 
       & $ 0.27\pm 0.16$ 
       & $ 0.04\pm 0.10$ 
       & $-0.08\pm 0.10$ 
       & $ 0.04\pm 0.08$ 
                    & -- 
                    & -- 
                    & -- 
                    & -- 
       & $-0.14\pm 0.10$ 
 \\
       Sk-71~50
       & $-0.04\pm 0.23$ 
       & $ 0.12\pm 0.12$ 
       & $ 0.02\pm 0.09$ 
       & $-0.12\pm 0.09$ 
       & $-0.05\pm 0.07$ 
                    & -- 
                    & -- 
       & $ 0.19\pm 0.15$ 
       & $ 0.10\pm 0.20$ 
                    & -- 
 \\
\hline
\hline
\end{tabular}
\label{tab LMC XFe_nucl}
\end{table}
\end{landscape}

\begin{landscape}
\begin{table}
\centering
\caption{Same as Table \ref{tab LMC XFe_nucl}, but for the SMC.}
\begin{tabular}{l | c c c c c c c c }
\hline
\hline

            ID 
  & [Mg/Fe]$_{\rm nucl}$ 
  & [Si/Fe]$_{\rm nucl}$ 
  & [Cr/Fe]$_{\rm nucl}$ 
  & [Ni/Fe]$_{\rm nucl}$ 
  & [Zn/Fe]$_{\rm nucl}$ 
  & [Ti/Fe]$_{\rm nucl}$ 
   & [S/Fe]$_{\rm nucl}$ 
  & [Mn/Fe]$_{\rm nucl}$ 
\\ \hline
         AzV~18
 & $ 0.38\pm 0.12$
             & -- 
 & $ 0.10\pm 0.16$
 & $-0.07\pm 0.14$
 & $-0.07\pm 0.12$
 & $ 0.38\pm 0.16$
             & -- 
 & $-0.47\pm 0.12$
 \\
         AzV~26
 & $ 0.14\pm 0.09$
 & $ 0.29\pm 0.09$
 & $ 0.04\pm 0.10$
 & $-0.05\pm 0.09$
 & $ 0.03\pm 0.08$
 & $ 0.23\pm 0.10$
             & -- 
 & $-0.37\pm 0.08$
 \\
         AzV~47
 & $ 0.28\pm 0.15$
 & $ 0.22\pm 0.10$
 & $ 0.05\pm 0.12$
 & $-0.08\pm 0.11$
 & $ 0.02\pm 0.10$
 & $ 0.28\pm 0.12$
 & $ 0.35\pm 0.11$
 & $-0.35\pm 0.10$
 \\
         AzV~78
 & $ 0.24\pm 0.19$
             & -- 
 & $ 0.30\pm 0.21$
 & $-0.08\pm 0.20$
             & -- 
 & $ 0.31\pm 0.23$
             & -- 
 & $-0.35\pm 0.18$
 \\
         AzV~80
 & $ 0.35\pm 0.16$
             & -- 
 & $ 0.04\pm 0.18$
 & $-0.06\pm 0.18$
 & $-0.15\pm 0.13$
 & $ 0.38\pm 0.20$
             & -- 
 & $-0.43\pm 0.15$
 \\
         AzV~95
 & $ 0.19\pm 0.17$
 & $ 0.32\pm 0.10$
 & $ 0.07\pm 0.12$
 & $-0.10\pm 0.11$
 & $-0.03\pm 0.10$
 & $ 0.30\pm 0.13$
 & $ 0.31\pm 0.11$
 & $-0.37\pm 0.10$
 \\
        AzV~104
 & $ 0.43\pm 0.23$
 & $ 0.38\pm 0.24$
 & $-0.04\pm 0.30$
 & $-0.07\pm 0.28$
 & $-0.07\pm 0.25$
             & -- 
 & $ 0.32\pm 0.21$
             & -- 
 \\
        AzV~207
             & -- 
 & $ 0.29\pm 0.41$
 & $ 0.05\pm 0.46$
             & -- 
 & $-0.15\pm 0.31$
 & $ 0.34\pm 0.51$
             & -- 
 & $-0.40\pm 0.39$
 \\
        AzV~216
 & $ 0.39\pm 0.12$
             & -- 
 & $ 0.02\pm 0.13$
 & $-0.03\pm 0.12$
 & $-0.01\pm 0.11$
             & -- 
             & -- 
             & -- 
 \\
        AzV~229
 & $ 0.32\pm 0.12$
 & $ 0.39\pm 0.09$
 & $ 0.11\pm 0.13$
 & $ 0.10\pm 0.10$
 & $-0.14\pm 0.10$
 & $ 0.42\pm 0.11$
 & $ 0.41\pm 0.08$
 & $-0.29\pm 0.09$
 \\
        AzV~242
 & $ 0.34\pm 0.12$
 & $ 0.36\pm 0.09$
 & $-0.09\pm 0.23$
 & $ 0.01\pm 0.10$
 & $ 0.31\pm 0.19$
 & $ 0.38\pm 0.10$
 & $ 0.39\pm 0.09$
 & $-0.42\pm 0.09$
 \\
        AzV~321
             & -- 
 & $ 0.32\pm 0.12$
             & -- 
 & $ 0.15\pm 0.16$
 & $ 0.02\pm 0.13$
 & $ 0.41\pm 0.11$
 & $ 0.42\pm 0.09$
 & $-0.31\pm 0.15$
 \\
        AzV~332
             & -- 
 & $ 0.39\pm 0.10$
 & $ 0.02\pm 0.11$
 & $-0.14\pm 0.15$
 & $ 0.02\pm 0.09$
             & -- 
 & $ 0.41\pm 0.10$
 & $-0.22\pm 0.10$
 \\
        AzV~388
 & $ 0.60\pm 0.16$
 & $ 0.38\pm 0.12$
             & -- 
             & -- 
 & $-0.08\pm 0.11$
 & $ 0.45\pm 0.17$
 & $ 0.44\pm 0.10$
 & $-0.37\pm 0.14$
 \\
        AzV~456
             & -- 
 & $ 0.41\pm 0.23$
             & -- 
             & -- 
 & $ 0.41\pm 0.20$
 & $ 0.41\pm 0.32$
             & -- 
 & $-0.25\pm 0.25$
 \\
        AzV~476
             & -- 
             & -- 
 & $ 0.03\pm 0.14$
 & $-0.10\pm 0.15$
 & $ 0.23\pm 0.19$
 & $ 0.24\pm 0.15$
             & -- 
 & $-0.33\pm 0.11$
 \\
         Sk~190
             & -- 
 & $ 0.61\pm 0.15$
             & -- 
 & $ 0.09\pm 0.19$
             & -- 
 & $ 0.49\pm 0.20$
 & $ 0.47\pm 0.12$
             & -- 
 \\
         Sk~191
             & -- 
 & $ 0.37\pm 0.15$
             & -- 
             & -- 
             & -- 
 & $ 0.41\pm 0.20$
 & $ 0.43\pm 0.14$
 & $-0.29\pm 0.17$
 \\
        AzV~238
             & -- 
 & $ 0.35\pm 0.06$
 & $ 0.09\pm 0.11$
             & -- 
 & $ 0.07\pm 0.07$
             & -- 
             & -- 
             & -- 
 \\
        AzV~327
             & -- 
 & $ 0.66\pm 0.17$
 & $-1.20\pm 1.04$
             & -- 
 & $ 0.15\pm 0.24$
             & -- 
             & -- 
             & -- 
 \\
           Sk~9
             & -- 
 & $ 0.20\pm 2.10$
 & $ 0.00\pm 2.10$
             & -- 
 & $ 1.50\pm 2.10$
             & -- 
             & -- 
             & -- 
 \\
         Sk~116
             & -- 
 & $ 0.30\pm 0.06$
 & $ 0.00\pm 0.06$
             & -- 
 & $ 0.03\pm 0.05$
             & -- 
             & -- 
             & -- 
 \\

\hline
\hline
\end{tabular}
\label{tab SMC XFe_nucl}
\end{table}
\end{landscape}

\appendix

\section{On the use of [Zn/Fe] as dust tracer}

The main processes that shape the abundance patterns in the neutral ISM are visualized in Figure \ref{fig abu}. The different metals are used to determine the chemical properties of the ISM (see Sect. \ref{sec comparison}). The $x$-axis there refers to a general refractory index, which is a way to measure how heavily each element is typically depleted into dust. In our analysis, we use the $B2_X$ coefficients for this purpose, which are determined by \citet{DeCia16} and \citet{Konstantopoulou22} preferentially using [Zn/Fe] to trace the amount of dust in the neutral ISM. The fact that the observed abundances in the neutral ISM overall line up linearly in the observed abundance patterns both in the Milky Way \citep{DeCia21} and the Magellanic Clouds (Figs. \ref{fig xy LMC} and \ref{fig xy SMC}) indicates that the values of $B2_X$ are, to a first approximation, reliable. The metallicities resulting from the use of the $B2_X$ coefficients (i.e. the 'relative method' based on the relative abundances of metals) do overall agree with those resulting from an analogous but independent method that does not make preferential use of [Zn/Fe] \citep{DeCia21}, which is based on $F*$ to trace the strength of dust depletion \citep{Jenkins09}. 

The reasons for the choice of [Zn/Fe] to trace dust is that Zn and Fe are depleted into dust grains in very different ways, they are often well observed, and their nucleosynthesis overall follow each other in the metallicity regime of interest. This last point is crucial, specially given that stellar observations of [Zn/Fe] can show large scatter and deviations from solar values.

The overall distribution of [Zn/Fe] with [Fe/H] in stars in the Milky Way is rather flat and near solar, in the metallicity range $-2.0 < $ [Fe/H] $ < +0.5$ and strongly deviates from solar at lower and higher metallicities \citep[e.g.][]{Sneden91,Primas00,Nissen07,Saito09}. Recently, \citet{Sitnova22} study the distribution of [Zn/Fe] with [Fe/H] in Milky Way stars (in the halo, thick and thin disk) in the metallicity regime $-2.0 < $ [Fe/H] $ < +0.2$, using non-LTE measurements, and find [Zn/Fe] enhancement of up to 0.2~dex in low-metallicity stars (in the halo) and under-abundance of down to $-0.2$~dex in high-metallicity stars (in the thin disk). We explore this possibility further in Sect. \ref{sec caveats}. Stellar abundances of stars younger than a few Gyrs in the thin disk of the Milky Way show also a large scatter in [Zn/Fe], and values down to $-0.2$~dex at solar and super-solar metallicities \citep{DelgadoMena19}. The stellar [Zn/Fe] in halo stars in the Milky Way is close to zero for metallicities in the range $-2.0 < $ [Fe/H] $ < -1.0$ and increases to higher values (by 0.1--02~dex) at lower metallicities, and up to $+0.5$~dex at [Fe/H] $\sim -3$ \citep{Nissen07}. In the bulge of the Milky Way, a gas-poor environment with no star-formation activity, the values of [Zn/Fe] go down to $-0.6$~dex at super-solar metallicities \citep{Barbuy15,Duffau17,daSilveira18}. In nearby dwarf spheroidal galaxies stellar measurements of [Zn/Fe] also show a large scatter, with value down to about $-1$~dex \citep{Skuladottir17}. Dwarf spheroidals are fundamentally different from DLAs because they do not currently have gas and should have lost their gas content by $z\sim2.3$ \citep{Salvadori12}.

In the neutral ISM in local and distant galaxies there is no evidence so far for large intrinsic (nucleosynthetic) deviations of [Zn/Fe]. The existence of depletion sequences with scatter $<0.2$~dex \citep{DeCia16,Konstantopoulou22} limits the possible [Zn/Fe] deviations due to nucleosynthesis. We speculate that any Zn deviations due to nucleosynthesis should be lower than $0.2$~dex. In particular, in DLAs we observe values down to [Zn/Fe] $=0$, and never strongly negative values. In this regime, dust depletion is negligible. When [Zn/Fe] $=0$, DLAs show enhancement of $\alpha$ elements of about 0.3~dex, and under-abundance of Mn by about 0.3~dex \citep{DeCia16}, similar to the [$X$/Fe] expected from observations in the Local Group and chemical evolution of galaxies (see Sect. \ref{sec alpha}). In addition, [Zn/Fe] correlates well with [Si/Ti], [O/Si], and other dust tracers. This excludes strong deviations of Zn from Fe due to nucleosynthesis. More details about why [Zn/Fe] can be considered a reliable dust tracer, to a first approximation, are discussed in \citet{DeCia18a}. Future studies of dust depletion based on relative abundances may be able to expand the framework and avoid any preferential use of [Zn/Fe] as a dust tracer.

\section{Choosing different sets of metals in the analysis of the abundance patterns}
\label{sec comparison}

\begin{figure}
   \centering
   \includegraphics[width=0.5\textwidth]{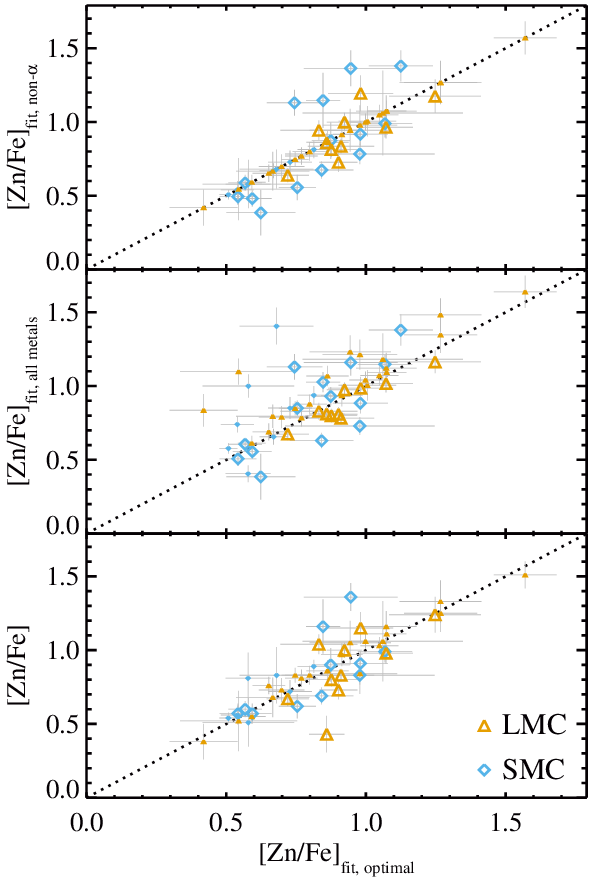}
      \caption{Comparison between the overall strength of the dust depletion [Zn/Fe]$_{\rm fit}$ from the fit to the abundance patterns for the optimal choice of metals (see Sect. \ref{sec method}) with the observed [Zn/Fe] (bottom panel), the fit to all metals (middle panel), and the fit to only the non-$\alpha$ elements. Orange triangles show the measurements for the LMC and blue diamonds those for the SMC. The dotted curve is the $x=y$ line. Larger symbols highlight the golden sample.}
         \label{fig [Zn/Fe] comparison}
   \end{figure}

\begin{figure}
   \centering
   \includegraphics[width=0.5\textwidth]{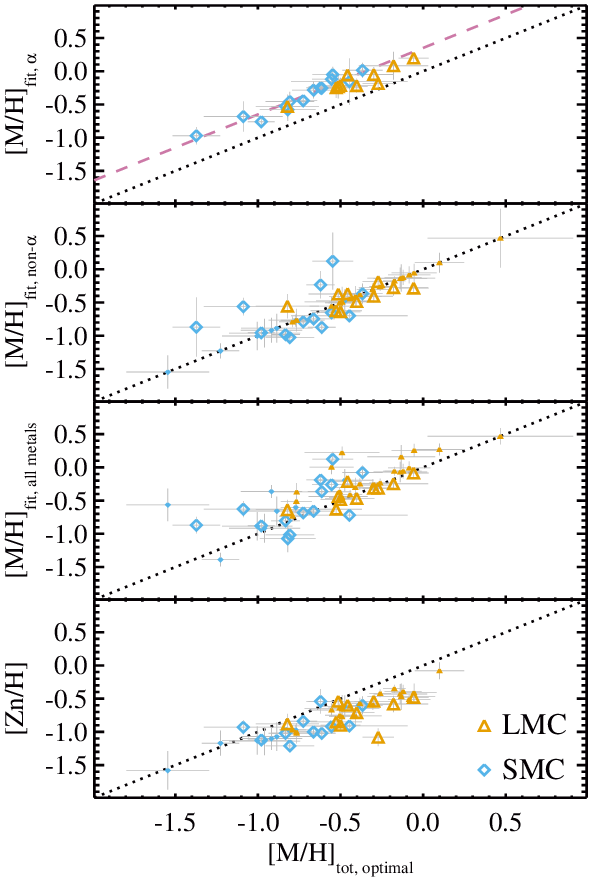}
      \caption{Comparison between the total metallicity [M/H]$_{\rm tot}$ from the fit to the abundance patterns for the optimal choice of metals (see Sect. \ref{sec method}) with the observed [Zn/H] (bottom panel), the fit to all metals (third panel), and the fit to only the non-$\alpha$ elements (second panel) and the fit to only the $\alpha$ elements (top panel). Orange triangles show the measurements for the LMC and blue diamonds those for the SMC. The dotted curve is the $x=y$ line. The dashed line on the top panel is not a fit to the data, but marks the $x=y+0.35$~dex line, the enhancement in the abundances of $\alpha$ elements. Larger symbols highlight the golden sample.}
         \label{fig [M/H] comparison}
   \end{figure}

We compare the results obtained from different choices of metals as reference for the linear fit to the abundance patterns.   
First, we use all available metals except Mn, and derive the metallicity $[{\rm M/H}]_{\rm tot, all\,metals}$ and strength of dust depletion $[{\rm Zn/Fe}]_{\rm fit, all\,metals}$. This basic approach already shows $\alpha$-element enhancement and Mn under-abundance  in several systems. The choice of the exclusion of Mn is driven by the Mn under-abundance observed in the ISM of the Magellanic Clouds \citep{Jenkins17,Konstantopoulou22}. Indeed, including Mn in the fit to the abundance patterns produces sometimes non-physical negative slopes for cases when not many relatively volatile elements are available. Second, we consider the fit to all metals but excluding all the $\alpha$ elements and Mn, to avoid misinterpreting the effects of dust, nucleosynthesis, and metallicity. We denote the resulting metallicity and strength of depletion as $[{\rm M/H}]_{\rm tot, non-\alpha}$ and $[{\rm Zn/Fe}]_{\rm fit, non-\alpha}$, respectively, and use them as reference. In the three cases where neither Zn nor other non-$\alpha$ volatile elements are constrained (namely, SK-67 14, SK~190 and SK~191), we consider the fit to all metals including a posteriori an additional 0.32~dex\footnote{The 0.32~dex is chosen here to be in-between the typical values of $\alpha$-element enhancement in the Milky Way \citep[$\sim0.35$~dex][]{McWilliam97} and DLAs \citep[$\sim0.3$~dex][]{DeCia16}.} in the uncertainty on their total metallicity to account for the paucity of data and potential nucleosynthesis effects causing over- or underestimate of the metallicity. Third, we include only the $\alpha$ elements, with the exception of O, and linearly fit the abundance patterns only for systems that have a constrained measurement of Ti, to ensure a large enough dynamical range in the refractory index. This is possible for a limited number of systems (the golden sample, with larger symbols in Figs. \ref{fig [Zn/Fe] comparison} and \ref{fig [M/H] comparison}.). We denote this resulting metallicity as $[{\rm M/H}]_{\rm tot, \alpha}$ and the strength of dust depletion as $[{\rm Zn/Fe}]_{\rm fit, \alpha}$. Oxygen is observed in only three LMC systems, and its exclusion from the fit is driven by the potential presence of ISM mixture at least in one of them, which we discuss in Section \ref{sec ISM mix}. 

The reference values $[{\rm Zn/Fe}]_{\rm fit}$ and $[{\rm M/H}]_{\rm tot}$ are those from the optimal approach, which is described in Sect. \ref{sec method}.  Here we compare these with the results from the same analysis but using three different sets of metals: only the $\alpha$ elements, only the non-$\alpha$ elements, and all the metals. Figure \ref{fig [Zn/Fe] comparison} shows a comparison between the strength of depletion $[{\rm Zn/Fe}]_{\rm fit}$ (i.e. the slope of the linear fits to the abundance patterns) for the optimal case with the case where only the non-$\alpha$ elements were used (top panel), or all metals (mid panel), or the observed [Zn/Fe] (bottom panel). The optimal $[{\rm Zn/Fe}]_{\rm fit}$ correlates reasonably well with all the three quantities, with notable exceptions.

In a similar way, Figure \ref{fig [M/H] comparison} shows a comparison between the total metallicity $[{\rm M/H}]_{\rm tot}$ (i.e. the normalization of the linear fits to the abundance patterns) for the optimal case and the case where only the $\alpha$ elements were used (top panel), or only the non-$\alpha$ elements (second panel), or all metals (third panel), or the observed [Zn/H] (bottom panel). Clearly, the observed [Zn/H] underestimates the actual metallicity, and this is likely because of dust depletion. The values of $[{\rm M/H}]_{\rm tot, non-\alpha}$ and $[{\rm M/H}]_{\rm tot, all\,metals}$ correlate well with the optimal total metallicity, however with some deviations. The values of $[{\rm M/H}]_{\rm tot, \alpha}$ are mostly higher than the optimal metallicity, and this is likely due to $\alpha$-element enhancement. The dashed purple line on the top panel of Fig. \ref{fig [M/H] comparison} is not a fit to the data, but the $x=y$ line increased by 0.35~dex, as a proxy for the overall level of $\alpha$-element enhancement in the Milky Way \cite[e.g.][]{Lambert87}. Overall, because of the likely $\alpha$-element enhancement, it is not optimal to choose derive the metallicity from the fit to the abundance patterns of the $\alpha$ elements only. Indeed, we refer to the total metallicity $[{\rm M/H}]_{\rm tot}$ as the metallicity of the Fe-peak elements, after dust correction.

\end{document}